\documentclass[aps, prb, twocolumn, amsmath, amssymb, nofootinbib]{revtex4-2}
\usepackage{hyperref}
\hypersetup{hidelinks}
\usepackage{bm}
\usepackage{xcolor}
\usepackage{amsmath}
\usepackage{amsfonts}
\usepackage{dsfont}
\usepackage{graphics}
\usepackage[caption=false]{subfig}
\usepackage{subfig}           
\usepackage{overpic}
\usepackage[capitalise]{cleveref}
\usepackage{tikz}
\usepackage{simpler-wick}
\usepackage{physics}
\usepackage{braket}
\usepackage{ulem}

\begin{document}
\title{Topological effect on order-disorder transitions in U(1) sigma models}
\author{Ryuichi Shindou}
\affiliation{International Center for Quantum Materials, Peking University, Beijing 100871, China}
\author{Pengwei Zhao}
\affiliation{International Center for Quantum Materials, Peking University, Beijing 100871, China}
\date{\today}
\begin{abstract}
U(1) non-linear sigma model (NLSM) with a one-dimensional (1D) Berry phase is studied by a renormalization group theory. Order-disorder transition in U(1) NLSMs in $D \!\ (\ge 2)$-dimensional space ($d+1$-dimensional spacetime, $d\ge 1$) is instigated by the proliferation of vortex excitations, where the 1D Berry phase term confers finite phase factors upon those vortex excitations that have finite projection in a subspace complementary to a topological direction with the 1D Berry phase. Due to a destructive interference effect caused by the phase factors, a partition function near the order-disorder transition point can be dominated by vortex excitations polarized along the topological direction with the Berry phase. The proliferation of the polarized vortex excitations helps to develop an extremely anisotropic correlation of the order parameter, which has a divergent correlation length along the topological direction with the Berry phase, and a finite correlation length along the other directions. In order to explore such a possibility in $D=3$, we develop a perturbative renormalization group theory of a 3D model of vortex loops, in which loop segments interact via a $1/r$ Coulomb interaction, and the 1D Berry phase confers the phase factor upon each vortex loop. We derive renormalization group (RG) equations among vortex-loop fugacity, Berry phase term, and the Coulomb potential. The RG equations analyzed with approximations show that a characteristic size of the vortex loop along the topological direction becomes anomalously large near an order-disorder transition point, while the characteristic loop size within the other directions remains finite. Utilizing a duality mapping to a lattice model of a type-II superconductor under a magnetic field, we also argue that a global phase diagram of the 3D U(1) sigma model with 1D Berry phase should have an intermediate quasi-disordered phase between ordered and disordered phases.   
\end{abstract}

\maketitle
\section{Introduction}
An U(1) nonlinear sigma model (NLSM) with a Berry phase term is an effective continuum model that describes phase transitions associated with spontaneous symmetry breaking of a global U(1) symmetry,
\begin{eqnarray}
Z = \int D\theta(x) e^{- \int d^Dx \!\ \Big\{ \frac{1}{2T}
{\bm \nabla} \theta(x) \cdot {\bm \nabla} \theta(x) + i\chi \nabla_{0} \theta(x) \Big\}}, \label{Eq:I-1}
\end{eqnarray}
with $x\equiv(x_1,\cdots, x_d,x_0)^T$, $D \equiv d+1$, and a U(1) phase variable $\theta(x) \in [0,2\pi)$. Relevant phase transitions and physical systems include superfluid systems~\cite{popov1973,wiegel1973,banks1977,nelson1981,williams1987}, three-dimensional (3D) type-II superconductors under an external magnetic field~\cite{einhorn1978,peskin1978,savit1980,dasgupta1981,blatter1994,nguyen1999,nguyen1998a,nguyen1998b,ryu1998}, and disordered systems of free particles in Hermitian chiral symmetry classes~\cite{gadeAndersonLocalizationSublattice1993,gadeReplicaLimitModels1991,konigMetalinsulatorTransitionTwodimensional2012,altlandSpectralTransportProperties2001,altlandQuantumCriticalityQuasiOneDimensional2014,altlandTopologyAndersonLocalization2015,luoCriticalBehaviorAnderson2020,wang2021,luoUnifyingAndersonTransitions2022,karcher2023a,karcher2023b,mondragon-shemTopologicalCriticalityChiralSymmetric2014,claes2020,xiaoAnisotropicTopologicalAnderson2023,pwz2024}. The  NLSM in $D=d+1$ dimension represents a zero-temperature partition function of $d$-dimensional superfluid systems where superfluid phase $\theta(x)$ fluctuates in space $x_{\perp} \equiv (x_1,\cdots,x_d)$ and imaginary time $x_0$, while superfluid amplitude is constrained around a finite value by certain physical means. Thereby, the Berry phase term $\chi$ along the imaginary time direction originates from a quantum-mechanical commutation relation between superfluid amplitude and phase~\cite{herbutModernApproachCritical2007,tanakaShortGuideTopological2015}. Previous studies on the superfluid systems discovered  Bose glass~\cite{fisher1989,fallani2007,choi2016} and Griffiths~\cite{griffiths1969,bray1987,dsfisher1992,Vojta2006,vazquez2011} phases. In these glassy phases, the superfluid correlation time is divergent, while the superfluid correlation length remains finite.

A dual lattice model of the U(1) NLSM in $D=3$ portrays magnetostatics of the 3D type-II superconductors in the $x_1$-$x_2$-$x_0$ space, where the order and disordered phases of the U(1) NLSM describe normal (``Maxwell") and superconducting (``Meissner") phases, respectively~\cite{einhorn1978,peskin1978,savit1980,dasgupta1981,kiometzis95,kiometzis1994,nguyen1999,herbutModernApproachCritical2007}. Thereby, the Berry phase term $\chi$ in the NLSM originates from an external magnetic field applied along the $x_0$ direction in the superconductors. The dual description identifies a correlation function of the U(1) phase variable $e^{i\theta({x})}$ in the NLSM with a correlation function of magnetic monopole fields in the superconductors. It has been known that the type-II superconductors under the magnetic field have novel mixed phases, such as 3D vortex lattice or vortex liquid phases. In these mixed phases, the monopole-field correlation function has a divergent correlation length along the field ($x_0$) direction, and a finite correlation length along the others ($x_1$ and $x_2$) directions.

The U(1) NLSM in $D=d+1$ dimension is also relevant to the $D$-dimensional Anderson transition of disordered Hermitian in the chiral symmetric classes~\cite{gadeAndersonLocalizationSublattice1993,gadeReplicaLimitModels1991,konigMetalinsulatorTransitionTwodimensional2012,karcher2023a,karcher2023b}. The ordered and disordered phases in the U(1) NLSM correspond to metal and localized phases in  the Anderson transition of the chiral symmetric models, respectively. Thereby, the Berry phase term $\chi$ originates from one-dimensional (1D) band topology along the $x_0$ direction in the chiral symmetric Hamiltonians~\cite{altlandQuantumCriticalityQuasiOneDimensional2014,altlandTopologyAndersonLocalization2015,mondragon-shemTopologicalCriticalityChiralSymmetric2014,claes2020}, and such Hamiltonians can be realized in semimetal models~\cite{luoCriticalBehaviorAnderson2020,xiaoAnisotropicTopologicalAnderson2023,pwz2024}. Recent numerical studies on the semimetal models in $D=2,3$ clarified that the 1D topology universally induces a quasi-localized phase between metal and Anderson localized phases~\cite{xiaoAnisotropicTopologicalAnderson2023,pwz2024}. In the quasi-localized phase, an exponential localization length is divergent along the 1D topological ($x_0$) direction, while it is finite along the other directions.

A recent theory discussed the universal emergence of the quasi-localized phase in the 2D disordered chiral symmetric systems through the lens of the NLSM with the 1D Berry phase term~\cite{pwz2024}. The 2D Anderson transition in the chiral symmetry classes is driven by a spatial proliferation of vortex-antivortex pairs associated with a U(1) phase $\theta(x)$ degree of freedom of a $Q$-field in a matrix-formed NLSM~\cite{konigMetalinsulatorTransitionTwodimensional2012,kosterlitz1973,kosterlitz1974,Berezinskii71,Berezinskii72}. The 1D band topology along the $x_0$ direction confers a complex phase factor upon such pairs, and the phase for each pair is proportional to a projected length $m_1$ of a dipole vector $m \equiv (m_0,m_1)$ of the pair along the $x_1$ direction~\cite{altlandSpectralTransportProperties2001,altlandQuantumCriticalityQuasiOneDimensional2014,tanakaShortGuideTopological2015,pwz2024}. In a partition function, such phase factor induces destructive interference among those pairs with a finite angle against the $x_0$ axis and different dipole lengths $|m|$. Consequently, the partition function near the transition point is expected to be dominated by vortex-antivortex pairs polarized along the $x_0$ axis ($m_1=0$). An introduction of a vortex-antivortex pair generally adds on a relative U(1) phase $\delta\theta(y,z) \equiv \theta(y)-\theta(z)$ between two spatial points, $y$ and $z$. Such add-on phase winds up the $2\pi$ phase when the dipole length $|m|$ of the pair with a perpendicular geometry  $m\perp y-z$ changes from 0 to $\infty$. For the other geometry ($m\parallel y-z$), the additive phase does not. This indicates that the proliferation of the polarized pairs results in an emergence of the quasi-localized phase where a correlation function $\langle e^{i\theta(x)-i\theta(y)}\rangle$ of the U(1) phase has a divergent correlation length along the topological ($x_0$) direction, and finite correlation length along the other ($x_1$) direction. To uphold this physical picture, the theory further studied a U($N$) NLSM for chiral unitary classes with the 1D weak topology, and derived RG equations among vortex fugacity, 1D topology parameter, and conductivities. The RG equations have a stable fixed point, and a stable fixed region. The stable point is characterized by divergent vortex fugacity, {\it finite} conductivity along the topological ($x_0$) direction, and vanishing conductivity along the other ($x_1$) direction.  The stable region is characterized by vanishing vortex fugacity, and finite conductivities along both spatial directions.  These two describe the quasi-localized phase and (critical) metal phase, respectively. The paper shows a direct transition between these two phases, supporting the universal emergence of the quasi-localized phase next to the (critical) metal phase in the 2D chiral unitary class.

        An order-disorder transition of the U(1) NLSM in general $D$ dimension is also induced by the spatial proliferation of vortex excitations~\cite{popov1973,wiegel1973,savit1978,einhorn1978,peskin1978,savit1980,dasgupta1981,nguyen1999,williams1987,shenoy1989,herbutModernApproachCritical2007}. Such vortex excitation takes the form of a closed loop of a vortex line in $D=3$, and a closed surface of a vortex sheet in $D=4$. The Berry phase term along $x_0$ direction generally endows each of these vortex excitations with a complex phase factor. For $D=3$ and $4$, the phase is proportional to an area inside the vortex loop projected on the $x_1$-$x_2$ plane, and a volume inside the vortex surface projected on the $x_1$-$x_2$-$x_3$ space, respectively~\cite{tanakaShortGuideTopological2015}. Such a complex phase factor induces destructive quantum interference among vortex excitations with finite projections in a space complementary to the $x_0$ axis. Meanwhile, vortex excitations with no projection in the complementary subspace are free from the destructive interference effect, dominating the partition function near the transition point. As in $D=2$, the proliferation of such polarized vortex excitations may result in the emergence of a quasi-disordered phase with a divergent correlation length along the $x_0$ direction, and finite correlation length along the others.     

\subsection{highlight of this paper}
       In order to establish a theory of such quasi-disordered phases in $D=3$, this paper aims to 
develop a renormalization group (RG) theory of the $D=3$ U(1) NLSM with the 1D Berry phase. We first introduce a model of the vortex loops, where loop segments interact with each other through the $1/|r|$ Coulomb interaction~\cite{popov1973,wiegel1973,savit1978,peskin1978,fradkin1978}, 
\begin{align}
S =\frac{\pi}{2T} \sum_{j,m}\oint_{\Gamma_j} \oint_{\Gamma_m} \frac{dx^j \cdot dy^m}{|x^j-y^m|} 
+ 2\pi i \!\ \chi \sum_j S^{j}_{12}. 
\end{align}
Here $dx^j$ and $dy^m$ are tangential vectors of the $j$th and $m$th vortex loops $\Gamma_j$ and $\Gamma_m$ at $x^j$ and $y^m$, respectively. The 1D Berry phase term endows each vortex loop with a complex phase factor proportional to its projected area onto the 2D plane complementary to the topological ($x_0$) direction: $S^j_{12}$ is the projected area of the $j$th vortex loop onto the $1$-$2$ plane. 
 
\begin{figure}[t]
\centering
    \includegraphics[width= 1.0\linewidth]{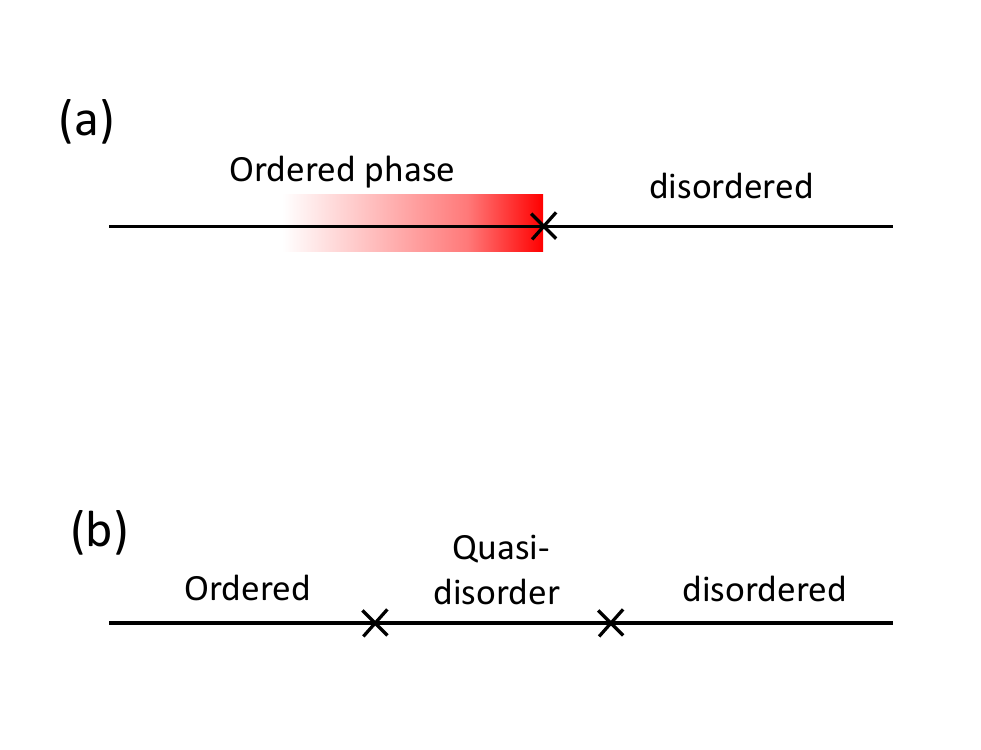} 
\caption{Schematic phase diagrams of the U(1) sigma model with 1D Berry phase term; (a) phase diagram obtained from the RG analysis in 
Section IV; (b) phase diagram deduced from the duality argument in Section V. In (a), the red shaded area with gradient in the ordered phase represents that the characteristic size of vortex loops in the topological ($x_0$) direction diverges toward the order-disorder transition point [denoted 
by the cross mark], while a size of vortex loops in the other directions remain finite. The phase diagram in (b) has an intermediate quasi-disorder phase between ordered and disordered phases, in which the correlation function of the U(1) phase variable has a divergent correlation length along the topological direction, and it has a finite correlation length along the other directions.}
\label{Fig0}
\end{figure}

    By recursively accounting for the screening effect of smaller vortex loops onto the interaction between larger vortex loops, we derive RG equations among a vortex fugacity term, Berry phase term, and Coulomb interaction potential. We clarify that the complex phase factor in the action generally suppresses the screening effect of {\it unpolarized} vortex loops, while leaving intact the screening effect of {\it polarized} vortex loops. As a result, the screening effect of the smaller loops generates two other types of {\it anisotropic} Coulomb interactions, 
\begin{align}
\delta S & = \sum_{j,m}\oint_{\Gamma_j} \oint_{\Gamma_m} dx^j \cdot {\cal F}_2(x^j-y^m)
\cdot dy^m  \nonumber \\
& \ \ + \sum_{j,m} \oint_{\Gamma_j} \oint_{\Gamma_m} dx^j \cdot {\cal G}(x^j-y^m)
\cdot dy^m.  
\end{align}
These interactions are characterized by 3 by 3 matrices in the tangential vector space,
\begin{align}
\begin{aligned}
&{\cal F}_{2}(r) = \frac{{\cal A}_2}{|r|} \left(\begin{array}{ccc}
1 & & \\
& 1 & \\
& & -2 \\
\end{array}\right) \equiv \frac{{\cal A}_2}{|r|} \eta_2,  \\
&{\cal G}(r) 
=   {\cal A}_0 \!\ \eta_2 \cdot \bigg(\frac{1}{|r|} - \frac{r r^T}{|r|^3} \bigg) \cdot \eta_2,
\end{aligned} \label{Eq:I-2}
\end{align}
with ${\cal A}_2>0$ and ${\cal A}_0 < 0$. Notably, both of these interactions help the longer vortex loops to be confined in planes parallel to the topological ($x_0$) direction. Namely, due to the traceless diagonal matrix $\eta_2$,  the short-distance part of the ${\cal F}_2(r)$ interaction with the positive ${\cal A}_2$ favors straight vortex lines polarized along the $x_0$ axis, while the ${\cal G}(r)$ interaction with the negative ${\cal A}_0$  -- dipole-dipole interaction modulated by $\eta_2$ -- favors vortex loops curving within the $x_1$-$x_0$ or $x_2$-$x_0$ planes over vortex loops curving within the $x_1$-$x_2$ plane. Upon the recursive inclusion of the screening effect of the smaller loops, ${\cal F}_1(r)\equiv \pi/(2T|r|)$, ${\cal F}_2(r)$ and ${\cal G}(r)$ further generate a spatial anisotropy between $r_0$ and $r_{\perp}=(r_1,r_2)$ with $r=(r_1,r_2,r_0)$. Consequently, the RG equations take functional forms of the three functions of $r_0$ and $|r_{\perp}|$. By analyzing the functional RG equations with approximations, we demonstrate that vortex loops near the order-disorder transition point are not only polarized along the topological direction by the ${\cal F}_2(r)$ and ${\cal G}(r)$ interactions, but also anomalously {\it stretched} along the topological direction [see Fig.~\ref{Fig0}(a)].   

    To deduce a global phase diagram of the 3D U(1) sigma model, we also exercise a complementary argument based on a duality mapping. Thereby, a dual lattice model of the sigma model  describes 3D type-II superconductors, and the correlation function of the U(1) phase variable in the sigma model becomes a correlation function of magnetic monopole fields. Notably,  the 1D Berry phase term along the topological ($x_0$) direction becomes an external magnetic field applied along $x_0$ in the superconductor model. Such a mapping suggests that the generic phase diagram of the sigma model with the 1D Berry phase has a quasi-disordered phase between ordered and disordered phases [See Fig.~\ref{Fig0}(b)]. In the quasi-disordered phase, the exponential correlation length of the U(1)-phase correlation function is divergent along the topological ($x_0$) direction, while it is finite along the other directions. In the paper, we also discuss a cause of the discrepancy between the RG result [Fig.~\ref{Fig0}(a)] and the duality argument [Fig.~\ref{Fig0}(b)], and a possible remedy to rescue the RG approach. 

        The rest of the paper is organized as follows. In the next section, we introduce a 3D vortex loop model, where loop segments interact via the $1/|r|$ Coulomb interaction. In section III, we develop a perturbative RG theory of the vortex loop model without the 1D Berry phase term. In section IV, we study the loop model with the Berry phase term by the RG theory. In section IVA, we first argue that the proliferation of the polarized vortex loops helps to develop the spatially anisotropic coherence of the U(1) phase variable. In section IVB and IVC, we derive the functional RG equations of the loop model with the Berry phase term. In section IVD, we approximate the functional RG equations into simpler equations. In section IVE, we show numerical solutions of the approximate RG equations. Thereby, we demonstrate that the characteristic vortex-loop size is divergent along the topological ($x_0$) direction near the order-disorder transition point, while the vortex-loop size along the other directions remains finite [Fig.~\ref{Fig0}(a)]. In Section V, we employ a duality-mapping argument and argue that a global phase diagram of the U(1) sigma model with the 1D Berry phase term must have the intermediate quasi-disordered phase between ordered and disordered phases [Fig.~\ref{Fig0}(b)]. The appendices cover useful details for understanding the main text. In Appendix A, we discuss a linear confining potential between two endpoints of an open vortex line, and a form of associated dipolar interaction. Appendices B, C, and D derive the renormalizations of the fugacity parameter, Berry phase term, and three types of the Coulombic potentials, respectively. Appendix E explains details of the approximation used in Section IVD.

\section{From U(1) models to Coulomb loop gas models}
Let us consider a partition function of Eq.~(\ref{Eq:I-1}) in a $L_1\times L_2\times L_0$ system with a periodic boundary condition. The phase variable $\theta(x)$ modulo $2\pi$ respects the Born-von Karman boundary condition, $\theta(x)=\theta(x+L_a)+2\pi \mathbf{Z}$.  Since Eq.~(\ref{Eq:I-1}) takes a quadratic action of its gradient vector, a spin-wave fluctuation around a uniform configuration $\theta(x)=\theta_0$ comprises only an action of a free theory; the spin-wave fluctuation on its own cannot drive an order-disorder transition. Thereby, topological excitations play a primary role in the phase transition of the U(1) NLSM. The U(1) phase has a 2D configuration with a pair of vortex and antivortex. In the 3D system, the vortex excitation forms a closed line -- vortex loop --. A line integral of the gradient vector around the vortex line is quantized to $2\pi$, 
$2\pi = \oint \nabla \theta \cdot dl = \int \nabla \times \nabla \theta \!\ \cdot dn$,  relating a rotation of the gradient vector 
with a configuration of a vortex loop,
\begin{align}
\nabla \times \nabla\theta(x) &=2\pi \sum^n_{j=1} \int_{\Gamma_j} dx^j \!\ \delta(x-x^j) \nonumber \\
&= 2\pi \sum^n_{j=1} \int^{l_j}_{0} d\lambda \!\ \frac{dx^j}{d\lambda} \!\  
\delta(x-x^j(\lambda)). \label{Eq:II-1}
\end{align}
In Eq.~(\ref{Eq:II-1}), we consider a general case with $n$ closed vortex loops $\Gamma_j$ ($j=1,\cdots,n$). A spatial coordinate of the $j$-th vortex loop is given by a vector field $x^j\equiv (x^j_1,x^j_2,x^j_0)$. $\delta(x-x^j)$ is a delta function in 3D, $\delta(x-x^j) \equiv \delta(x_1-x^j_1)\delta(x_2-x^j_2)\delta(x_0-x^j_0)$. $dx^j$ is a tangential vector of the $j$-th vortex loop at $x^j$. The right-hand side is a parametric representation of Eq.~(\ref{Eq:II-1}) with a 1D length-scale parameter $\lambda$. $l_j$ is a length of the $j$-th vortex loop, $dx^j/d\lambda$ is the normalized tangential vector. Since all the vortex lines form closed loops, $x^j(0)=x^j(l_j)$ for $j=1,\cdots,n$. 

The 1D Berry phase term confers a complex phase factor upon each of these vortex loops, and the phase for 
each vortex loop is proportional to a projected area of the loop onto the $x_1$-$x_2$ plane~\cite{tanakaShortGuideTopological2015},
\begin{align}
S_1 \equiv i\chi \int d^3x \nabla_0 \theta (x) = 2\pi i \chi \sum^n_{j=1} S^j_{12}.  \label{Eq:II-2}
\end{align}
$S^j_{12}$ denotes the 2D projected area of the $j$-th loop. Here all the vortex loops are considered to be generated from the vacuum, when we contract the closed loops back into points by reducing their projected areas, $\theta(x)$ gets back to the uniform configuration. The sign of $S^j_{12}$ is determined by a sign of the vorticity of the vortex loop. To see Eq.~(\ref{Eq:II-2}) with the sign, one can start with a vortex loop $\Gamma$ and its projected area $S$. Choose the vorticity of the vortex loop to be anticlockwise when the projected area $S$ on the $x_1$-$x_2$ plane is seen from the $x_0>0$ side. Then, the 1D line integral of the gradient vector along the $x_0$ axis takes 
$2\pi$ and $0$, when $(x_1,x_2)$ is inside and outside the projected area $S$, respectively,
\begin{align}
\int^{L_0}_{0} \nabla_0 \theta(x_1,x_2,x_0) \!\ \!\ dx_0 = \begin{cases}
   2\pi  &\text{if } (x_1,x_2) \in S \\
   0  &\text{if } (x_1,x_2) \notin S.
\end{cases} \label{Eq:II-3}
\end{align}
As the winding number of the 1D Berry phase is additive with respect to an addition of the vortex loops, Eq.~(\ref{Eq:II-3}) readily gives Eq.~(\ref{Eq:II-2}) for the general $n$ vortex-loops case.    

A vortex-loop segment interacts with others via the $1/|r|$ Coulomb interaction~\cite{popov1973,wiegel1973,williams1987}. To introduce the interaction, one can decompose the gradient vector into longitudinal $u_L$ and transverse components $u_T$, $\nabla \theta=u_L+u_T$, $\nabla\times u_{L}=0$, and $\nabla \cdot u_T=0$. This also decomposes the gradient term into longitudinal and transverse parts. Here the longitudinal part will be omitted, since it is only associated with the spin-wave fluctuation part $Z_{\rm sw}$ of the partition function [see Eq.~(\ref{Eq:II-7})], 
\begin{align}
S_0 &\equiv \frac{1}{2T}\int \nabla \theta \cdot \nabla \theta \!\ d^3x  
= \frac{1}{2T}\int u^2_T 
\!\ d^3 x+ \frac{1}{2T}\int u^2_L \!\ d^3x \nonumber \\
&= \frac{T}{2} \int H^2 \!\ d^3 x  + i \int H\cdot u_T \!\ d^3 x + \cdots  \nonumber \\
&= \frac{T}{2} \int (\nabla\times A)^2 \!\ d^3 x  - 2\pi i \int A(x) \cdot v(x) \!\ d^3 x + \cdots. \label{Eq:II-3a} 
\end{align}
In the second line, the transverse part is decoupled in terms of a Stratonovich-Hubbard (SH) field $H$. The SH field thus introduced is a divergence-free vector field. Thus, a magnetic vector potential $A$ can be further defined from the SH field,  $H = \nabla \times A$. The vector potential here is divergence-free, as its longitudinal part would have nothing to do with the SH field. As all the vortex loops considered here are generated from the vacuum, trivial boundary conditions can be imposed on the vector potential, e.g. $A(x)=0$ at the boundary of the system. Thereby, in the second term at the last line, we can drop a surface term after a partial integral. From eq.~(\ref{Eq:II-1}), a vector $v(x) \equiv \frac{1}{2\pi} \nabla \times u_T$ is given by a sum of the quantized flux lines associated with the vortex loops, 
\begin{align}
v(x) = \sum_j \int_{\Gamma_j} dx^j \!\ \delta(x-x^j).  \label{Eq:II-4}
\end{align}
We henceforth call $v(x)$ vortex vector.

The Coulomb interaction between the flux lines is obtained by the Gaussian integration over the vector potential. Completing the square in the momentum space yields:
\begin{widetext}
\begin{align}
S_0& = \int_{k\ne 0} \frac{dk^3}{(2\pi)^3} \bigg\{
\frac{Tk^2}{2} \Big(A(k) - \frac{2\pi i}{Tk^2} v(k)\Big)^T  \Big(1 -  \frac{k k^T}{k^2} \Big)  
\Big(A(-k) - \frac{2\pi i}{Tk^2} v(-k)\Big)  + \frac{4\pi^2}{2T k^2} v^T(k) 
\Big(1- \frac{k k^T}{k^2}\Big) v(-k)\bigg\}. \label{Eq:II-5}
\end{align}
As all the vortex loops considered here are closed loops, the vortex vector is also divergence-free,  
$\nabla \cdot v(x)=0$, and the integral over $A$ gives the $1/|r|$ Coulomb 
interaction between the flux lines~\cite{popov1973}, 
\begin{align}
S_0 &= \frac{\pi}{2T} \int d^3x \int d^3y \frac{v(x) \cdot v(y)}{|x-y|} = \frac{\pi}{2T} \sum^n_{i,j=1} \oint_{\Gamma_i} \oint_{\Gamma_j} 
\frac{dx^i\cdot dy^j}{|x^i-y^j|}.  \label{Eq:II-6}
\end{align}
In summary, the 3D U(1) NLSM can be described by a vortex-loop model, Eqs.~(\ref{Eq:II-2},\ref{Eq:II-6}), 
\begin{align}
Z &= Z_{\rm sw} Z_{\rm v}, \!\ Z_{\rm sw} = \int du_L(x) \!\ \exp 
\Big[-\frac{1}{2T}\int u^2_L \!\ d^3x \Big], \label{Eq:II-7} \\
Z_{\rm v} &= 1 + \sum^{\infty}_{n=1} \frac{1}{n!} \prod^n_{j=1} 
\Bigg(\int^{\infty}_{a} \bigg(\frac{dl_j}{a_0}\bigg) \!\ \!\ t^{l_j} \int \bigg(\frac{d^3 R_j}{a^3_0}\bigg) 
\int {D} \Omega_j(\lambda) \Bigg) \exp\bigg[-\frac{\pi}{2T} 
\sum^n_{i,j=1}
\oint_{\Gamma_i}  \oint_{\Gamma_j} 
\frac{d x^i \cdot dy^j }{|x^i-y^j|}  -2\pi i \chi \sum^n_{i=1} S^i_{12} \bigg]. \label{Eq:II-8}
\end{align}
\end{widetext}
Here $Z_{\rm sw}$ and $Z_{\rm v}$ denote the longitudinal (spin-wave) part and transverse (vortex excitations) part of the partition function, respectively. Since the order-disorder transition in the U(1) NLSM is primarily driven by the proliferation of the vortex loops, this paper studies only $Z_{\rm v}$. 

In $Z_{\rm v}$, we introduced a fugacity parameter $``t"$ of vortex loop segments per unit length, a chemical potential of the loop segment with unit length is given by $\ln t$. We also introduced an ultraviolet (UV) cutoff $``a"$ for an integral over the vortex-loop length $l_j$. The UV cutoff is on the order of a lattice constant $a_0$ of an underlying lattice model. $R_j$ stands for a center-of-mass coordinate of the $j$-th loop. The integrals over the loop length $l_j$ and over the center-of-mass coordinate $R_j$ have the dimensions of length and volume, respectively. To make the partition function $Z_{\rm v}$ to be dimensionless, we divided them by the lattice constant $a_0$, and by a unit volume $a^3_0$, respectively. For simplicity of the notation, we omit these normalizations, 
$dl_j/a_0 \rightarrow dl_j$, $d^3 R_j/a^3_0 \rightarrow d^3R_j$, while they will be recovered in the end of RG calculations.

$\Omega^j(\lambda)\equiv dx^j(\lambda)/d\lambda$ is a normalized tangential vector along the $j$-th loop at a segment $\lambda$. A path integral over the normalized tangential vectors for $\lambda \in [0,l_j)$ takes a summation over all possible shapes of the closed loop with a fixed length $l_j$,
\begin{align}
\int D \Omega^j(\lambda) \equiv \lim_{\varepsilon\rightarrow 0} \prod^{l_j / \varepsilon}_{M=1}
\bigg(\int_{\int^{l_j}_{0}\Omega^j(\lambda) \!\ d\lambda=0} d\Omega^j(\lambda=M\varepsilon)  \bigg). \label{Eq:II-9}
\end{align}
As $\Omega^j(\lambda)$ is a unit vector, the path integral over $\Omega^j(\lambda)$ is independent of the length scale. A factor $1/n!$ in Eq.~(\ref{Eq:II-8}) is a symmetric factor that sets off double counting of an identical configuration of the $n$ vortex loops.

The $1/|r|$ Coulomb interaction between the vortex-loop segments also needs a UV cutoff. We use the lattice constant $a_0$ as the UV cutoff for the Coulomb interaction length,   
\begin{align}
\oint_{\Gamma_i} \oint_{\Gamma_j} 
\frac{d x^i\cdot dy^j}{|x^j-y^j|} \equiv 
\oint \oint_{|x^i-y^j|>a_0}  
\frac{d x^j\cdot dy^j}{|x^j-y^j|}. 
\label{Eq:II-10}
\end{align}
A ratio between $``a"$ and $``a_0"$ is generally model-dependent. In section III, we will explain our choice of the UV model in this paper. 

In the next section, we first develop a renormalization group (RG) study of $Z_{\rm v}$ without the Berry phase ($\chi=0$). We derive coupled RG equations between the fugacity parameter $t$ and stiffness parameter $1/T$. The RG equations have a strong coupling fixed point with divergent fugacity (disordered phase), a weak coupling fixed point with vanishing fugacity (ordered phase), and a saddle-point fixed point between these two. A scaling analysis around the saddle-point fixed point gives us an estimate of a critical exponent of an order-disorder transition of the 3D U(1) NLSM without the Berry phase term. 

\section{RG analysis of a 3D Coulomb loop gas model}
   A renormalization comprises an integration over short-distance degrees of freedom (DOF) 
and a rescaling of the length scale~\cite{jose1977,williams1987,shenoy1989,cardy1996}. The integration is carried out perturbatively in $1/T$,  giving 1-loop renormalization to the two coupling constants, $T$ and $t$. The integration also changes the UV cutoffs $``a"$ and $``a_0"$  into  $``a b"$ and $``a_0 b"$, respectively, where an infinitesimally small positive $\ln b$ plays the role of an RG rescaling factor. The subsequent length rescaling puts the UV cutoffs back to the original values. 
\begin{align}
\begin{aligned}
& x^i \rightarrow x^{i \!\ \prime} = x^i b^{-1}, \!\  y^j \rightarrow y^{j \!\ \prime} = y^j b^{-1}, \!\  l_j \rightarrow l^{\prime}_{j} 
= l_j b^{-1}, \\
& R_j \rightarrow R^{\prime}_j = R_j b^{-1}, \!\  \lambda \rightarrow 
\lambda^{\prime} = \lambda b^{-1}. 
\end{aligned} \label{Eq:III-1}
\end{align}
The inverse temperature and the chemical potential of the vortex loop segment have their tree-level scaling dimensions to be $1$ ,
\begin{align}
\frac{1}{T} \rightarrow \frac{1}{T^{\prime}} = \frac{1}{T} \!\ b, \!\ \!\ 
\ln t \rightarrow \ln t^{\prime} = \ln t \!\ b,
\end{align}
$[[1/T] = 1$, $[\ln t] = 1]$.

       To integrate over the short-distance DOF~\cite{jose1977,williams1987}, we decompose an integral of the loop length $l$ into an integral over its short-length region $(a<l<ab)$ and the integral over its 
long-length region ($ab<l$), $\int^{\infty}_a dl = \int^{ab}_{a} dl + \int^{\infty}_{ab} dl$. 
Substitute it into Eq.~(\ref{Eq:II-8}), and keep up to the first order in small $\ln b$,
\begin{widetext}
\begin{align}
Z_{\rm v} = &1 + \bigg(\int^{ab}_{a} dl t^l \int d^3R \int D\Omega(\lambda)\bigg) \!\ \!\ e^{ - s_0(\Gamma,\Gamma)} 
+ \sum^{\infty}_{n=1}\frac{1}{n!} \prod^n_{j=1}
\bigg(\int^{\infty}_{ab} dl_j t^{l_j} \int d^3R_j \int D\Omega_j(\lambda)\bigg) \!\ \!\ e^{-\sum^n_{i,j=1}s_{0}(\Gamma_i,\Gamma_j)} \nonumber \\
& \hspace{-0.8cm} + \bigg(\int^{ab}_{a} dl \!\ t^l 
\int d^3R \int D\Omega(\lambda)\bigg)  \sum^{\infty}_{n=1}\frac{1}{n!} \prod^n_{j=1} 
\!\ \bigg(\int^{\infty}_{ab} dl_j \!\ t^{l_j} \int d^3R_j \int D\Omega_j(\lambda)\bigg) 
\!\ \!\ e^{-s_{0}(\Gamma,\Gamma)-2\sum^n_{j=1}s_0(\Gamma,\Gamma_j)-\sum^n_{i,j=1} s_0(\Gamma_i,\Gamma_j)} 
\label{Eq:III-2}
\end{align}
Here $\Gamma$ is the shortest closed loop with its length being in $a<l<ab$. 
$\int d^3R\int D\Omega(\lambda)$ stands for a configurational integral of the shortest 
loop, $R$ and $\Omega(\lambda)$ are a center-of-mass coordinate and the tangential 
vectors of the shortest loop. $s_0(\Gamma_i,\Gamma_j)$ is the Coulomb interaction between the $i$-th and $j$-th loops,
\begin{align}
s_0(\Gamma_i,\Gamma_j) = \frac{\pi}{2T} \oint_{\Gamma_i}  \oint_{\Gamma_j} 
\frac{d x^i \cdot dy^j }{|x^i-y^j|}.  \label{Eq:III-3}
\end{align}
To carry out the configurational integral over the shortest loop, we expand the Coulomb 
interaction $s_0(\Gamma,\Gamma_j)$ between the shortest loop and the other loops,
\begin{align}
Z_{\rm v} = &1 + \bigg(\int^{ab}_{a} dl \!\ t^l \int d^3R \int D\Omega(\lambda)\bigg)  
\!\ e^{-s_0(\Gamma,\Gamma)}
+ \sum^{\infty}_{n=1}\frac{1}{n!} \prod^n_{j=1}
\bigg(\int^{\infty}_{ab} dl_j \!\ t^{l_j} \int d^3R_j \int D\Omega_j(\lambda)\bigg) \!\ \!\ e^{-\sum^n_{i,j=1}s_{0}(\Gamma_i,\Gamma_j)} \nonumber \\
 & +   \sum^{\infty}_{n=1}\frac{1}{n!} \prod^n_{j=1} 
\!\ \bigg(\int^{\infty}_{ab} dl_j \!\ t^{l_j} \int d^3R_j \int D\Omega_j(\lambda)\bigg) 
\!\ \!\ e^{-\sum^n_{i,j=1} s_0(\Gamma_i,\Gamma_j)} \nonumber \\ 
& \hspace{1cm} \times \bigg(\int^{ab}_{a} dl \!\ t^{l}
\int d^3R \int D\Omega(\lambda)\bigg) \!\ e^{-s_0(\Gamma,\Gamma)}   
\bigg\{1 - \Big(\sum^n_{j=1} 2s_0(\Gamma,\Gamma_j)\Big) + \frac{1}{2} 
\Big(\sum^n_{j=1} 2s_0(\Gamma,\Gamma_j)\Big)^2 + \cdots \bigg\} \label{Eq:III-4}
\end{align}
\end{widetext}
The expansion is perturbative in $1/T$, and it is justified a posteriori by an observation 
that an inverse temperature around the saddle fixed point is small. Note that the temperature $T$ in this paper has a dimension of length, so that the 
temperature shall be compared to the UV cutoff length $a_0$ in the observation. 

After the configurational integral of the shortest loop, the fourth term in 
Eq.~(\ref{Eq:III-4}) gives a renormalization to the temperature 
in the third term. More specifically, the first order in $s_0(\Gamma,\Gamma_j)$ 
vanishes after the configurational sum, while the second-order in $s_0(\Gamma,\Gamma_j)$ 
induces a renormalization to the Coulomb interaction among the rest of the other 
loops. The renormalization is nothing but a screening effect caused by the shortest loop. 
The second term in Eq.~(\ref{Eq:III-4}) is extensive because of  
its $R$ integral, while it is also on the order of $\ln b$. Thus, the second term can be 
included as an inhomogeneous part of a free-energy renormalization. Since we are primarily interested 
in the RG equations among the coupling constants, we do not delve ourselves into the free-energy renormalization in this paper. 

The Coulomb interaction $s_0(\Gamma_i,\Gamma_j)$ in the third term needs to be also decomposed into  
short-distance part and long-distance part~\cite{jose1977,williams1987},
\begin{align}
&\oint_{\Gamma_i} \oint^{|x^i-y^j|>a_0}_{\Gamma_j} \frac{dx^i\cdot dy^j}{|x^i-y^j|} \nonumber \\
& \ \ =\bigg( \oint_{\Gamma_i} \oint^{a_0b>|x^i-y^j|>a_0}_{\Gamma_j}  +  \oint_{\Gamma_i} \oint^{|x^i-y^j|>a_0b}_{\Gamma_j} \bigg) \!\ \frac{dx^i\cdot dy^j}{|x^i-y^j|} 
\end{align}
The short-distance part of the Coulomb interaction within a same loop $\Gamma_i$ gives a renormalization to the fugacity for the loop. 
\begin{align}
&\oint_{\Gamma_i} \oint^{a_0 b|x^i-y^j|>a_0}_{\Gamma_i} \frac{dx^i\cdot dy^i}{|x^i-y^i|} 
= 2 \ln b \!\ l_i  + \cdots. \label{Eq:III-5}
\end{align}
Here $``\cdots"$ in the right-hand side stands for renormalization to other physical parameters associated with a vortex loop.  Such parameters include an elastic energy parameter $u$ of the vortex loop. To see Eq.~(\ref{Eq:III-5}), one can use the parametric representation of the loop and expand the normalized tangential vectors in terms of the short-distance $a_0$,
\begin{align}
&\oint_{\Gamma_i} \oint^{a_0b>|x^i-y^i|>a_0}_{\Gamma_i} 
\frac{dx^i\cdot dy^i}{|x^i-y^i|} \nonumber \\
&= \int^{l_i}_{0} d\lambda 
\int_{ba_0>|\lambda^{\prime}-\lambda|>a_0} d\lambda^{\prime} \!\ \frac{\Omega_i(\lambda)\cdot \Omega_i(\lambda^{\prime})}{|\lambda-\lambda^{\prime}|} 
\nonumber \\
&=  2 \ln b \!\ l_i + \delta u \ln b \int^{l_i}_0 d\lambda \!\  
\bigg|\frac{\partial \Omega_i}{\partial \lambda}\bigg|^2 + \cdots  
\end{align}
Here $\delta u = - \frac{1}{2} a^2_0$ is the renormalization to the elastic energy parameter $u$. Since the elastic energy parameter $u$ has a negative tree-level scaling dimension ($[u]=-1$), we will omit this effect in this paper, and 
consider only the renormalization to the vortex fugacity parameter, i.e. $2\ln b\!\ l_i$. The short-distance part of the Coulomb interaction between different loops ($i\ne j$) may induce renormalization to physical parameters of a vortex loop when the two loops merge at some vortex segments. Unlike the fugacity renormalization, however, such merging events happen only occasionally upon the renormalization. Thus, we ignore by hand the short-distance part of the Coulomb interaction between different loops.

To calculate the renormalization to the temperature, we need to specify the shortest loop in detail. To this end, let us be inspired by a lattice-regularized model on a cubic lattice. A lattice point of the cubic lattice accommodates the U(1) phase variable $\theta(x)$, and its lattice constant $a_0$ is the UV cutoff for the Coulomb interaction length in Eq.~(\ref{Eq:II-10}).  A dual lattice is also a cubic lattice. Nearest neighboring sites of the dual lattice are connected by the vortex vector $v(x)$, and the square plaquette of the dual lattice forms the shortest vortex loop. As a counterpart of such square-plaquette loop, we choose a symmetric circle with its diameter $a_0$ as the shortest loop in the continuum theory. The choice of the shortest loop sets the UV cutoff $a$ for the loop length in Eq.~(\ref{Eq:II-8}),  $a=a_0 \pi$. Here, the shortest loops with other shapes, e.g. elliptical circle, non-coplanar closed loop, are not considered as physical loops. For example, an ellipse with its circumference $a_0\pi$ has its diameter along its minor axis to be shorter than $a_0$. We do not consider such elliptical circles as physical because two flux lines with the opposite vorticity cannot be closer than the lattice constant $a_0$ on the dual lattice. 

The shortest loop thus introduced is always coplanar, so that a unit vector $\Omega$ normal to the plane, and a center of the circle entirely parameterize the configuration of the loop.  The configuration sum of the shortest loop comprises of an integral over the center coordinate $R$ over a whole volume $V$, and a 2D integral with respect to $\Omega$ over a unit sphere $S_2$, 
\begin{align}
\int^{ab}_a dl \int d^3R \int D\Omega(\lambda) 
= a\ln b \int_V d^3 R \int_{S_2} d^2\Omega. \label{Eq:III-6}
\end{align}
The spatial coordinate $x$ and tangential vector $dx$ of the loop segment in the short loop can be parameterized by an angle $\alpha \in [0,2\pi)$,  
\begin{align}
x=R+\frac{a_0}{2} n(\alpha), \!\ \!\ 
dx = \frac{a_0}{2} \frac{dn}{d\alpha} 
\equiv  \frac{a_0}{2} m(\alpha), 
\end{align}
with $n(\alpha) = \cos\alpha (-e_2) + \sin\alpha  \!\ e_1$, 
and $m(\alpha)=\cos\alpha \!\ e_1 + \sin\alpha \!\ e_2$. Here $e_1$ and $e_2$ are 
orthonormal vectors on the coplanar plane, and $e_1$, $e_2$  and $\Omega$ form an orthonormal basis frame with $e_1\times e_2 = \Omega$. 

To evaluate the screening effect by the short vortex loop, note first that the UV cutoff $a_0$ is much smaller than the temperature, so that we could also expand $s_{0}(\Gamma,\Gamma_j)$ in powers of the diameter of the circular loop $\Gamma$,
\begin{align}
&2s_0(\Gamma,\Gamma_j) = \frac{\pi}{T} \int_{\Gamma} \int_{\Gamma_j} 
\frac{dx\cdot dy^j}{|x-y^j|} \nonumber \\
& = 
\frac{\pi}{T}\frac{a_0}{2} \int^{\pi}_{0} d\alpha 
\int_{\Gamma_j} 
\frac{m(\alpha)\cdot dy^j} {|R+\frac{a_0}{2} n(\alpha) - y^j|} \nonumber \\
& \hspace{1cm}  - 
\frac{\pi}{T}\frac{a_0}{2} \int^{\pi}_{0} d\alpha 
\int_{\Gamma_j} 
\frac{m(\alpha)\cdot dy^j} {|R-\frac{a_0}{2} n(\alpha) - y^j|} \nonumber \\
&= \frac{\pi a^2_0}{2T} \int_{\Gamma_j} dy^j_{\nu} \int^{\pi}_{0} 
d\alpha   
\!\ m_{\nu}(\alpha) \!\ n_{\mu}(\alpha) \!\   
\nabla_{R_{\mu}}\bigg(\frac{1}{|R-y^j|}\bigg) + \cdots \nonumber \\
&= \frac{\pi^2 a^2_0}{4T} \int_{\Gamma_j} dy^j_{\nu} \!\ \Omega_{\lambda} \!\ 
\epsilon_{\lambda\mu\nu} \nabla_{R_{\mu}} 
\bigg(\frac{1}{|R-y^j|}\bigg) + \cdots  \label{Eq:III-7}
\end{align}
Upon a substitution of this into Eq.~(\ref{Eq:III-4}), the integral over $\Omega$ makes the first order in $s_0(\Gamma,\Gamma_j)$ 
vanish in Eq.~(\ref{Eq:III-4}), while the second order term in $s_0(\Gamma,\Gamma_j)$  yields the 
effective Coulomb interaction among the other loops,
\begin{align}
&\int_V d^3R \int_{S_2} d^2\Omega \!\  
\bigg\{1  + \frac{1}{2} 
\Big(\sum^n_{j=1} 2s_0(\Gamma,\Gamma_j)\Big)^2 + \cdots \bigg\} \nonumber \\
 & = 4\pi V + \frac{2\pi}{3}\Big(\frac{\pi^2 a^2_0}{4T}\Big)^2 
\sum^n_{i,j=1}  \int d^3 R\nonumber \\
&\bigg\{
\oint_{\Gamma_i} dx^i_{\lambda} \oint_{\Gamma_j} dy^j_{\lambda} 
\nabla_{R_{\mu}} \Big(\frac{1}{|R-x^i|}\Big) 
\nabla_{R_{\mu}}\Big(\frac{1}{|R-y^j|}\Big) - \nonumber \\
&  \oint_{\Gamma_i} dx^i_{\lambda} \oint_{\Gamma_j} dy^j_{\mu} 
 \nabla_{R_{\mu}} \Big(\frac{1}{|R-x^i|}\Big) 
\nabla_{R_{\lambda}}\Big(\frac{1}{|R-y^j|}\Big) \bigg\} + \cdots, \nonumber \\
&= 4\pi V + \frac{2\pi^2}{3}\Big(\frac{\pi^2 a^2_0}{2T}\Big)^2 
\sum^n_{i,j=1} \oint_{\Gamma_i} \oint_{\Gamma_j} 
\frac{dx^i\cdot dy^j}{|x^i-y^j|} + \cdots \label{Eq:III-8}
\end{align}
Here we took an integration by parts with respect to $R_{\mu}$ in the second line, and used 
$\nabla^2_R (1/|R-x|)=-4\pi \delta(R-x)$ for the first term. We also note that the loop 
is closed so that $\oint_{\Gamma_j} dy^j_{\mu} \!\ \nabla_{R_{\mu}}f(R-y^j) =-\oint_{\Gamma_j} dy^j_{\mu} \!\ \nabla_{y^j_{\mu}}f(R-y^j)=0$ for the second term in the second line. 
 
A substitution of Eqs.~(\ref{Eq:III-5},\ref{Eq:III-8}) into Eq.~(\ref{Eq:III-4}) yields an 
effective action only for the longer vortex loops up to the first order in $\ln b$,
\begin{align}
&Z_{\rm v} = e^{4\pi V a \!\ t^a \!\ \ln b} \bigg\{ 1 + \sum^{\infty}_{n=1}\frac{1}{n!} \nonumber \\
& \prod^n_{j=1}
\bigg(\int^{\infty}_{ab} dl_j  \!\ \big(\overline{t}\big)^{l_j}
\int d^3R_j \int D\Omega_j(\lambda)\bigg) \!\ \!\ 
e^{ -\sum^n_{i,j=1} \overline{s}_{0}(\Gamma_i,\Gamma_j)} \bigg\},  \label{Eq:III-9}
\end{align}
where
\begin{align}
\overline{s}_0(\Gamma_i,\Gamma_j) = 
\frac{\pi}{2\overline{T}} \oint_{\Gamma_i} 
\oint^{|x_i-y_y|>a_0 b}_{\Gamma_j} \frac{dx^i\cdot dy^j}{|x^i-y^j|}, 
\end{align}
with 
\begin{align}
\ln \overline{t} &= \ln t - \frac{\pi}{T} \!\ \ln b, \label{Eq:III-10} \\ 
\frac{\pi}{2\overline{T}} 
& = \frac{\pi}{2T} - a t^{a} \frac{2\pi^2}{3} \Big(\frac{\pi^2 a^2_0}{2T}\Big)^2 \!\ \ln b,  
\label{Eq:III-11}
\end{align}
and $a=a_0\pi$.  Note that $s_0(\Gamma,\Gamma) \propto \ln b$ is omitted in Eq.~(\ref{Eq:III-9}) because its contribution appears in the partition function with the higher power in $\ln b$.  Combining these renormalization with the length rescaling in Eq.~(\ref{Eq:III-1}), we finally 
obtain
\begin{align}
\ln t^{\prime} &= \ln \overline{t} \!\ b = \ln t + \ln t \ln b - \frac{\pi}{T} \ln b, \label{Eq:III-12} \\
\frac{\pi}{2T^{\prime}} &= \frac{\pi}{2\overline{T}} \!\ b \nonumber \\
&= \frac{\pi}{2T} + \frac{\pi}{2T} \ln b - a t^{a} \frac{2\pi^2}{3} \Big(\frac{\pi^2 a^2_0}{2T}\Big)^2 \!\ \ln b.  \label{Eq:III-13}
\end{align}
These two yield coupled RG equations between $T$ and $\ln t$, 
\begin{align}
&\frac{d\ln t}{d\ln b} = \ln t - \frac{\pi}{T},  \label{Eq:III-14a} \\
&\frac{d}{d\ln b} \bigg(\frac{\pi}{2T}\bigg) 
= \frac{\pi}{2T} - a_0e^{a_0 \pi \ln t} \frac{2\pi^3}{3} \bigg(\frac{\pi^2 a^2_0}{2T}\bigg)^2. \label{Eq:III-14b}  
\end{align}
Before proceeding to a fixed-point analysis of the RG equation, let us note that the length rescaling in $\int dl_j \int d^3 R_j$ induces an additional factor of $b^4$ for each vortex loop. This factor can be included into a renormalization of another type of vortex fugacity parameter $v$ that does not depend on the loop length,  $t^{l_j} \rightarrow t^{l_j} v$, together with $v^{\prime} = b^4 v$~\cite{williams1987}. The scaling equation of $v$ gives   
\begin{align}
\frac{d\ln v}{d\ln b} = 4.  \label{Eq:III-14c}
\end{align}
A comparison between Eqs.~(\ref{Eq:III-14a},\ref{Eq:III-14c}) suggests that $t$ decreases/increases exponentially in the RG scale factor $
\ln b$, while $v$ increases only in the power of $b$. Such $t$ always dominates over $v$, when $t$ decreases as well as increases. Thus, the order and disordered phases can be determined only by divergent and vanishing $t$, respectively, and we ignore the scaling equation of $v$ henceforth.  Eqs.~(\ref{Eq:III-14a},\ref{Eq:III-14b},\ref{Eq:III-14c}) are consistent with a set of RG equations that Williams and Shenoy derived previously in studies of the superfluid $\lambda$ transition~\cite{williams1987,shenoy1989,nelson1977,williams1993,williams1999,williams2004}. The RG equations were also applied to the studies of cosmic strings in the early universe~\cite{antunes1998a,antunes1998b} and high-Tc superconductors~\cite{kiometzis1994,nguyen1998a,nguyen1998b,ryu1998,williams1999}. For completeness,  
we will give a fixed-point analysis of Eqs.~(\ref{Eq:III-14a},\ref{Eq:III-14b}) below. 

\begin{figure}[t]
\centering
    \includegraphics[width= 0.9\linewidth]{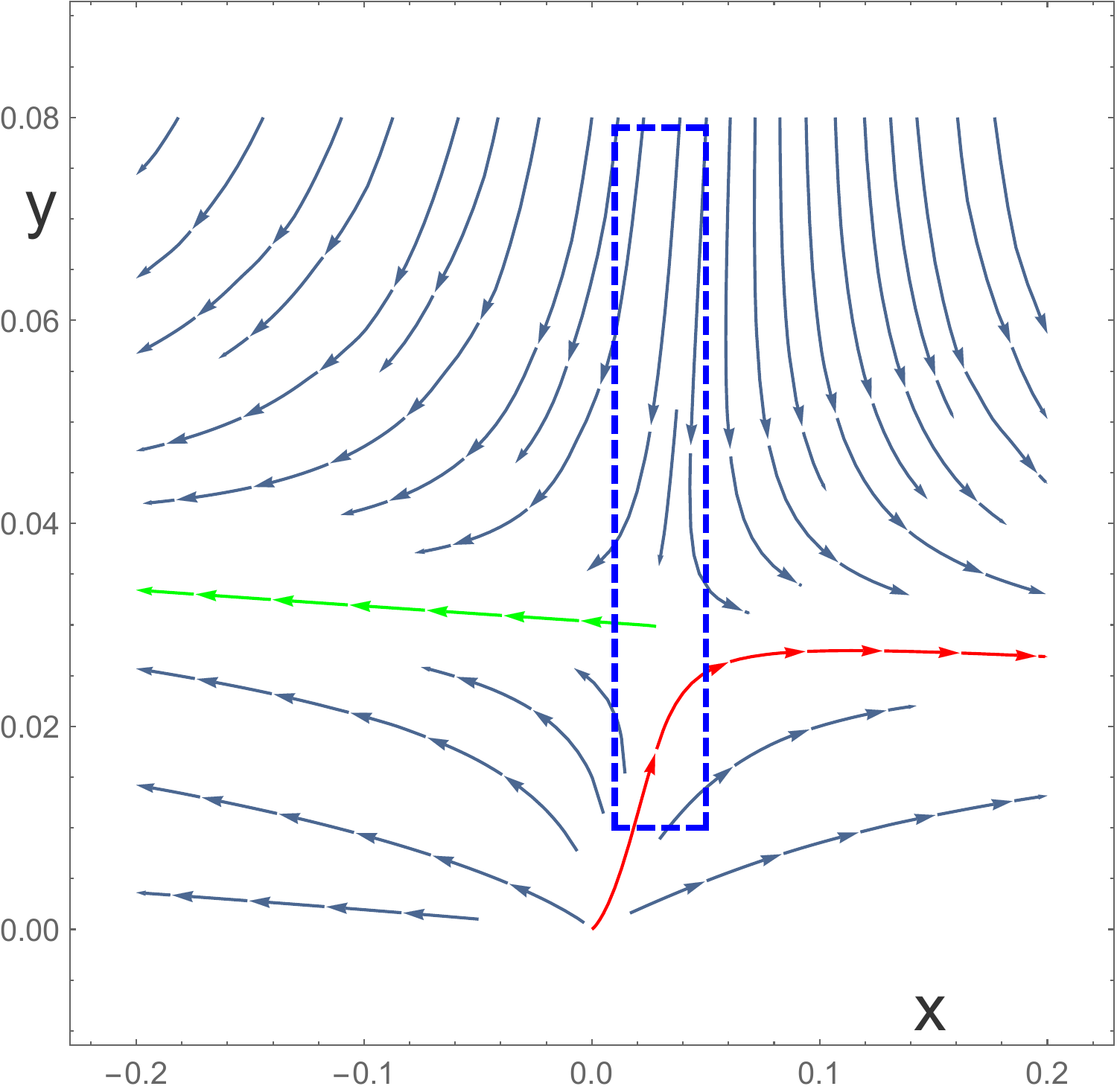} 
\caption{An RG flow diagram of 3D Coulomb loop gas model without the 1D Berry phase term. It is obtained from Eqs.~(\ref{Eq:III-14a},\ref{Eq:III-14b}), where the horizontal and vertical axes are $\pi x = \pi a_0 \ln t$, and $2\pi y = \pi^2 a_0 /T$ respectively. The green flow goes to a low-$T$ fixed point at $(t,T)=(0,0)$, and red flow goes to a high-$T$ fixed point at $(t,T)=(\infty,\infty)$. A saddle fixed point at $(\pi x,2\pi y)=0.0299...(1,1)$ determines the criticality of the order-disorder transition.The blue dotted rectangular region stands for a $x$-$y$ parameter region for a phase diagram with the Berry phase term depicted in Fig.~\ref{Fig3}.}
\label{Fig1}
\end{figure}

To analyze fixed points of Eqs.~(\ref{Eq:III-14a},\ref{Eq:III-14b}) , we first normalize $T$ and $\ln t$ by the UV cutoff length scale $a_0$. Besides, we will also recover the normalization factor of the integrals over $l_j$ and $R_j$ for each loop [see a text between Eqs.~(\ref{Eq:II-8},\ref{Eq:II-9})]. This gives the dimensionless RG equations for $x \equiv a_0\ln t$  and $y\equiv \frac{\pi a_0}{2T}$ as follows,  
\begin{align}
\frac{dx}{d\ln b} = x-2y, \!\ \!\  \frac{dy}{d\ln b} = y - \frac{2\pi^5}{3} e^{\pi x} y^2. \label{Eq:III-15}
\end{align}
The RG equations have a high-$T$ fixed point with divergent fugacity parameter $t$ at $(x,y)=(\infty,0)$, a low-$T$ fixed point with vanishing fugacity parameter at $(x,y)=(-\infty,\infty)$ , and a saddle fixed point at $(x,y)=y_0(2,1)$ with $y_0=0.00485\cdots$. The fixed point with divergent fugacity $t$ and divergent $T$ characterizes the disordered phase of the NLSM, while the fixed point with vanishing fugacity $t$ and vanishing $ T$ is for the ordered phase. A linearization around the saddle fixed point 
\begin{align}
\frac{d}{d\ln b} \left(\begin{array}{c}
x \\
y \\
\end{array}\right) = \left(\begin{array}{cc} 
1 & -2 \\
- y_0 \pi & -1 \\
\end{array}\right) \left(\begin{array}{c}
x \\
y \\
\end{array}\right) 
\end{align}
gives a scaling dimension $y_t$ of the relevant scaling variable around the saddle fixed point to be $y_t\simeq 1+ y_0 \pi = 1.0153\cdots$ and critical exponent $\nu \equiv 1/y_t$ for the order-disorder transition to be $\nu=0.984\cdots$. Note that a small value of $y_0 \simeq 0.005$ at the saddle fixed point justifies a posteriori the perturbative expansion with respect to $a_0/T$ in Eq.~(\ref{Eq:III-4}). 
 
     The value of $y$ at the saddle-point fixed point depends on a choice of the shortest loop. To make the perturbative expansion in $a_0/T$ truly controlled as in the Wilson-Fisher theory~\cite{wilson1972,wilson1974}, one may generalize the 3D model into a $D$-dimensional model with $1/ |r|^{D-2}$ Coulomb interaction~\cite{nelson1977}. In the $D$-dimensional model, the vortex excitation has a $(D-2)$-dimensional volume, where the fugacity parameter $\ln t$  is introduced as a chemical potential of the vortex excitation per [length]$^{(D-2)}$. Such $\ln t$ as well as the inverse temperature are normalized by $a^{-(D-2)}_0$. Under the scaling of Eq.~(\ref{Eq:III-1}), the 1-loop RG equations for these normalized coupling constants, $x$ and $y$, take the following forms of 
\begin{align}
\begin{aligned}
&\frac{dx}{d\ln b} = (D-2)x- A y,  \\
&\frac{dy}{d\ln b} = (D-2) y - B e^{C x} y^2. 
\end{aligned} \label{Eq:III-16}
\end{align}
Note that the RG equation for the normalized fugacity parameter $x$ physically makes sense only in $D>2$, and Eq.~(\ref{Eq:III-16}) does not reproduce the Berezinskii-Kosterlitz-Thouless (BKT) transition in $D=2$~\cite{Berezinskii71,Berezinskii72,kosterlitz1973,kosterlitz1974}.  Nonetheless, a close resemblance between the $D>2$ theory and the BKT theory at $D=2$  suggests that `$B$' in the RG equation for the inverse temperature $y$ can be positive and continuous function of the spatial dimension $D$ at $D \ge 2$. In that case, by $D=2+\epsilon$ , the value of $y$ at the saddle fixed point might be on the order of  $\epsilon$, allowing the  perturbative treatment of $y$ for small $\epsilon$.

\section{RG analysis of a 3D Coulomb loop gas with the 1D Berry phase term} 

      In the previous section, we have developed a perturbative RG theory of the 3D Coulomb loop gas model. In this section, we use the perturbative RG theory to study the loop gas model with the 1D Berry phase term. According to the partition function Eq.~(\ref{Eq:II-8}), the 1D Berry phase factor for each vortex loop  is proportional to an area within the loop projected onto the $x_1$-$x_2$ plane. Such complex phase factor  induces a destructive interference among the vortex loops with finite projections onto the $x_1$-$x_2$ plane. Meanwhile, those loops that are confined in planes parallel to the $x_0$-axis are free from the destructive interference, dominating the partition function near the order-disorder transition point. In the next subsection, we first discuss how the proliferation of such polarized vortex loops leads to an extremely spatially anisotropic correlation of the U(1) order parameter.    

\subsection{spatially anisotropic phase coherence induced by the interference effect}

\begin{figure*}[t]
        \subfloat[The shortest circular loop is introduced far from $y$ and $z$]{%
            \includegraphics[width=.49\linewidth]{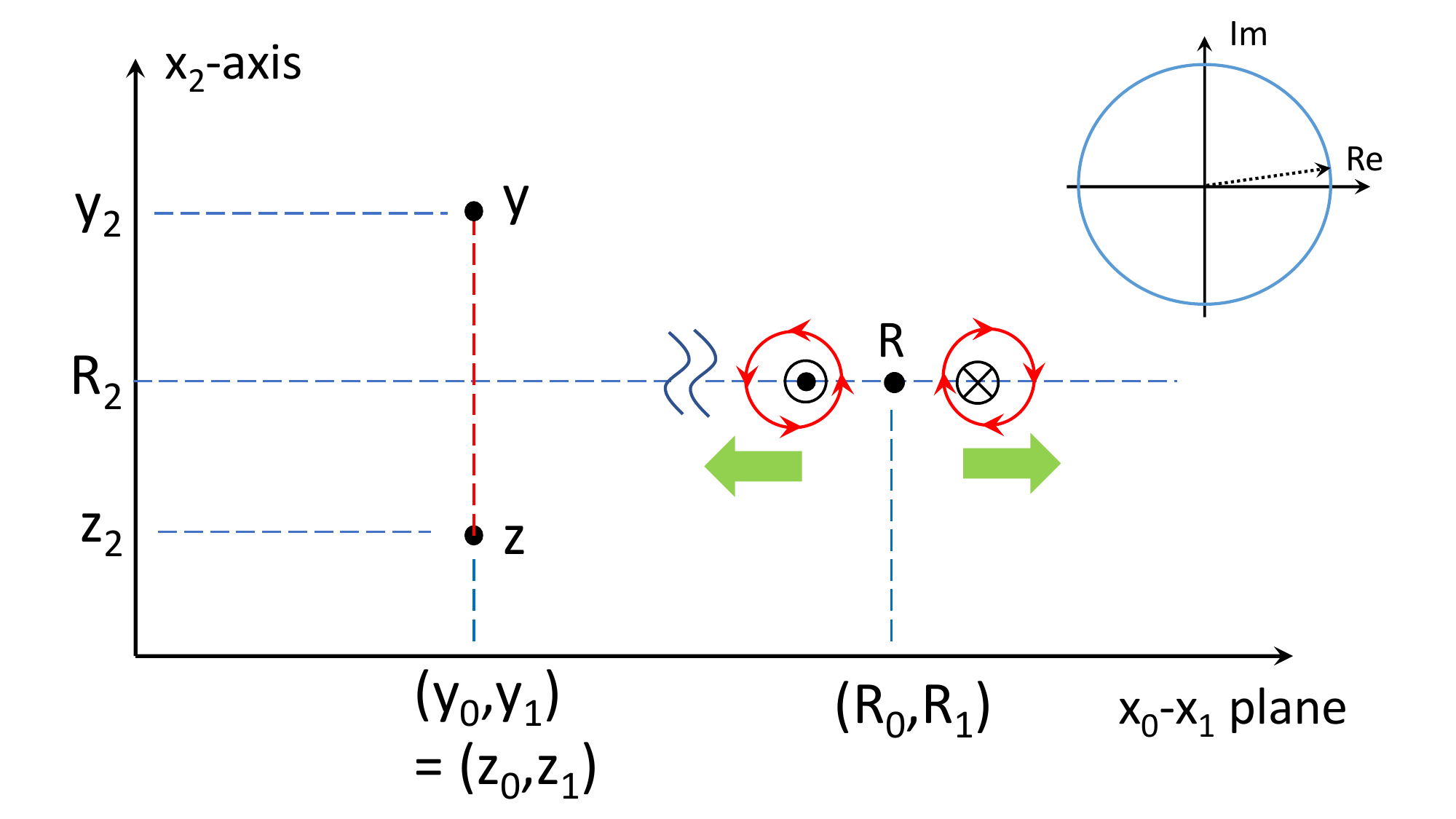}%
            \label{Subfig:2a}%
        }\hfill
        \subfloat[Some segment of the loop is nearing $y$ and $z$]{%
            \includegraphics[width=.49\linewidth]{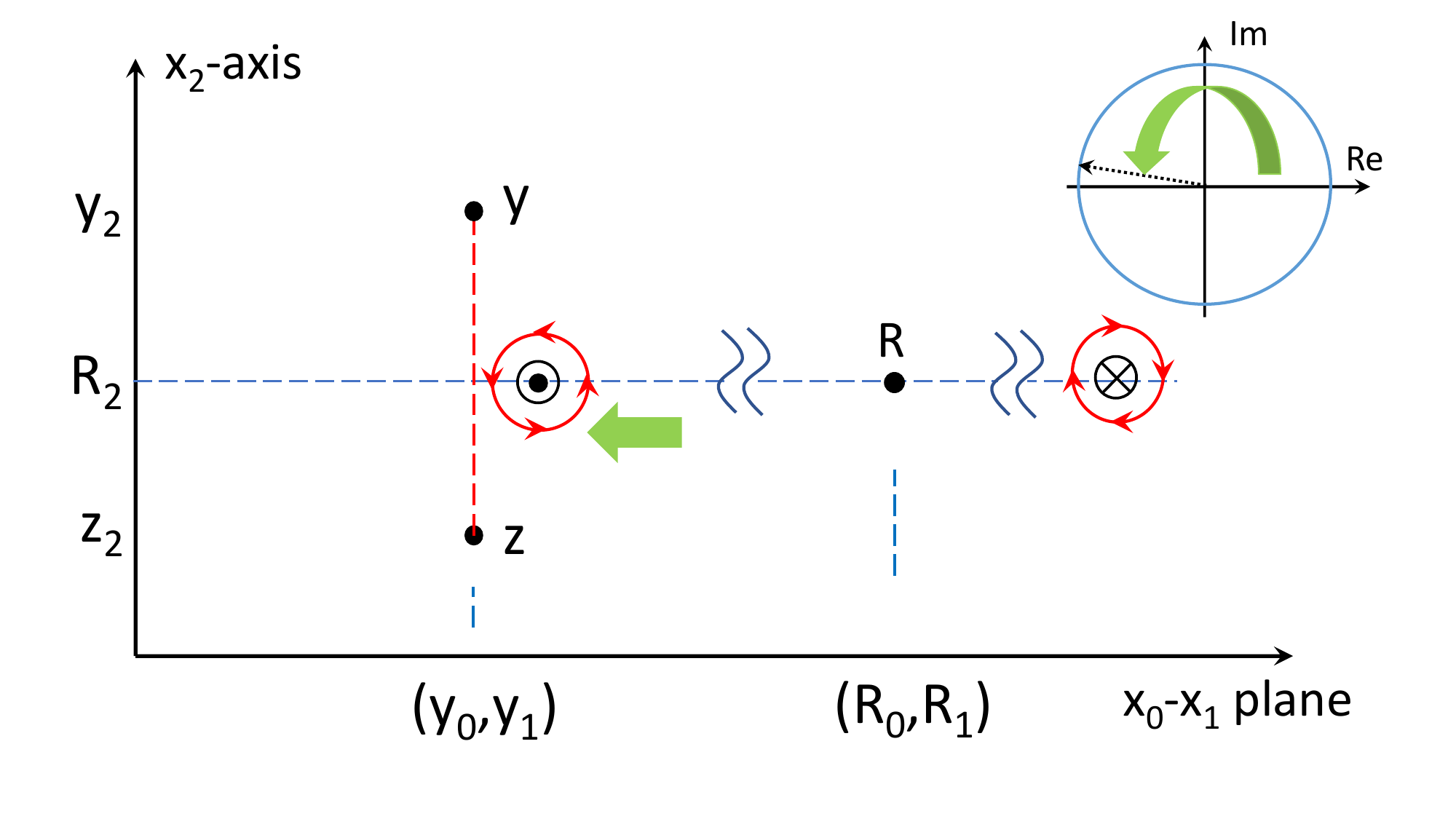}%
            \label{Subfig:2b}%
        }\\
        \subfloat[The segment crosses the integral line in Eqs.~(\ref{Eq:IV-0-1},\ref{Eq:IV-0-2})]{%
            \includegraphics[width=.49\linewidth]{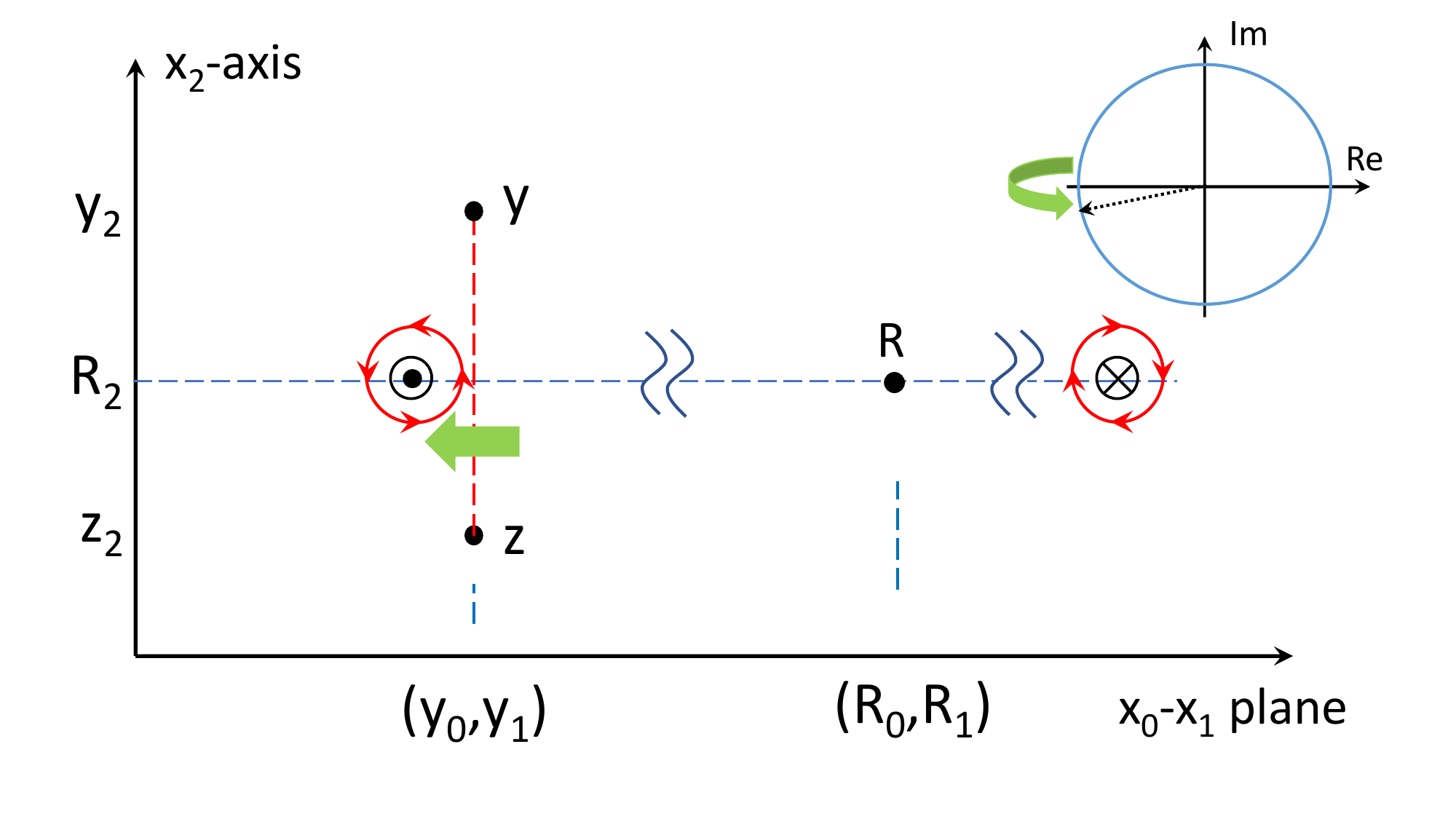}%
            \label{Subfig:2c}%
        }\hfill
        \subfloat[The loop becomes large enough]{%
            \includegraphics[width=.49\linewidth]{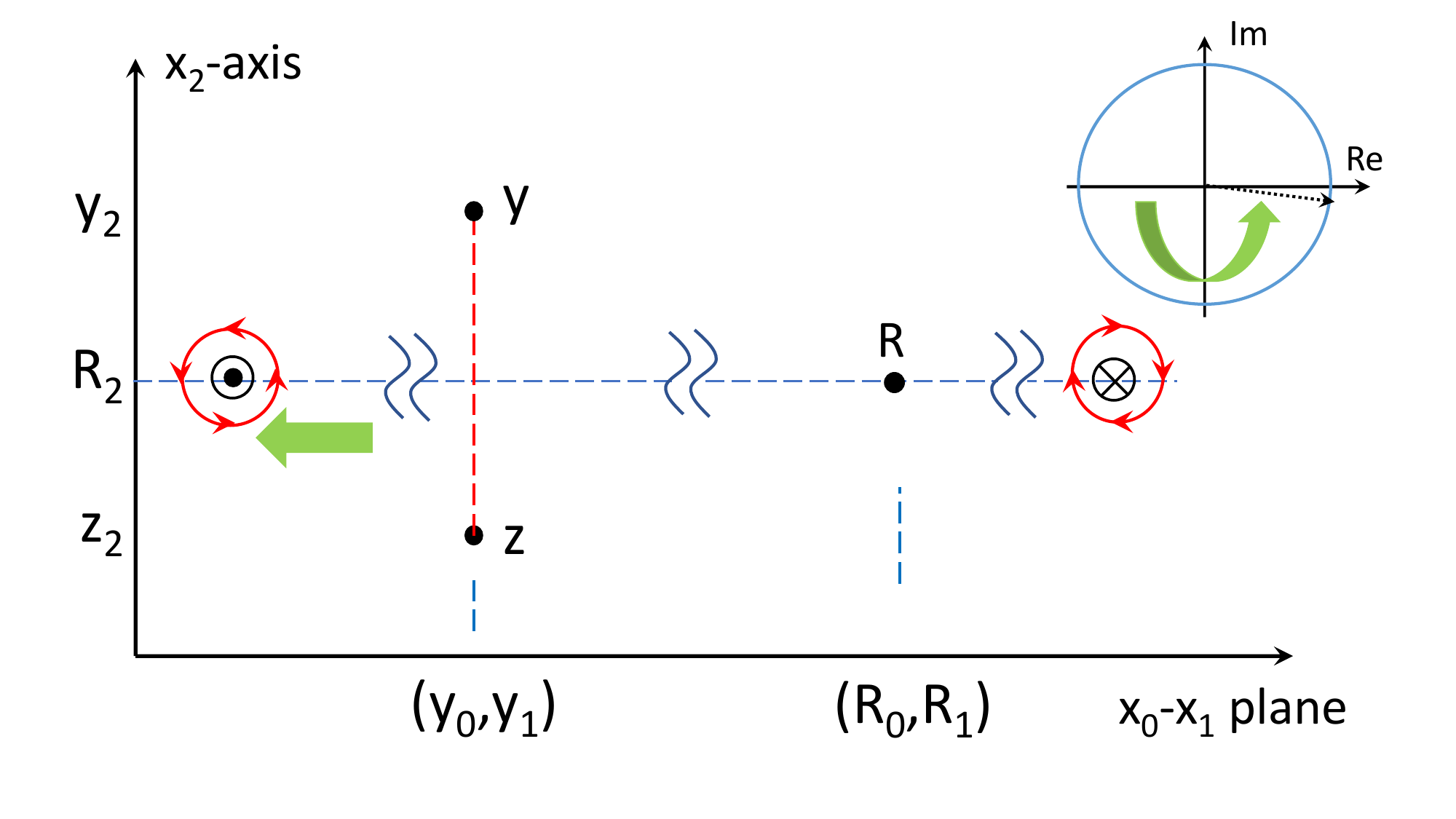}%
            \label{Subfig:2d}%
        }
        \caption{Four schematic figures explain how $e^{i\delta \theta(y,z)}$ moves along the unit circle in the complex plane when a vortex loop extends from a shorter loop to a larger loop in the perpendicular geometry [$y-z \perp $ the $x_0$ axis]. Small figure in the upper right corner of each figure shows the motion of $e^{i\delta\theta(y,z)}$ along the unit circle. Here, $y-z$ is along the $x_2$ axis, and a circular and coplanar vortex loop is placed on a $x_0$-$x_1$ plane between $y$ and $z$. A cross section of the vortex loop is denoted by $\odot$ and $\otimes$. Red lines with arrows denote a distribution of the magnetic field $B$ around the vortex loop. The center-of-mass coordinate $R$ of the loop is fixed far from the test points, $y$, $z$, while the diameter of the circular loop changes from a shorter length to a larger length. The integral line of Eqs.~(\ref{Eq:IV-0-1},\ref{Eq:IV-0-2}) is shown by a vertical dashed red line connecting $y$ and $z$. (a) When the loop is small, so is the magnetic field near the test points, $e^{i\delta\theta(y,z)}\simeq 0$. (b) When the loop gets larger and some segment of the loop is approaching the test points, the magnetic field near the test points is along $-x_2$ direction, and $\delta \theta(y,z)$ in Eq.~(\ref{Eq:IV-0-1}) increases from $0$ to $\pi-0$:  $e^{i\theta(y,z)}$ rotates from $0$ to $-1+i0$. (c) When the segment crosses $(y_0,y_1,R_2)$, the magnetic field in Eq.~(\ref{Eq:IV-0-1}) changes its sign, and  $\delta\theta(y,z)$ changes from $\pi$ to $-\pi$ : $e^{i\delta\theta(y,z)}$ moves from $-1+i 0$ to $-1-i0$. (d) When the loop becomes large enough, the magnetic field near the test points becomes smaller again: $e^{i\theta(y,z)}$ comes back to $+1$. }
        \label{Fig2}
\end{figure*}

      The proliferation of the vortex loops polarized along the $x_0$ axis renders a correlation of $e^{i\theta(x)}$ within the $x_1$-$x_2$ plane to be strongly disordered, while leaving the correlation along the $x_0$ axis intact.  To see this anisotropy, let us consider a relative U(1) phase between the two "test" points $y$ and $z$, $\theta(y,z) \equiv \theta(y)-\theta(z)$, and see how much an add-on phase $e^{i\delta\theta(y,z)}$ the relative phase $e^{i\theta(y,z)}$ acquires from an introduction of a polarized vortex loop.  A vortex loop $\Gamma$ induces a magnetic scalar potential $\theta_m(x)$:  $\nabla \times \nabla \theta_m(x) = \oint_\Gamma d x^{\prime} \!\ \delta(x-x^{\prime})$, and the magnetic scalar potential changes the relative U(1) phase, $\theta(y,z) \rightarrow \theta(y,z) + \theta_m(y)-\theta_m(z)$. Such phase change is given by a line integral of a magnetic field $B(x) = - \nabla \theta_m(x)$ along an arbitrary line connecting $y$ and $z$,
\begin{align}
\delta \theta(y,z) = \int^z_{y} dx \cdot B(x). \label{Eq:IV-0-1}
\end{align}
$\delta\theta(y,z)$ is determined up to a multiple of $2\pi$. Arbitrary multiple of $2\pi$ comes from a choice of the integral line, while $\exp [i\delta \theta(y,z)]$ is free from the choice of the integral line,
\begin{align}
e^{i\delta \theta(y,z)} = \exp \bigg[i\int^z_y dx \cdot B(x) \bigg]. \label{Eq:IV-0-2}
\end{align}
The Biot-Savart law gives $B(x)$ as a "magnetic induction" generated by a "quantized electric current" along the loop~\cite{jackson1999},   
\begin{align}
B(x) = \frac{1}{2} \oint_{\Gamma} dx^{\prime} \times \nabla \Big(\frac{1}{|x-x^{\prime}|}\Big).  \label{Eq:IV-0-3}
\end{align} 
       In the perpendicular geometry [$y-z \perp $ the $x_0$ axis], the add-on phase $e^{i\delta\theta(y,z)}$ always rotates 360 degrees around the origin in the complex plane when the polarized vortex loop extends from a shorter loop to a  larger loop. In the parallel geometry  [$y-z \parallel$  the $x_0$ axis], however, it does not make the rotation around the origin. To see the phase winding in the perpendicular geometry, let us place the two test points $y$ and $z$ along the $x_2$ axis, $y_0=z_0$, $y_1=z_1$, and $y_2 > z_2$ [Fig.~\ref{Fig2}],  and choose the $x_2$ axis as an integral line in Eqs.~(\ref{Eq:IV-0-1},\ref{Eq:IV-0-2}), 
\begin{align}
\delta \theta(y,z) = \int^{z_2}_{y_2} dx_2 B_2(y_0,y_1,x_2). \label{Eq:IV-0-4}
\end{align} 
Consider that a small circular loop is introduced in a $x_1$-$x_0$ plane between $y$  and $z$,  $y_2<R_2<z_2$, and its center-of-mass coordinate $R$ is fixed far away from $y$ and $z$, $(y_2-z_2)^2 \ll (R_0-y_0)^2+(R_1-y_1)^2$ 
[see Fig.~\ref{Fig2}]. The vorticity of the loop is counterclockwise when viewed from $y$ .

        When the circular loop extends from a smaller loop to a larger loop, the complex unit number  $e^{i\delta \theta(y,z)}$ always makes one full counterclockwise rotation around the origin  [Fig.~\ref{Fig2}]. For the smaller loop, the magnetic field around the test points is small: so is $\delta \theta(y,z)$ in Eq.~(\ref{Eq:IV-0-4}),  $\delta \theta(y,z) \simeq 0$. When the loop gets larger, some segment of the loop is nearing the test points, which increases the magnetic field  $B(y_0,y_1,R_2)$ at $(y_0,y_1,R_2)$ along the $-x_2$ direction. When the segment crosses $(y_0,y_1,R_2)$, the magnetic field $B(y_0,y_1,R_2)$ changes its direction from $-x_2$ direction into the $+x_2$ direction. Importantly, $\delta \theta(y,z)$ defined in eq.~(\ref{Eq:IV-0-4}) takes values of $\pi-0$ and $-\pi+0$ before and after the segment crosses $(y_0,y_1,R_2)$. When the loop becomes even larger,  the field strength at the test points decreases with the power of the distance between the segment and the test points.  Thus, the unit complex number $e^{i\delta\theta(y,z)}$  makes a full counterclockwise rotation around the origin in the complex plane during the change of the loop size.

     The argument so far does not depend on the detailed shapes of the vortex loop, as well as its center-of-mass coordinate $R$. It holds in other geometry with a finite angle between $y-z$ and the $x_0$ axis, as far as a loop segment crosses the line between $y$ and $z$ during the size change of the polarized loop. Namely, the winding of $e^{i\delta\theta(y,z)}$ around the  unit circle depends only on how many times a vortex segment crosses the line between $y$ and $z$. In a geometry with a finite angle between $y-z$ and the $x_0$ axis, there always exist polarized vortex loops, whose segment crosses the line between $y$ and $z$ during the change of the loop size. Due to the phase windings, the add-on phase $\langle e^{i\delta\theta(y,z)} \rangle$ averaged over different sizes of the polarized vortex loops with the same fugacity ($[\ln t=0]$) always reduces to zero. On the other hand, in the parallel geometry [$y-z \parallel$ the $x_0$ axis], no coplanar vortex loop confined in planes parallel to the $x_0$ axis can cross the line between $y$ and  $z$ during its size change. Thus, an add-on phase averaged over the polarized loops with different sizes remains finite. This suggests that the proliferation of the polarized loop only destroys the phase coherence within the $x_1$-$x_2$ plane.

     Ultimately, the anisotropic phase coherence near the order-disorder transition point helps the emergence of a quasi-disordered phase between ordered and disordered phases. In the quasi-disordered phase, the U(1) order parameter has a divergent correlation length along the $x_0$ direction, and a finite correlation length within the other directions. To aim at a theoretical description of such a quasi-disordered phase, we develop, in the remaining part of this section, the perturbative RG theory for the loop-gas model with the 1D Berry phase term.

\subsection{Generic form of the 3D Coulomb interactions}
      In the presence of the Berry phase term, the Coulomb interaction potential acquires emergent anisotropies both in space and in vortex-vector space. A generic form of the emergent interaction potential is determined by  symmetries of a bare action in Eqs.~(\ref{Eq:II-2},\ref{Eq:II-6}). Suppose that the interaction among vortex-loop segments is described by a two-body interaction potential between vortex vectors at $x$ and at $y$,
\begin{align}
{\cal S}_0[v(x)] \equiv  
\int d^3 x \int d^3 y \!\ \!\ \!\ v(x) \cdot  H(x-y) \cdot v(y).  
\end{align}
As the two vectors are commutable, a 3 by 3 matrix-formed potential $H(x-y)$ is symmetrized,  $H(x-y)=H^T(y-x)$. The bare action in Eqs.~(\ref{Eq:II-2},\ref{Eq:II-6}) 
 has an inversion symmetry, $
v(x) \rightarrow - v(-x)$, as well as a continuous rotational symmetry around the $x_0$ axis, 
\begin{align}
v(x) \rightarrow \overline{v}(\overline{x}) &\equiv \left(\begin{array}{ccc}
\cos \varepsilon & \sin\varepsilon &  \\
-\sin \varepsilon & \cos\varepsilon & \\
& & 1 \\
\end{array}\right)  v(x), \nonumber \\
\overline{x} &\equiv \left(\begin{array}{ccc}
\cos \varepsilon & \sin\varepsilon &  \\
-\sin \varepsilon & \cos\varepsilon & \\
& & 1 \\
\end{array}\right) x. \label{Eq:IV-A-1}
 \end{align}
So does the interaction potential,  ${H}(x-y) = {H}(y-x) = {H}^T(y-x)$ and 
\begin{align}
 &H(x-y) = \nonumber \\
 &\left(\begin{array}{ccc}
\cos \varepsilon & -\sin\varepsilon &  \\
\sin \varepsilon & \cos\varepsilon & \\
& & 1 \\
\end{array}\right) \!\ {H}(\overline{x}-\overline{y}) \!\ \left(\begin{array}{ccc}
\cos \varepsilon & \sin\varepsilon &  \\
-\sin \varepsilon & \cos\varepsilon & \\
& & 1 \\
\end{array}\right).  \label{Eq:IV-A-2}
\end{align}

   The interaction potential in this section is most conveniently analyzed in terms of its Fourier transform $H(q)$, 
\begin{align}
H(x-y) \equiv \int_{q\ne 0} \frac{d^3 q}{(2\pi)^3} \!\  H(q) \!\ \!\ e^{-i q(x-y)},
\end{align}
with momentum $q\equiv (q_1,q_2,q_0)$. In the right-hand side, we take a principal value around $q=0$, because $\int d^3x \!\ v(x) =0$. The symmetries of $H(x-y)$ determine a form of 3 by 3 matrix $H(q)$ as,
\begin{align}
{H}(q) =& F_1(q)  + F_2(q) \!\ \eta_2 \nonumber \\
& + G_0(q)\big(b_1(q) +\eta_2\big) \!\ \hat{q} \!\ \hat{q}^T  \!\ \big(b_1(q)  + \eta_2\big), \label{Eq:IV-A-3}  
\end{align}
with a traceless diagonal matrix $\eta_2$, 
\begin{align}
\eta_2 \equiv \left(\begin{array}{ccc}
1 & & \\
& 1 & \\
& & -2\\
\end{array}\right).  \label{Eq:IV-A-4}
\end{align}
Here $\hat{q}\equiv q/|q|$ is a normalized $q$ vector, and a trace of a symmetric matrix $\hat{q} \!\ \hat{q}^T$ is 1. $F_1$, $F_2$, $G_0$ and $b_1$ are scalar functions of $q_{\perp} \equiv (q_1,q_2)$ and $q_0$. They are symmetric under the continuous rotation and inversion: $F_1(q) \equiv F_1(|q_{\perp}|,|q_0|)$,   $F_2(q) \equiv F_2(|q_{\perp}|,|q_0|)$, $ G_0(q) \equiv G_0(|q_{\perp}|,|q_0|)$, $b_1(q) \equiv b(|q_{\perp}|,|q_0|)$.  In the partition function with closed vortex loops, $b_1(q)$ terms in $H(q)$ do not contribute to ${\cal S}_0$, because they all appear in the action with the divergence of the vortex vector, 
\begin{align}
&{\cal S}_0[v(x)] = \cdots + \nonumber \\
&\  2\int d^3x \int d^3 y \big(\nabla\cdot v(x)\big) \!\ 
\big(h^{\prime}(x-y) \cdot \eta_2 \cdot v(y)\big)  \nonumber \\
& \ \ +\int d^3 x \int d^3 y \!\ \big(\nabla\cdot v(x)\big) \!\ h^{\prime\prime}(x-y) \!\ \big(\nabla \cdot v(y)\big),
\end{align}
Here $h^{\prime}(x-y)$, and $h^{\prime\prime}(x-y)$ are from the $b_1$ terms in $H(q)$,  
\begin{align}
\begin{aligned}
&h^{\prime}(x-y) \equiv \int_{q \ne 0} \frac{d^3 q}{(2\pi)^3} \!\ \frac{ib_1 G_0 \hat{q}}{q} e^{-iq(x-y)},  \\
&h^{\prime\prime}(x-y) \equiv \int_{q \ne 0} \frac{d^3 q}{(2\pi)^3} \!\ 
\frac{b^2_1 G_0}{q^2} e^{-iq(x-y)}.
\end{aligned}
\end{align}

      Thus, the 3D vortex loop model with the 1D Berry phase term can be generally studied by the following partition function, 
\begin{widetext}
\begin{align}
Z &= 1 + \sum^{\infty}_{n=1} \frac{1}{n!} \prod^n_{j=1} 
\Bigg(\int^{\infty}_{a} \bigg(\frac{dl_j}{a_0}\bigg) \!\ \!\ t^{l_j} \int \bigg(\frac{d^3 R_j}{a^3_0}\bigg) 
\int {D} \Omega_j(\lambda) \Bigg) \exp\bigg[-2\pi i \chi \sum^n_{i=1} S^i_{12} \nonumber \\
& \hspace{2cm} - \sum^n_{i,j=1}
\oint_{\Gamma_i}  \oint_{\Gamma_j}  
\Big \{ d x^i \cdot dy^j \!\ F_1(x^i-y^j) +  d x^i \cdot \eta_2 \cdot dy^j \!\ F_2(x^i-y^j) + dx^i \cdot G(x^i-y^j) \cdot dy^j\Big\} 
\bigg]. \label{Eq:IV-A-6}
\end{align}
\end{widetext}
The scalar functions $F_1(r)$ and $F_2(r)$ are the Fourier transforms of $F_1(q)$ and $F_2(q)$. A 3 by 3 symmetric matrix $G(r)$ is given by the scalar function $G_0(q)$,
\begin{align}
G(r) \equiv \int \frac{d^3 q}{(2\pi)^3}  \!\ G_0(q) \!\ \eta_2 \!\  \hat{q} \!\  \hat{q}^T  \eta_2 \!\ e^{-i q(x-y)}.  \label{Eq:IV-A-7}
\end{align}
$F_1(r)$, $F_2(r)$ and $G(r)$ that emerge from the isotropic Coulomb potential [$F_1(r) \propto 1/|r|$, $F_2=G=0$] must be also algebraic functions of $|r_{\perp}|$ and $r_0$ with the same exponent. Such interaction potentials need the  UV cutoff for the interaction length. We choose this UV cutoff in the same way as in Section II [see Eq.~(\ref{Eq:II-10}) ], 
\begin{align}
& \oint_{\Gamma_i} \oint_{\Gamma_j} \Big\{dx^i \cdot dy^j \cdots + dx^i\cdot \eta_2 \cdot dy^j \cdots + \cdots \Big\} \nonumber \\
& = \oint \oint_{|x^i-y^j|>a_0}\Big\{dx^i \cdot dy^j \cdots + dx^i\cdot \eta_2 \cdot dy^j \cdots + \cdots\Big\}. \label{Eq:IV-A-8}
\end{align}

\subsection{Renormalization group equations}
     To analyze the partition function in Eq.~(\ref{Eq:IV-A-6}) by the renormalization group method, we decompose the loop length integral in Eq.~(\ref{Eq:IV-A-6}) into short-length region ($a<l_j<ab$) and long-length region ($ab<l_j$), and decompose the interaction in Eq.~(\ref{Eq:IV-A-8}) into the short-distance part ($a_0<|x^i-y^j|<a_0b$) and the long-distance part ($a_0b<|x^i-y^j|$).  We expand the partition function up to the first order in the small $\ln b$. As in Section III, the vortex loop $\Gamma$ in the short-length region is represented by a coplanar symmetric circle with its diameter $a_0$, $a=a_0\pi$. To integrate out such short-loop DOF explicitly,  we Taylor-expand the action in the power of the interaction between the short loop and others, while leaving the 1D Berry phase term on the shoulder of the exponential function,   
\begin{widetext}
\begin{align}
Z_{\rm v} = &1 + \bigg(\int^{ab}_{a} dl \!\ t^l \int_V d^3R \int_{S_2} d^2\Omega\bigg)  
\!\ e^{-s_1(\Gamma)} \nonumber \\
& \ \  \ \ \ \ + \sum^{\infty}_{n=1}\frac{1}{n!} \prod^n_{j=1}
\bigg(\int^{\infty}_{ab} dl_j \!\ t^{l_j} \int d^3R_j \int D\Omega_j(\lambda)\bigg) \!\ \!\ e^{-\sum^n_{i=1} s_1(\Gamma_i) -\sum^n_{i,j=1}s_{0}(\Gamma_i,\Gamma_j) } 
\bigg\{1 - \sum^n_{j=1} \delta s_0(\Gamma_j) \bigg\} \nonumber \\
 & +   \sum^{\infty}_{n=1}\frac{1}{n!} \prod^n_{j=1} 
\!\ \bigg(\int^{\infty}_{ab} dl_j \!\ t^{l_j} \int d^3R_j \int D\Omega_j(\lambda)\bigg) 
\!\ \!\ e^{-\sum^n_{i=1}s_1(\Gamma_i) 
-\sum^n_{i,j=1} s_0(\Gamma_i,\Gamma_j)} \nonumber \\ 
& \hspace{1cm} \times \bigg(\int^{ab}_{a} dl \!\ t^{l}
\int_V d^3R \int_{S_2} d^2\Omega\bigg) \!\ e^{-s_1(\Gamma)}   
\bigg\{1 - \Big(\sum^n_{j=1} 2s_0(\Gamma,\Gamma_j)\Big) + \frac{1}{2} 
\Big(\sum^n_{j=1} 2s_0(\Gamma,\Gamma_j)\Big)^2 + \cdots \bigg\} \label{Eq:IV-A-9}
\end{align}
Here $s_1(\Gamma_j)$ and $s_{0}(\Gamma_i,\Gamma_j)$ are the 1D Berry phase term for the $j$-th loop, and the Coulombic interaction between the $i$-th and $j$-th loops, respectively,
\begin{align}
s_1(\Gamma_j) &\equiv 2\pi i \!\ \chi \!\ S^j_{12}, \label{Eq:IV-A-10b} 
\end{align}
\begin{align}
s_0(\Gamma_i,\Gamma_j) &\equiv \oint_{\Gamma_i}  \oint^{a_0b<|x^i-y^j|}_{\Gamma_j}  
\Big \{ d x^i \cdot dy^j \!\ F_1(x^i-y^j) + 
d x^i \cdot \eta_2 \cdot dy^j \!\ F_2(x^i-y^j) + dx^i \cdot G(x^i-y^j) \cdot dy^j\Big\}.  \label{Eq:IV-A-10}
\end{align}
$\delta s_0(\Gamma_j)$ is the short-distance part of the Coulomb interaction within the $j$-th loop, 
\begin{align}
&\delta s_0(\Gamma_j) \equiv \oint_{\Gamma_j} \oint_{a_0<|x^j-y^j|<a_0b} \Big \{ d x^j \cdot dy^j \!\ F_1(x^j-y^j) \!\  + 
d x^j \cdot \eta_2 \cdot dy^j \!\ F_2(x^j-y^j) + dx^j \cdot G(x^j-y^j) \cdot dy^j\Big\}. \label{Eq:IV-A-11}
\end{align}
As in section III, $s_0(\Gamma,\Gamma)\propto\ln b$ is omitted here, as its contribution appears with the higher power in $\ln b$. $\delta s_0(\Gamma_j)$ is on the order of $\ln b$, so that we expand the partition function only up to the first order in $\delta s_0(\Gamma_j)$.

 $\delta s_0(\Gamma_j)$ in  Eq.~(\ref{Eq:IV-A-9}) gives rise to 1-loop renormalization to the vortex fugacity parameter $\ln t$  [Appendix B]. After an integration over the short-loop DOF [$\int_V d^3 R \int_{S_2} d^2\Omega$], the first-order expansion term in $s_0(\Gamma,\Gamma_j)$ in the fourth term of Eq.~(\ref{Eq:IV-A-9}) generates 1-loop renormalization to the Berry phase term $s_{1}(\Gamma_j)$ in the third term of  Eq.~(\ref{Eq:IV-A-9}) [Appendix C]. The second-order expansion term in $s_0(\Gamma,\Gamma_j)$ in the fourth term of Eq.~(\ref{Eq:IV-A-9}) produces 1-loop renormalization to the interaction potentials $s_{0}(\Gamma_i,\Gamma_j)$ in the third term of Eq.~(\ref{Eq:IV-A-9})[Appendix D]. Note also that the second term as well as the zeroth-order expansion term in $s_0(\Gamma,\Gamma_j)$ in the fourth term in Eq.~(\ref{Eq:IV-A-9}) can be included as an inhomogeneous part of the free-energy renormalization. Consequently, the integration of the short-loop DOF yields the following partition function for the long-loop DOF,
\begin{align}
&Z_{\rm v} = e^{4\pi V \frac{\sin\nu}{\nu} a \!\ t^a \!\ \ln b} \Bigg( 1 + \sum^{\infty}_{n=1}
\frac{1}{n!} \prod^n_{j=1}\bigg(\int^{\infty}_{ab} dl_j \!\ 
\int d^3R_j \int D\Omega_j(\lambda) \bigg) e^{-i2\pi \overline{\chi}\sum^n_{j=1} S^j_{12} - \sum^n_{i,j=1} 
\overline{s}_0(\Gamma_i,\Gamma_j) + \sum^n_{j=1} l_j \ln \overline{t}} \Bigg). \label{Eq:IV-C-23} 
\end{align}
\end{widetext}

     The renormalized fugacity $\bar{t}$ in Eq.~(\ref{Eq:IV-C-23}) is generally given by the statistical average over multiple vortex-loops configurations [see Appendix B]. In the leading-order expansion in the power of $a_0/T$, however, the average can be taken over {\it single} vortex-loop configuration,  
\begin{align}
&\ln \overline{t} = \ln t - 2a_0 \ln b \!\ \Big\{ \big\langle F_1(\Omega \!\ a_0)\big\rangle
\nonumber \\
&\hspace{0.5cm} 
+ \big\langle  \Omega\cdot \eta \cdot \Omega \!\ F_2(\Omega \!\ a_0) \big\rangle 
+ \big\langle \Omega \cdot G(\Omega \!\ a_0) \cdot \Omega\big\rangle \Big\}, \label{Eq:IV-X-6a} 
\end{align}
with 
\begin{align}
  \big\langle f(\Omega a_0) \big\rangle &\equiv  \frac{\int dl \!\ t^{l}  \int D\Omega(\lambda) \!\ e^{-s_1(\Gamma) } \!\ f\big(\Omega (\lambda) \!\ a_0\big)}{\int dl \!\ t^{l} \int D\Omega(\lambda) \!\ e^{-s_1(\Gamma) } }.  \label{Eq:IV-X-7}
 \end{align}
   
    The renormalized Berry phase $\overline{\chi}$ in Eq.~(\ref{Eq:IV-C-23}) is given by
\begin{align}
\overline{\chi} &\equiv \chi - \ln b \!\ \!\  a \!\ t^a \!\ a^2_0\pi \frac{\partial}{\partial \nu} \Big(\frac{\sin \nu}{\nu}\Big) \nonumber \\
& \int_{V} d^3 R \!\ \big(\nabla^2_{R_1} + \nabla^2_{R_2}\big) \big(F_1(R) + F_2(R)\big) , \label{Eq:IV-B-4}  
\end{align}
with $a=a_0\pi$ and $\nu \equiv \frac{a^2_0 \pi^2}{2}\chi$ [Appendix C]. For the isotropic Coulomb interaction [$F_1(R)= 1/|R|$, $F_2(R)=G(R)=0$], $(\nabla^2_{R_1}+\nabla^2_{R_2})F_1(R) = - \frac{8\pi}{3} \delta(R)$,  where the 1-loop renormalization to $\chi$ takes a negative value for small $\chi$ . For slightly generalized forms of $F_1(R)$ and $F_2(R)$ [see, for example, Eq.~(\ref{Eq:IV-E-2a})], the 1-loop renormalization to $\chi$ is also negative for small $\chi$. Thanks to the negative value, the 1D Berry phase term is always renormalized to zero near a high-$T$ fixed point with divergent fugacity $t$ [see Section IVE]. 

      $\overline{s}_0(\Gamma_i,\Gamma_j)$ in Eq.~(\ref{Eq:IV-C-23}) includes the three types of effective Coulomb interactions among the longer loops,   
\begin{widetext}
\begin{align}
\overline{s}_{0}(\Gamma_i,\Gamma_j) \equiv 
\oint_{\Gamma_i} \oint^{a_0b<|x^i-y^j|}_{\Gamma_j} 
\Big\{ dx^i \cdot dy^j \!\ \overline{F}_1(x^i-y^j) + dx^i \cdot\eta_2 dy^j \!\ \overline{F}_2(x^i-y^j) 
+ dx^i \cdot \overline{G}(x^i-y^j)\cdot dy^j\Big\}, \label{Eq:IV-C-25}
\end{align}
The renormalizations to these interaction potentials are most conveniently given in terms of their Fourier transforms, 
\begin{align}
\overline{F}_1(q) & \equiv F_1(q) - a \!\ t^a \!\ \big(c_{111}(q) + c_{121}(q) + c_{221}(q) \big) \!\ \ln b, \label{Eq:IV-C-26} \\ 
\overline{F}_2(q) & \equiv F_2(q) - a\!\ t^a \!\ \big(c_{112}(q) + c_{122}(q) + c_{222}(q)\big) \!\ \ln b, \label{Eq:IV-C-27} \\
 \overline{G}_0(q) & \equiv G_0(q) - a \!\ t^a \!\ \big(c_{120}(q) + c_{220}(q) + c_{100}(q) + c_{200}(q) + c_{000}(q) \big) \!\ \ln b. \label{Eq:IV-C-28}
\end{align}
\end{widetext}
Here $c_{ijk}(q)$ is the renormalization to $\overline{F}_{k}(q)$ due to the screening effect mediated by the $F_i(q)$ and $F_{j}(q)$ interactions.  $c_{ijk}(q)$ play a similar role as the operator product expansion coefficient in the perturbative renormalization group~\cite{cardy1996}. The detailed expressions of all the $c_{ijk}(q)$ are given in Appendix D [Eqs.~(\ref{Eq:IV-C-5a},\ref{Eq:IV-C-6a},\ref{Eq:IV-C-9a},\ref{Eq:IV-C-10a},\ref{Eq:IV-C-11a},\ref{Eq:IV-C-14a},\ref{Eq:IV-C-15a},\ref{Eq:IV-C-16a},\ref{Eq:IV-C-18},\ref{Eq:IV-C-20},\ref{Eq:IV-C-22})].

          After the length rescaling of Eq.~(\ref{Eq:III-1}) together with, 
\begin{align}
\begin{aligned}
&q^{\prime} = q \!\ b, \  \ S^{j\  \prime}_{12} = S^j_{12} \!\ b^{-2}, \ \ 
 \chi^{\prime} = \overline{\chi}\!\ b^{2},  \\
& F^{\prime}_j(q^{\prime}) = \overline{F}_j(q) \!\ b^{-1},  \ 
 \  F^{\prime}_j(r^{\prime}) = \overline{F}_j(r) \!\ b^2, 
 \end{aligned}   \label{Eq:IV-C-29}
\end{align}
we obtain closed RG equations for $F_1(q)$, $F_2(q)$, $G_0(q)$, $\ln t$ and $\chi$. To put the RG equations with proper normalizations, let us multiply $F_j(q)$ ($j=1,2$) and $G_0(q)$ by $q^2 \equiv q^2_1+q^2_2 + q^2_0$, 
\begin{align}
\begin{aligned}
&F_j(q)  \equiv \frac{4\pi}{q^2} f_j(\hat{q}),  \\
&G_0(q) \equiv \frac{4\pi}{q^2}  g_0(\hat{q}). 
\end{aligned} \label{Eq:IV-C-30}
\end{align}
$f_j(\hat{q})$ and $g_0(\hat{q})$ thus introduced depend only on $\hat{q}\equiv q/|q|$, which correspond to $\frac{\pi}{2T}$ in Section III.  By a  multiplication by the cutoff length scale $a_0$, we define three dimensionless functions $y_j(\hat{q})$ as follows,  
\begin{align}
\begin{aligned}
 &y_j(\hat{q}) \equiv a_0 f_j(\hat{q}),  \\
 &y_0(\hat{q}) \equiv a_0 g_0(\hat{q}). 
 \end{aligned}  \label{Eq:IV-C-32}
\end{align}
They share the same tree-level scaling dimension with $\frac{\pi a_0}{2T}$ in Section III: $ y^{\prime}_j(\hat{q}) = \overline{y}_j(\hat{q}) \!\ b$. For convenience, let us also define three functions $Y_j(r)$ ($j=1,2,0$) from $y_j(\hat{q})$,  
\begin{align}
\begin{aligned}
&Y_j(r) \equiv \int \frac{d^3 q}{(2\pi)^3} \frac{4\pi} {q^2} \!\  y_j(\hat{q}) e^{-iqr},  \\ 
&Y_0(r) \equiv \int \frac{d^3 q}{(2\pi)^3} \frac{4\pi} {q^2} \!\  y_0(\hat{q}) 
\!\ \eta_2 \!\ \hat{q}\!\ \hat{q}^T \!\ \eta_2\!\ e^{-iqr}, 
\end{aligned} \label{Eq:IV-C-33}
\end{align}
with $Y_j(r) = a_0 F_j(r)$ ($j=1,2$) and $Y_0(r) = a_0 G_0(r)$.

       The functional RG equations for normalized fugacity parameter $x \equiv a_0 \ln t$, normalized Berry phase term $\nu \equiv \frac{a^2_0\pi^2}{2} \chi$, and normalized Coulomb potentials $y_j(\hat{q})$ and $Y_j(r)$ $(j=1,2,0)$ are given by,  
\begin{align}
\frac{d x}{d\ln b}& = x - 2 \!\ \big\{ \big\langle Y_1(\Omega) \big\rangle \nonumber \\
&\ \ + \big\langle \Omega \cdot \eta_2 \cdot \Omega \!\ Y_2(\Omega) \big\rangle 
+ \big\langle \Omega \cdot  Y_0(\Omega) \cdot \Omega \big\rangle \big\}, \label{Eq:IV-C-34} \\
\frac{d \nu}{d\ln b} &= 2 \nu - \frac{3 \Delta}{4} \!\ \frac{\partial}{\partial \nu} 
\Big(\frac{\sin \nu}{\nu}\Big) \nonumber \\
& \hspace{1cm} \times \int_V d^3 r \!\ 
\!\ \nabla^2_{r_{\perp}} \big(Y_1(r) + Y_2(r)\big), \label{Eq:IV-C-35} \\
\frac{d y_1}{d\ln b} & = y_1  -  \Delta 
\Big\{ \big( (d_1 - d_2 ) \hat{q}^2_0\!\  + (d_1+2d_2) \hat{q}^2_{\perp}\big) \!\  y^2_1 \nonumber \\
&\hspace{1cm} - 4d_2 y_1y_2 + \big(2d_1+ 6d_2 \hat{q}^2_0 \big) \!\ y^2_2 \Big\}, \label{Eq:IV-C-36} \\
\frac{dy_2}{d\ln b}& = y_2 - \Delta  \Big\{ 
- d_2 y^2_1 + \big(2d_1 + 6d_2 \hat{q}^2_0\big) \!\ y_1 y_2
\nonumber \\
&\hspace{1cm}- \big(d_1 + d_2 ( 2 + 3 \hat{q}^2_0) \big) \!\ y^2_2 \Big\}, \label{Eq:IV-C-37} \\
\frac{dy_0}{d\ln b}& = y_0 - \Delta \Big\{ 
2 d_2 y_1 y_2 - \big(d_1 + 2d_2\big) \!\ y^2_2 \nonumber \\
&\hspace{1cm}+ 2\big(d_1 + d_2\big) \!\ y_1 y_0 
- 2\big(d_1 + d_2\big) \!\ \big(2\hat{q}^2_{\perp} \nonumber \\
& \hspace{1cm} -\hat{q}^2_0\big) y_2y_0 +9\big(d_1 
+ d_2\big) \!\ \hat{q}^2_{\perp}\hat{q}^2_0 y^2_0 \Big\}, \label{Eq:IV-C-38}
\end{align}
Here $\Delta \equiv  \frac{2\pi^5}{3} e^{\pi x}$ , $\hat{q}^2_{\perp} \equiv \hat{q}^2_1 + \hat{q}^2_2$. $d_1$ and $d_2$ are functions of the normalized Berry phase parameter $\nu $, 
\begin{align}
\begin{aligned}
&d_1  \equiv \frac{\sin \nu}{\nu}, \\
&d_2 \equiv - \frac{\sin \nu}{\nu} + \frac{3 (\sin \nu - \nu \cos\nu)}{\nu^3}.  
\end{aligned}\label{Eq:IV-C-2}
\end{align}

According to the functional RG equation, in the presence of non-zero Berry phase term $\nu$, the initial  isotropic Coulomb interaction [$y_1=\frac{\pi a_0}{2T}$, $y_2=y_0=0$] leads to some functions of $\hat{q}^2_{\perp}$ and $q^2_0$ for $y_j(\hat{q})$ ($j=1,2,0$). The generic form of these dimensionless potentials is given by rational functions of  $z\equiv \hat{q}^2_0$, 
\begin{align}
y_j(\hat{q}) = \frac{P_j(z)}{Q_j(z)}, \label{Eq:IV-E-1}
\end{align}
where $Q_j(z)\ge 0$  in a range of $0\le z\le1$. 

\subsection{RG equations with approximations} 
     To gain useful insight from the functional RG equations, we solve the equations with two approximations. 
First, we solve Eqs.~(\ref{Eq:IV-C-36},\ref{Eq:IV-C-37},\ref{Eq:IV-C-38}) only at the equator of the unit sphere $[(|\hat{q}_{\perp}|,|\hat{q}_0|)=(1,0)]$ and at the poles of the sphere $[(|\hat{q}_{\perp}|,|\hat{q}_0|)=(0,1)]$, and interpolate intermediate values of $y_{j}(\hat{q})$ in terms of the following {\it ansatzes},
\begin{align}
\begin{aligned}
&y_1(\hat{q}) = \frac{A_1}{\hat{q}^2_\perp + b_1 \hat{q}^2_0}, \!\ \!\ y_2(\hat{q}) = \frac{A_2}{\hat{q}^2_\perp + b_2 \hat{q}^2_0},  \\
&y_0(\hat{q}) 
= \frac{A_0}{(\hat{q}^2_{\perp} + b_0 \hat{q}^2_0)^2},   
\end{aligned} \label{Eq:IV-E-2}
\end{align}
with $b_j\ge 0$ $(j=1,2,0)$. The values of $y_j(\hat{q})$ at the equator and poles for $j=1,2$ and those of $y_0(\hat{q})$ are given by $A_j$ and $B_j\equiv A_j/b_j$, and by $A_0$ and $B_0 \equiv A_0/b^2_0$, respectively. The ansatzes give the following simplest Coulomb interaction potentials in the coordinate space,   
\begin{align}
&Y_1(r) = \frac{A_1}{\sqrt{b_1 r^2_{\perp} + r^2_0}},  \ \ \ Y_2(r) = \frac{A_2}{\sqrt{b_2 r^2_{\perp} + r^2_0}}, \label{Eq:IV-E-2a} \\
&Y_0(r) = \frac{A_0}{2b_0} \nonumber \\ 
& \ \left(\begin{array}{ccc}
\sqrt{b_0} & & \\
& \sqrt{b_0} & \\
& & 1 \\
\end{array}\right) \eta_2 \frac{1- \hat{\sf r}\!\ \hat{\sf r}^T}{{|\sf r}|} \eta_2 \left(\begin{array}{ccc}
\sqrt{b_0} & & \\
& \sqrt{b_0} & \\
& & 1 \\
\end{array}\right), \label{Eq:IV-E-2b}
\end{align}
with ${\sf r} \equiv (\sqrt{b_0} r_{\perp}, r_0)$. Thereby, the three $b_j$ parameters $(j=1,2,0)$ can be regarded as effective metrics associated with the three types of the Coulomb potentials. Especially, the $Y_0(r)$ interaction in Eq.~(\ref{Eq:IV-E-2b}) can be considered as the dipolar interaction modulated by $\eta_2$ in the reframed coordinate [Eq.~(\ref{Eq:IV-F-1})].  Eq.~(\ref{Eq:IV-E-2a}) also simplifies Eq.~(\ref{Eq:IV-C-35}) , 
\begin{align}
\frac{d \nu}{d\ln b} &= 2\nu + 2\Delta \!\ \frac{\partial}{\partial \nu} \Big(\frac{\sin \nu}{\nu}\Big) A_+ \nonumber \\
&=2\nu \big( 1 - \frac{\Delta}{3}(d_1+d_2) A_+\big), \label{Eq:IV-E-3}
\end{align}
with $A_+ \equiv A_1+A_2$. Most notably, the approximation reduces the functional RG equations into differential equations among only 6 coupling constants, $A_j$, $B_j$ $(j=1,2,0)$. In terms of  $A_{+}\equiv A_1+A_2$, $A_{-}\equiv A_1-2A_2$, $B_+\equiv B_1+B_2$, and $B_{-} \equiv B_1-2B_2$,  they are particularly simplified, 
\begin{align}
&\frac{dA_+}{d\ln b} = A_+\!\ \big( 1 - \Delta (d_1-2d_2) A_{+}\big), \label{Eq:IV-E-4} \\
&\frac{dA_{-}}{d\ln b} = A_{-} \big( 1 - \Delta \!\ (d_1+d_2) A_{-}\big), \label{Eq:IV-E-5} \\
&\frac{dB_{+}}{d\ln b}  = B_{+} \big( 1 - \Delta (d_1+d_2) B_{+}\big), \label{Eq:IV-E-6} \\
&\frac{dB_-}{d\ln b}  = B_{-} \big(1 - \Delta \!\  (d_1+4d_2) B_{-}\big), \label{Eq:IV-E-7}  \\ 
&\frac{dA_0}{d\ln b}  = A_0 \big( 1 - 2\Delta (d_1+d_2) A_{-} \big) \nonumber \\
&\hspace{0.6cm} -\frac{\Delta}{9} \big(  2(d_1+d_2) A_{+} A_{-} \nonumber \\
&\hspace{0.7cm} -  (d_1-2d_2) A^2_+ 
- (d_1+4d_2) A^2_{-}\big), \label{Eq:IV-E-8} \\
&\frac{d B_0}{d\ln b}  = B_0 \big( 1 - 2 \Delta (d_1+d_2) B_+ \big)  \nonumber \\
&\hspace{0.6cm} -\frac{\Delta}{9} \big(  2(d_1+d_2) B_{+} B_{-} \nonumber \\
&\hspace{0.7cm} -  (d_1-2d_2) B^2_+ 
- (d_1+4d_2) B^2_{-}\big). \label{Eq:IV-E-9} 
\end{align}

       Second, to evaluate the renormalization to the normalized fugacity parameter $x\equiv a_0 \ln t $ in Eq.~(\ref{Eq:IV-C-34}),
       \begin{align}
       &\frac{dx}{d\ln b} = x - 2 \Big\{\big\langle Y_1(\Omega) \big\rangle 
      \nonumber \\
      &\ \ \ + \big\langle \Omega \cdot \eta_2 \cdot \Omega \!\ \!\ Y_2(\Omega) \big\rangle
       + \big\langle \Omega \cdot Y_0(\Omega) \cdot \Omega \big\rangle \Big\},\label{Eq:IV-E-10a}
       \end{align}
we replace the average over the single vortex-loop configurations by an average with an equal weight for $\Omega$ over the unit sphere $S_2$,  
\begin{align}
\big\langle Y(\Omega) \big \rangle &\equiv 
   \frac{\int dl \!\ t^{l}  \int D\Omega(\lambda) \!\ e^{-s_1(\Gamma) } \!\ f\big(\Omega (\lambda) \!\ a_0\big)}{\int dl \!\ t^{l} \int D\Omega(\lambda) \!\ e^{-s_1(\Gamma) } } \nonumber \\
& = \frac{1}{4\pi}\int_{S_2} d^2\Omega \!\ \!\ Y\big(\Omega\big)     + {\cal O}(\chi). \label{Eq:IV-E-10}
\end{align}
Together with Eq.~(\ref{Eq:IV-E-2},\ref{Eq:IV-E-2a},\ref{Eq:IV-E-2b}),  the second approximation gives out the followings,  
\begin{align}
&\big\langle Y_1(\Omega) \big \rangle 
= \frac{A_1}{\sqrt{b_1-1}} \arcsin\bigg[\sqrt{\frac{b_1-1}{b_1}}\bigg], 
\label{Eq:IV-E-11} \\
&\big\langle \Omega \cdot\eta_2  \cdot \Omega\!\ \!\  Y_2(\Omega) \big \rangle =\nonumber \\
&\frac{A_2}{2(b_2-1)}\Big( 3 - \frac{2+b_2}{\sqrt{b_2-1}} \arcsin\bigg[\sqrt{\frac{b_2-1}{b_2}}\bigg] \Big), \label{Eq:IV-E-12} \\
&\big\langle \Omega \cdot Y_0(\Omega) \cdot\Omega \big\rangle = 
 \frac{9A_0}{4(b_0-1)^2}\Big(-5 + 8\sqrt{b_0} \nonumber \\
 & \hspace{2cm}+ \frac{2-5b_0}{\sqrt{b_0-1}}  \arcsin\bigg[\sqrt{\frac{b_0-1}{b_0}}\bigg]\Big), \label{Eq:IV-E-13}
\end{align}
for $1<b_i$. For $0<b_i<1$, $\frac{1}{\sqrt{b_i-1}}\arcsin[\sqrt{1-\frac{1}{b_i}}]$ is replaced by $\frac{1}{\sqrt{1-b_i}}{\rm arcsinh} [\sqrt{\frac{1}{b_i}-1}]$ [see also Appendix E].

\subsection{phase diagram and emergent anisotropic correlations}

\begin{figure}
\includegraphics[width= 0.95\linewidth]{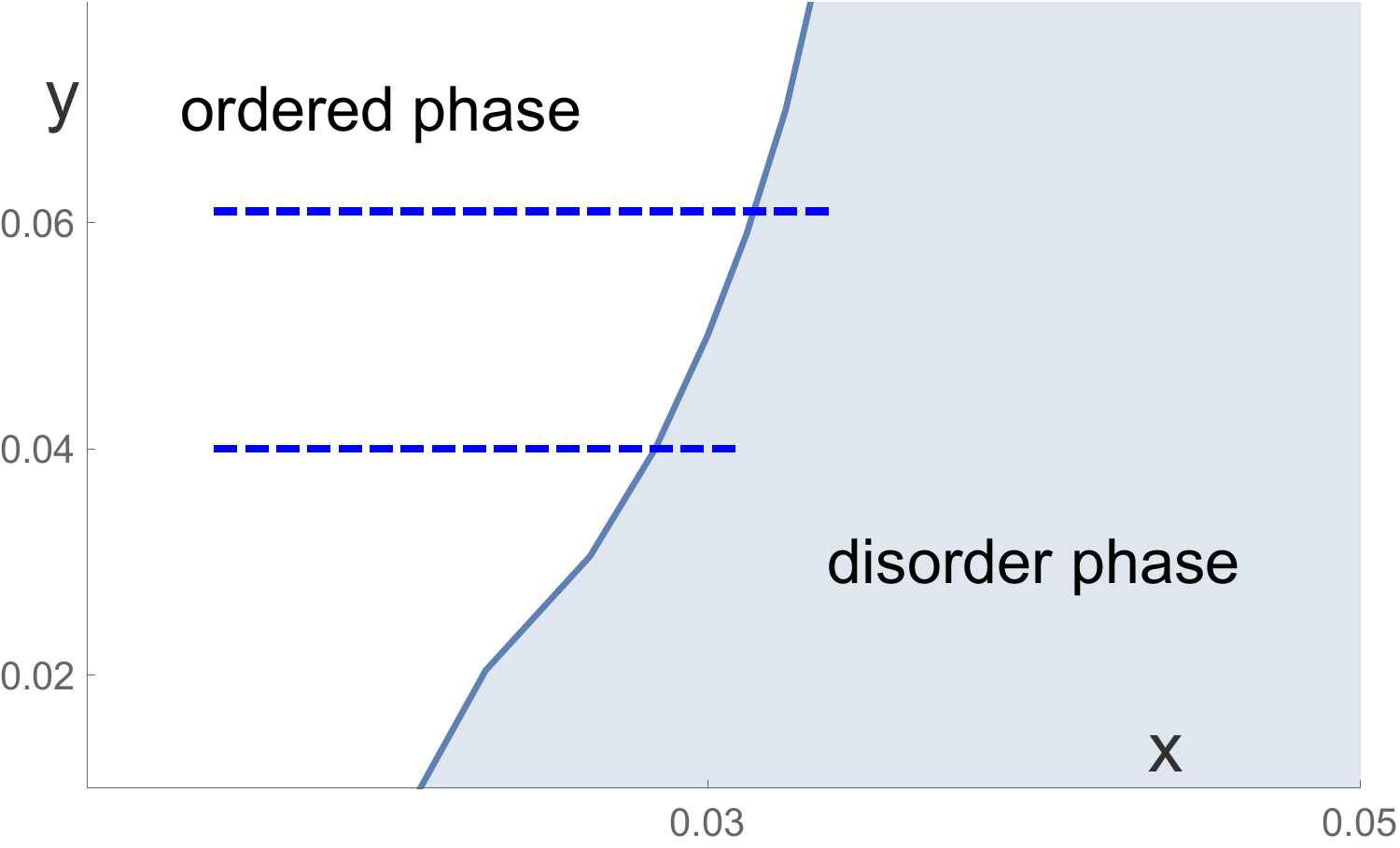} 
\caption{An RG phase diagram of 3D Coulomb loop gas model with the 1D Berry phase term $[\nu=0.5]$. The horizontal axis is the fugacity parameter $\pi x  \equiv \pi a_0 \ln t$, and vertical axis is $2\pi y \equiv  2\pi A_1=2\pi B_1$. The corresponding $x$-$y$ parameter region is shown  In Fig~\ref{Fig1}, as an area enclosed by blue dotted lines. Thermodynamics phase at a given $(x,y)$ is determined by the renormalized fugacity parameter  in a numerical solution of the RG equations, Eqs.~(\ref{Eq:IV-E-3}-\ref{Eq:IV-E-13}), with the parameter $(A_1,B_1,a_0\ln t,\nu)=(y,y,x,0.5)$ as the initial parameters $[A_2=B_2=0$ $(b_2=1)$, $A_0=B_0=0$ $(b_0=1)]$. In the ordered phase, $t$ vanishes and $A_j$ $(j=1,2,0)$ diverges in the IR limit [$\ln b \rightarrow \infty$], while in the disordered phase, $t$ diverges, and $A_j$ $(j=1,2,0)$ vanishes in the IR limit. Figs.~\ref{Fig4} show the renormalized $b_j$ ($j=1,2,0)$ as a function of $x$ along the blue dotted lines.}
\label{Fig3}
\end{figure}

        The approximate RG equations, Eqs.~(\ref{Eq:IV-E-3}-\ref{Eq:IV-E-13}), are numerically solved in a $x$-$y$ parameter region shown in Fig.~\ref{Fig3} with $x= a_0 \ln t$, $y \equiv A_{1}=B_1$, $A_2=B_2=A_0=B_0=0$ and $\nu=0.1$,  $0.5$, and $1.0$, where these parameters are used as initial values of the differential equations at $\ln b=0$. When the fugacity parameter $t$ diverges and vanishes in the IR limit [$\ln b \rightarrow \infty$], we consider that the initial parameters are in the disordered and ordered phases, respectively. The phase diagram thus determined has a disordered phase controlled by a high-$T$ fixed point with the divergent $t$, vanishing $\nu$ and vanishing $A_1$, and an ordered phase controlled by a low-$T$ fixed region with the vanishing $t$, divergent $\nu$ and divergent $A_1 >0$ [Fig.~\ref{Fig3}]. The high-$T$ fixed point is isotropic both in space and vortex-vector space, i.e. $b_1=b_2=b_0=1$, and $A_2/A_1=A_0/A_1=0$, where the dominant Coulomb interaction $A_1$ vanishes as  $A_1  \simeq  \frac{3}{2\pi^4} \!\ x e^{-\pi x}$. The disordered phase is essentially the same as the disordered phase without the 1D Berry phase term.

\begin{figure*}[t]
\centering
\begin{tabular}{@{}c@{}}
\subfloat{\includegraphics[width=0.6\linewidth]{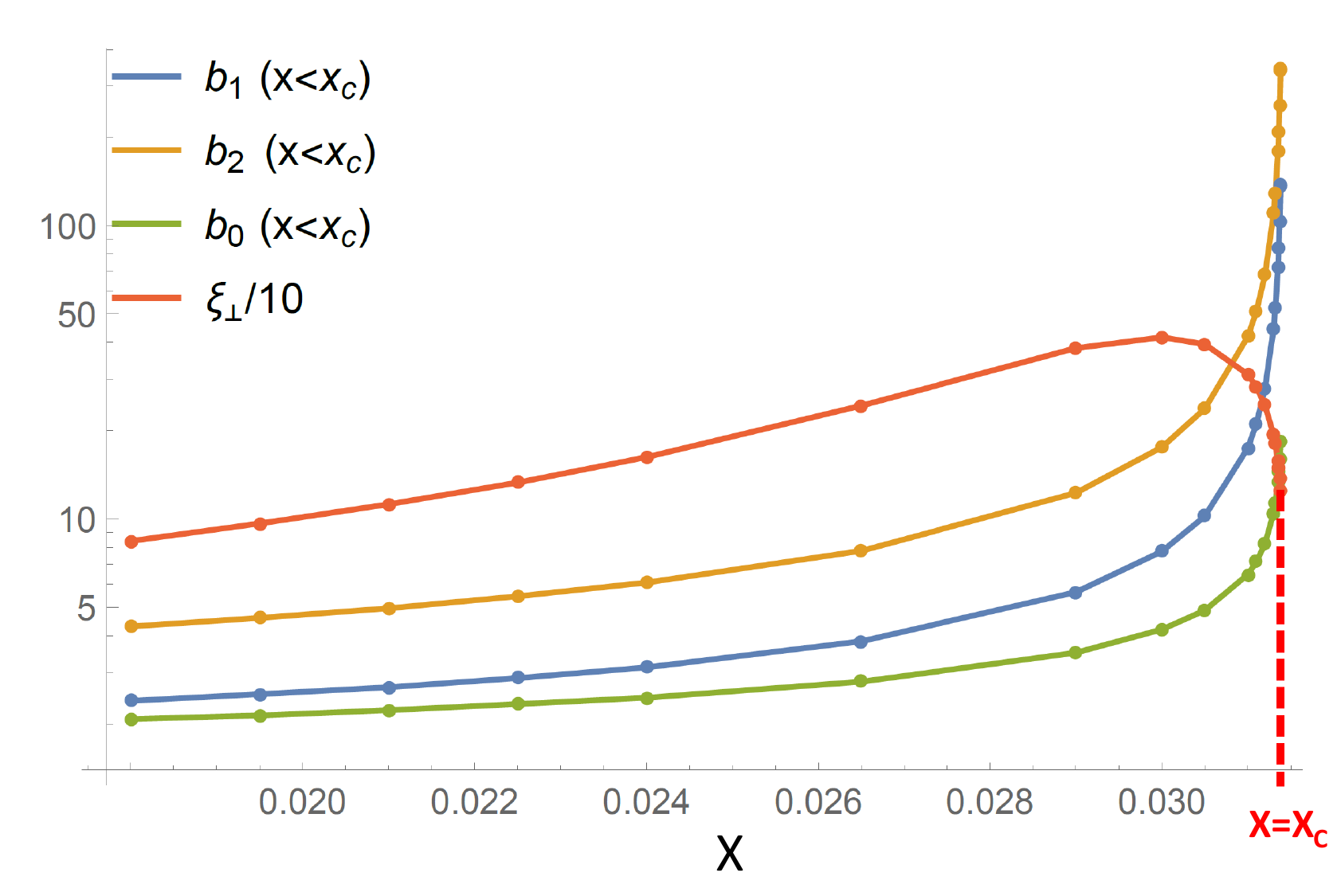}}\\ (a)
\end{tabular}\qquad 
\begin{tabular}{@{}c@{}}
\subfloat{\includegraphics[width=0.3\linewidth]{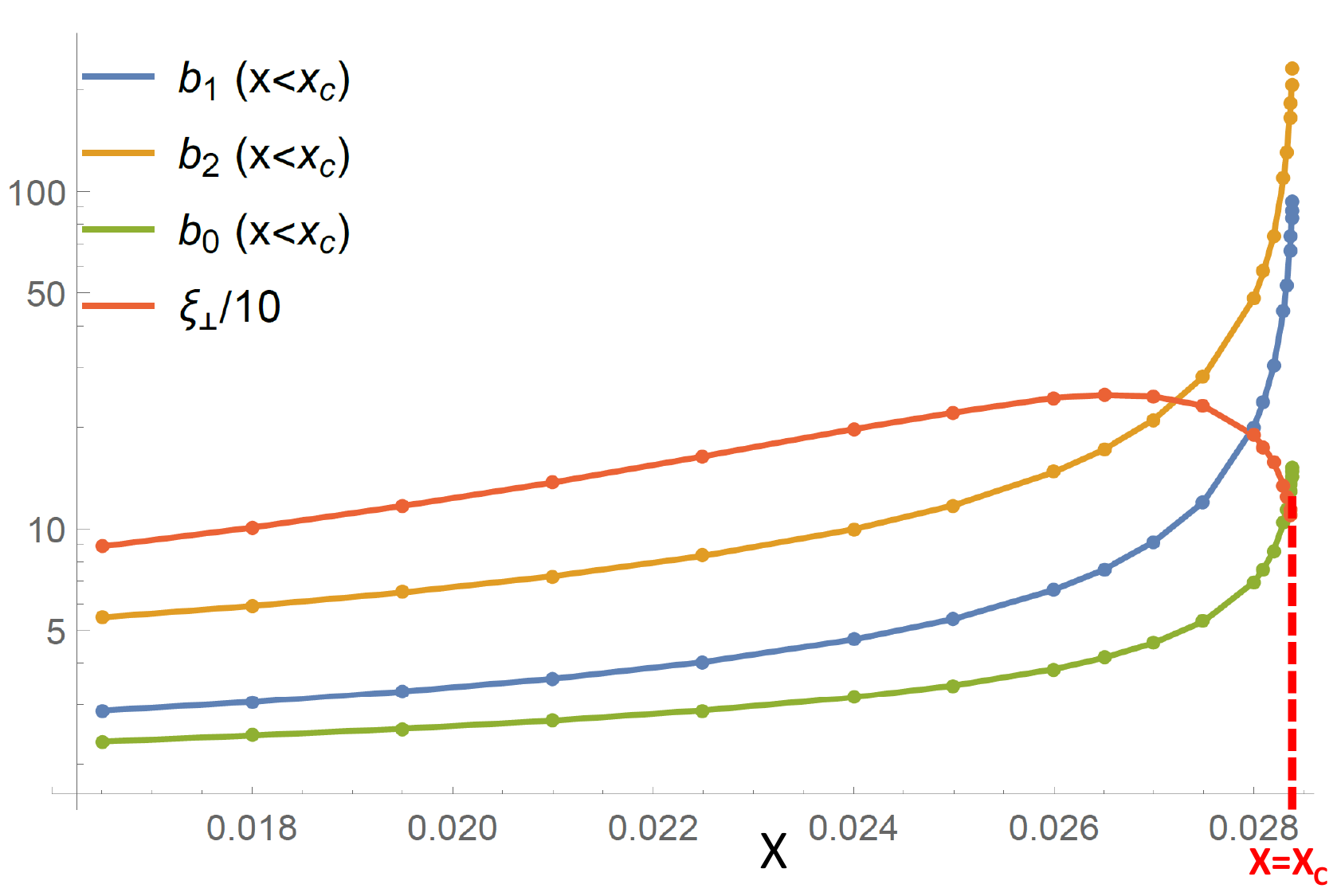}}\\ (b)
\\[0.1cm]
\subfloat{\includegraphics[width=0.3\linewidth]{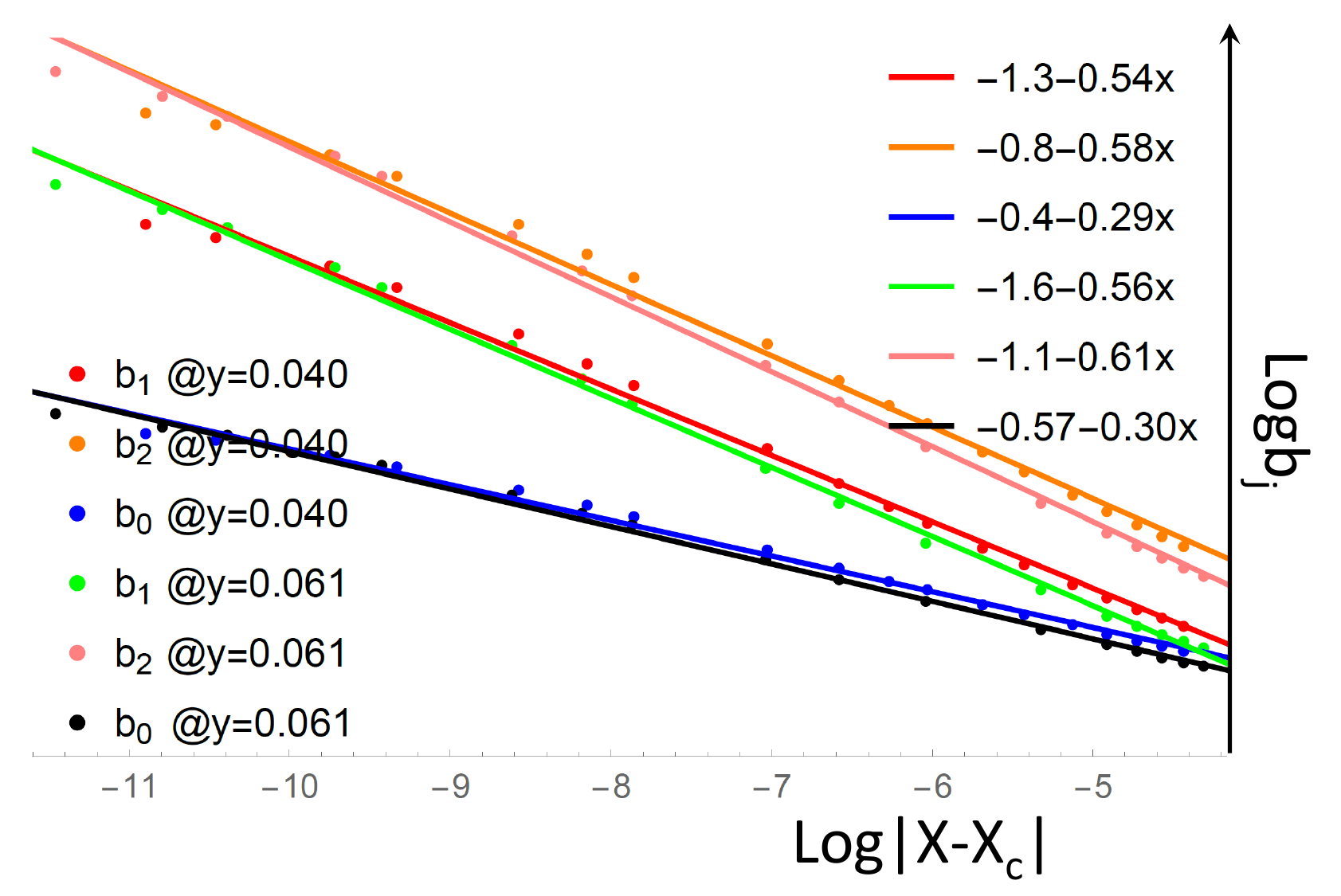}}\\ (c)
\end{tabular}
\caption{(a,b) renormalized $b_j$ and a length scale $\xi_{\perp}$ within the $1$-2 plane as a function of an initial parameter $x=a_0 \ln t$ for  (a) $y=0.061$, and (b) $y=0.040$.  The horizontal axis is the initial value of $x$. The values of $b_j$ in the vertical axis is a renormalized value of $b_j$ obtained from numerical solutions of the differential equations with the initial parameters of $(x,y)$, $\nu=0.5$, $A_1=B_1$  $A_2=B_2=A_0=B_0=0$,  and $b_1=b_2=b_0=1$ at $\ln b =0$. The renormalized value of $b_j$ is estimated at $x=-1$,} The value of $\xi_{\perp}$ in the vertical axis is an exponential of the RG scale factor $\ln b$ at $x=-1$, where the equations are solved numerically with the initial parameters of $(x,y)$, $\nu=0.5$, $A_1=B_1$  $A_2=B_2=A_0=B_0=0$,  and $b_1=b_2=b_0=1$ at $\ln b =0$.  (c) $\log b_j$ $(j=1,2,0)$ as a function of $\log|x-x_c|$ for $y=0.040$ (red, orange, blue) and $y=0.061$ (green, pink, black). Linear fitting curves is $\log b_j = c_0 + c_1 \log [|x-x_c|]$, where $c_1$ is around $-0.3$ for $j=0$, and ranges from $-0.5$ to $-0.6$ for $j=1,2$.
\label{Fig4}
\end{figure*}

               The low-$T$ fixed region is generally anisotropic both in space and in vortex-vector space, where, in the IR limit, all of $b_1$, $b_2$, and $b_0$ converge to finite positive values greater than $1$ [see, for example, Fig.~\ref{Fig4}(a,b)], and $A_2/A_1$ and $A_0/A_1$ approach finite positive and negative constants, respectively. $b_j>1$  for all $j=1,2,0$  indicates that a characteristic length scale $\xi_0$ along the topological ($x_0$) direction becomes longer than the length scale $\xi_{\perp}$ within the other two ($x_1$-$x_2$) directions. The characteristic length scale in the ordered phase of the sigma model is nothing but vortex-loop size. Thus $b_j>1$ means that the screening effect stretches vortex loops in the $x_0$ direction more than in the other two directions. In fact, the Berry phase term suppresses the screening effect of {\it unpolarized} vortex loops, and the screening effect of {\it polarized} vortex loops tends to reduce the isotropic Coulomb interaction within the  $x_1$-$x_2$ plane more than it reduces the Coulomb interaction along the $x_0$ direction. The induced spatial anisotropy in the Coulomb energy deforms vortex loops in such a way that they are elongated along $x_0$ relative to along the other two.

     The short-distance part of the $F_2(r)$ interaction determines the energetics of the {\it direction} of vortex-loop segments. Especially, the $F_2$ interaction with the positive $A_2$ favors a parallel alignment of neighboring vortex segments along the $x_0$ direction, yielding a straight vortex line along $x_0$. In fact, the screening effect of the polarized vortex loops tends to reduce the repulsive Coulomb interaction between two parallel vortex segments polarized along $x_0$ more than it does between two parallel segments polarized in the others.

      The $G(r)$ interaction with $b_0 \ne 1$ takes the form of the dipole-dipole interaction in a reframed coordinate, ${\sf x}\equiv (\sqrt{b_0} x_{\perp},x_0)$,  
\begin{align}
&\sum_{i,j}\oint_{\Gamma_i} \oint_{\Gamma_j} \!\ dx^i \cdot G(x-y) \cdot d y^j  
 =  \frac{A_0}{2a_0 b_0} \sum_{i,j} \nonumber \\
 & \hspace{1.4cm} \oint_{\Gamma_i} \oint_{\Gamma_j} 
 \frac{1}{ |{\sf r}_{ij}|}  \!\ \!\ d{\sf x}^i \cdot \eta_2 \cdot (1-\hat{\sf r}_{ij}\hat{\sf r}^{T}_{ij})\cdot \eta_2 \cdot d{\sf y}^j  \label{Eq:IV-F-1}
\end{align}
with ${\sf r}_{ij} \equiv {\sf x}^i-{\sf y}^j$. Notably, the $G(r)$ interaction between vortex-loop segments in the same vortex loop determines the energetics of the {\it curvature} of the vortex loop in the reframed coordinate space.  Due to the modulation by $\eta_2$, the $G$ interaction with the negative $A_0$ favors a vortex loop curving within ${\sf x}_0$-${\sf x}_1$ or ${\sf x}_0$-${\sf x}_2$ planes over a loop curving within ${\sf x}_1$-${\sf x}_2$ plane. This helps the vortex loop to be confined within a plane parallel to the ${\sf x}_0$ axis. With $b_0>1$, the polarized vortex loop is further stretched along the topological $(x_0)$ direction in the original spatial coordinate. 

    
        Interestingly, the numerical solutions also find that near the boundary between ordered and disordered phases, $b_j$ for all $j$ diverge with essential singular forms [Fig.~\ref{Fig4}(a,b)]. This suggests that a ratio between $\xi_0$ and $\xi_{\perp}$ diverges toward the boundary, $\xi_{0}/\xi_{\perp} \propto  \exp [-c/|x-x_c|^{\alpha}]$. From the fitting of $b_j$ for $j=1,2,0$, $\alpha$ is estimated around $\alpha=0.3\sim 0.6$ [Fig.~\ref{Fig4}(c)].

        The approximate RG equations have no saddle fixed points: a fixed point with the least number of relevant scaling variables turns out to have at least two relevant scaling variables. To deduce how  $\xi_{\perp}$ behaves near the ordered-disordered phase boundary, we regard that $\xi_{\perp}$ at $|\ln t|=1$ is on the order of the lattice constant $a_0$, and solve the equations inversely up to parameters near the boundary.  $\xi_{\perp}$ thus determined has no divergence around the boundary [Fig.~\ref{Fig4}(a,b)].

\section{Discussion}

      In this paper, we have developed a perturbative renormalization group theory for 3D U(1) sigma model with 1D Berry phase term. An ordered-disorder transition of the U(1) sigma model is induced by the spatial proliferation of vortex loops. The 1D Berry phase term confers a complex phase factor upon those unpolarized vortex loops that have finite projections in a space complementary to a topological direction with the 1D Berry phase. The complex phase factor suppresses screening effects of the unpolarized vortex loops, while the screening effects of the polarized vortex loops tend to confine other vortex loops to be within planes parallel to the topological direction. It also stretches the vortex loops in the topological direction more than in the other two directions. Numerical analyses of the approximate RG equations show that near a boundary between ordered and disordered phases, the length scale along the topological direction becomes anomalously larger, while the length scale within the other directions remains finite [Fig.~\ref{Fig0}(a)]. Ultimately, the extreme spatial anisotropy of the vortex-loop length may lead to an emergence of a quasi-disordered phase between ordered and disordered phases. In the quasi-disordered phase, the correlation length along the topological direction is divergent, while it is finite along the others.

      In fact, the emergence of the intermediate quasi-disordered phase is expected from a duality mapping~\cite{peskin1978,nguyen1999,herbutModernApproachCritical2007} between the U(1) sigma model and a lattice model of 3D type-II superconductor. To see this mapping, let us first start from Eqs.~(\ref{Eq:II-2}.\ref{Eq:II-3a}) and explain how to reach a partition function of a cubic-lattice model of the type-II superconductors,
      \begin{widetext}
\begin{align}
&1 +  \sum^{\infty}_{n=1} 
\frac{1}{n!} \prod^n_{j=1} \bigg(\int^{\infty}_{ab} \bigg(\frac{dl_j}{a_0}\bigg) \!\ t^{l_j} \int \bigg(\frac{d^3R_j}{a^3_0}\bigg) \int D\Omega_j\bigg)  \exp{\Big[-2\pi i \int A(x) \cdot v(x) \!\  d^3x -2\pi i \!\ \chi \sum_{j} S^j_{12}\Big]}  \nonumber \\
&\hspace{1cm} \rightarrow \prod^{j_1,j_2,j_0 \in \mathbf{Z}}_{{\bm j}=(j_1,j_2,j_0)} \bigg(\int^{2\pi}_{0} \frac{d\phi_{\bm j}}{2\pi}\bigg) 
\sum^{\infty}_{L=0} \frac{t^L}{L!} \Bigg(\sum_{{\bm m}=(m_1,m_2,m_0)} \sum_{\sigma =\pm} 
\sum_{\nu=1,2,0} e^{i\sigma \Big(\phi_{{\bm m}+\hat{\nu}} - \phi_{\bm m} 
+ 2\pi A_{{\bm m},\nu} + \pi \chi \big(\delta_{\nu,1} m_{2}-\delta_{\nu,2} m_1\big)\Big)}\Bigg)^L \nonumber \\
& \hspace{1.4cm} =  \prod_{\bm j} \bigg(\int^{2\pi}_{0} \frac{d\phi_{\bm j}}{2\pi}\bigg) 
{\rm exp} \Bigg[\!\ 2t \sum_{\bm m} \sum_{\nu} \cos \Big(\phi_{{\bm m}+\hat{\nu}} - \phi_{\bm m} 
+ 2\pi A_{{\bm m},\nu} + \pi \chi \big(\delta_{\nu,1} m_{2}-\delta_{\nu,2} m_1\big)\Big)\Bigg]. \label{Eq:V-1}
\end{align}
\end{widetext}
Here, ${\bm j}$, ${\bm m}$ and ${\bm m}+\hat{\nu}$ $(\nu=1,2,0)$ in the right-hand side are integer-valued 3D coordinate vectors of dual cubic lattice sites, and  $\hat{\nu}$  ($\nu=1,2,0$) stand for the three orthogonal unit vectors. In the lattice model, the vector potential $A(x)$ and  vortex vector $v(x)$ are represented by real-valued field $A_{{\bm m},\nu}$ and integer-valued field $l_{{\bm m},\nu} \in \mathbf{Z}$, respectively, both of which live on a dual-cubic-lattice link connecting ${\bm m}$ and ${\bm m}+\hat{\nu}$. Namely, we consider a unit plaquette of an original cubic lattice subtended by the two orthogonal unit vectors, e. g. $\hat{\mu}$ and $\hat{\lambda}$, and regard $\epsilon_{\mu\lambda\nu} l_{{\bm m},\nu}$ as a surface integral of $v(x)$ over the plaquette. Here, the dual-cubic-lattice link at $({\bm m},{\bm m}+\hat{\nu})$ penetrates the plaquette on the original cubic lattice. This allows us to enjoy the following translation, 
\begin{align}
\int A(x) \cdot v(x)\!\ d^3x \rightarrow \sum_{\bm m}\sum_{\nu=1,2,0} A_{{\bm m},\nu}  \!\ l_{{\bm m},\nu}. 
\end{align}
As $\nabla\cdot v(x)=0$, the integer-valued vector $l_{\bm m} \equiv (l_{{\bm m},1},l_{{\bm m},2},l_{{\bm m},0})$ also obeys a divergence-free condition on the lattice, $\sum_{\mu=1,2,0} (l_{{\bm m},\mu}-l_{{\bm m}-\hat{\mu},\mu})=0$.  In the right-hand side of Eq.~(\ref{Eq:V-1}), a sum over all possible configurations of $l_{\bm m}$ under the divergence-free condition is performed by multiple integrals over U(1) phase variables $\phi_{\bm m}$ defined on the dual cubic lattice sites. Thereby, $\sigma=\pm$ stands for $\pm1$ vorticity of the vortex segment on the link  $({\bm m},{\bm m}+\hat{\nu})$, and $\ln t$ stands for the chemical potential of the vortex loop with the unit vorticity and per the unit cubic-lattice constant. The integer $L$ in the right-hand side represents a sum of lengths of all the closed vortex loops on the dual cubic lattice. A trivial addition of the Maxwell term in Eq.~(\ref{Eq:II-3a}) into the right-hand side of Eq.~(\ref{Eq:V-1}) completes the duality mapping, 
\begin{align}
Z_{\rm v}  \rightarrow & \int DA_{\bm m} \int D \phi_{\bm m} \!\ \!\  {\rm exp} \bigg[-\frac{T}{2} \sum_{\bm m} \big(\nabla \times A_{\bm m}\big)^2 \nonumber \\
&  + 2t \sum_{{\bm m}} \sum_{\nu} \cos \Big(\phi_{{\bm m}+\hat{\nu}} - \phi_{\bm m} + 2\pi A_{{\bm m},\nu} \nonumber \\
& \hspace{2.3cm} + \pi \chi \big(\delta_{\nu,1} m_{2}-\delta_{\nu,2} m_1\big)\Big)\bigg].
\end{align}

      The dual lattice model thus obtained portrays the magnetostatics of the type-II superconductor under an external magnetic field. In the dual  model, the U(1) phase variable $\phi_{\bm m}$ stands for a phase of the superconducting order parameter on the dual lattice site $\bm m$, and the fugacity parameter $t$ plays the role of the Josephson coupling between neighboring superconducting order parameters. Thereby, the disordered phase with the divergent $t$ in the sigma model maps into the superconducting (Meissner) phase, and the ordered phase with vanishing $t$ is mapped into the normal (Maxwell) phase in the dual model. Importantly, the 1D Berry phase term $\chi$ becomes an external magnetic field applied along the topological ($x_0$) direction in the dual lattice model.

         It is well known that the type-II superconductors under the magnetic field along $x_0$ have intermediate mixed phase(s), where magnetic flux lines run along $x_0$, and they are separated by a finite distance within $x_1$-$x_2$ planes~\cite{blatter1994}. Thus, the mapping indicates that an intermediate phase must also appear between ordered and disordered phases in the U(1) sigma model with the 1D Berry phase term: the intermediate phase is nothing but the quasi-disordered phase [Fig.~\ref{Fig0}(b)]. In fact, the extremely spatially anisotropic correlation function expected in the quasi-disordered phase is consistent with the physical properties of the mixed phases in the type-II  superconductors.

         To see the correspondence between the quasi-disordered phase in the sigma model and the mixed phase in the type-II superconductor, let us start from the correlation function of $e^{i\theta(x)}$ in the sigma model, and disregard its spin-wave contribution. That is to say, the correlation function is solely given by a Dirac-string field $h(x)$~\cite{peskin1978,herbutModernApproachCritical2007},    
\begin{align}
&\langle e^{-i(\theta(y)-\theta(z))} \rangle_{\rm NLSM}  \nonumber \\
& \simeq \frac{1}{Z_{\rm v}} 
\int D u_T \int DH  \!\ \!\ e^{-\frac{T}{2} \int H^2  \  d^3 x  - i \int (H+ h)\cdot u_T  \  d^3 x},  \label{Eq:V-2}
\end{align}
with $\nabla\cdot h(x) = \delta(x-y)-\delta(x-z)$, and $Z_{\rm v} = \int Du_T \int DH e^{-\frac{T}{2} \int H^2 \!\ d^3 x - i \int H\cdot u_T \!\ \!\ d^3x}$.  The Dirac string field $h(x)$ is a magnetic flux that emanates from a pair of magnetic charges placed at the two test points, $y$ and $z$. The Dirac string field yields an additional vector potential $a(x)$ in the left-hand side of Eq.~(\ref{Eq:V-1}) as $A(x) \rightarrow A(x) + a(x)$  with $h(x) \equiv \nabla\times a(x)$. Thus, the duality transformation relates the correlation function with a ratio between partition functions of the lattice superconductor model with and without the magnetic charges~\cite{peskin1978,herbutModernApproachCritical2007},
\begin{align}
\langle e^{i\theta(y)-i\theta(z)} \rangle  \rightarrow Z_{\rm LS}[a]/Z_{\rm LS}[a=0]. \label{Eq:V-3}
\end{align}
Here $Z_{\rm LS}[a]$ stands for the partition function with the magnetic charges,   
\begin{align}
Z_{\rm LS}[a] \equiv & \int DA_{\bm m} \int D \phi_{\bm m} \!\ \!\  {\rm exp} \bigg[-\frac{T}{2} \sum_{\bm m} \big(\nabla \times A_{\bm m}\big)^2 \nonumber \\
& \ \ + 2t \sum_{\bm m} \sum_{\nu} \cos \Big(\phi_{{\bm m}+\hat{\nu}} - \phi_{\bm m} + 2\pi A_{{\bm m},\nu} \nonumber \\
& \ \ \ + 2\pi a_{{\bm m},\nu}+ \pi \chi \big(\delta_{\nu,1} m_{2}-\delta_{\nu,2} m_1\big)\Big)\bigg].  \label{Eq:V-4} 
\end{align}
 The lattice rotation of $a_{{\bm m}}$ describes the Dirac string field  $h_{\overline{\bm m}} \equiv (\nabla \times a_{\bm m})_{\nu} \equiv \epsilon_{\nu\lambda\mu} (a_{{\bm m},\lambda}-a_{{\bm m}+\hat{\mu},\lambda})$. The Dirac string field on the lattice can be depicted by a quantized flux line with an arrow that goes through the original cubic lattice sites from $\overline{\bm n}$ to $\overline{\bm l}$ [$\overline{\bm n}$ and $\overline{\bm l}$  denote the original cubic lattice sites that correspond to the two test points, $y$ and $z$, respectively]. To be specific, $ h_{\overline{\bm m},\nu} $ is equal to $+1$ and $-1$ when the original-cubic-lattice link $(\overline{\bm m},\overline{\bm m}+\hat{\nu})$ is on the Dirac string line, and $\hat{\nu}$ is  parallel and antiparallel to the arrow, respectively. $h_{{\bm m},\nu}=0$, otherwise. 

          In mixed phases with the magnetic flux lines along $x_0$, the right-hand side of Eq.~(\ref{Eq:V-3}) remains finite for large $|y-z|$ in the parallel geometry [$y-z \parallel$ the $x_0$ axis], while it decays exponentially in the distance in the perpendicular geometry [$y-z \perp$ the $x_0$ axis]. Namely, $Z_{\rm LS}[a]$  in the parallel geometry becomes independent of larger $|y-z|$, because the Dirac string field $h(x)$ emanating from the magnetic charges will be trapped by one of the magnetic flux lines near $y$ and $z$.  $h(x)$ trapped inside the flux line only feels the Maxwell term, so that the two magnetic charges attract each other by the 1/$|y-z|$ Coulomb interaction~\cite{herbutModernApproachCritical2007}. In the perpendicular geometry, however, $Z_{\rm LS}[a]$ decays exponentially in the larger distance $|y-z|$, because a superconducting region, to which the Dirac string field is exposed, and in which the $h(x)$ feels the Meissner mass, is inevitably proportional to $|y-z|$. These considerations together with Eq.~(\ref{Eq:V-3}) suggest that the quasi-disordered phase in the sigma model corresponds to the mixed phase in its dual model.
        
         In conclusion, the mapping together with the vortex physics in type-II superconductors indicates the emergence of the intermediate quasi-disordered phase in the U(1) sigma model with 1D Berry phase term. In the intermediate phase, the exponential correlation length is divergent along the topological ($x_0$) direction with the 1D Berry phase, while it is finite along the others. Contrary to the argument in this section, the result in Section IVE fails to find a fixed point for the intermediate quasi-disordered phase. In order to overcome the failure in the RG analysis, one may need to define {\it separately} the fugacity parameter $x_0$ along the topological direction, and the fugacity parameter $x_{\perp}$ along the other directions, e.g. 
\begin{align}
t^{l} &= \exp \Big[ x \int^{l}_0  \Omega^2(\lambda) \!\ \frac{d\lambda}{a_0} \Big]  \nonumber \\
&\rightarrow \exp \Big[x_0 \int^{l}_{0} \Omega^2_0(\lambda) \!\ \frac{d\lambda}{a_0} 
+ x_{\perp} \int^{l}_{0} \Omega^2_{\perp}(\lambda) \frac{d\lambda}{a_0} \Big].
\end{align}
Thereby, $x_0$ and $x_{\perp}$ must be renormalized by the $F_1$ and $F_2$ interactions differently, and the intermediate quasi-disorder phase is characterized by a fixed point with divergent $e^{x_0}$  and vanishing $e^{x_{\perp}}$. We leave this direction for future work.

\begin{acknowledgements} 
We thank Lingxian Kong, Yeyang Zhang, and Zhenyu Xiao for discussions. The work was supported by the National Basic Research Programs of China (No. 2019YFA0308401) and the National Natural Science Foundation of China (No. 11674011 and No. 12074008).
\end{acknowledgements}

\appendix
\section{Linear confining potential between monopole and antimonopole}

In the main text, we considered only vortex loops, while one could also consider a vortex line with open ends. In this appendix, we show that the two endpoints attract each other by a linear confining potential~\cite{peskin1978,herbutModernApproachCritical2007}. 

In the presence of the open ends, the divergence of the vortex vector is no longer zero, and it gives two point charges that can be regarded as a magnetic monopole and an anti-monopole, 
\begin{align}
\nabla \cdot v(x) &= \int^{x^j(l^j)=x_{\rm am}}_{x^j(0)=x_{\rm mm}} dx^j \cdot \nabla\delta(x-x^j) \nonumber \\
&= \delta(x-x_{\rm mm}) -\delta(x-x_{\rm am}). 
\end{align}
Here, two ends are at $x=x_{\rm mm}$ and $x=x_{\rm am}$, and vorticity is from $x=x_{\rm mm}$ to $x=x_{\rm am}$.  An interaction between the two point charges is encoded in the second term of Eq.~(\ref{Eq:II-5}),
\begin{align}
&- \frac{4\pi^2}{2T}  \int_{k\ne 0} \frac{dk^3}{(2\pi)^3} \frac{1}{k^4} 
\big(\nabla\cdot v\big)_k \big(\nabla\cdot v\big)_{-k} \nonumber \\
&\equiv - \frac{1}{2T} \int d^3x \int d^3y \!\ V(x-y) \!\ \big(\nabla\cdot v(x)\big) \big(\nabla\cdot v(y)\big). 
\label{Eq:A-1}
\end{align}
The interaction potential thus introduced is attractive and linear in the distance between the charges,
\begin{align}
V(r) &= \int_{k\ne 0} \frac{d^3k}{(2\pi)^3} \frac{4\pi^2}{k^4} e^{ikr} = \frac{1}{i|r|} 
\int^{\infty}_{\varepsilon} dk \!\ \frac{e^{ik|r|}-e^{-ik|r|}}{k^3}  \nonumber \\
& = -\frac{1}{i|r|} \Big[\frac{1}{\varepsilon^2} 
\int^0_{\pi} id\theta \!\ e^{-2i\theta} + \frac{i|r|}{\varepsilon} 
\int^0_{\pi}id\theta \!\ e^{-i\theta} \nonumber \\ 
& \hspace{2cm} - \frac{1}{2} |r|^2 \int^0_{\pi} id\theta 
+ {\cal O}(\varepsilon)\Big] \nonumber \\
&= \frac{2}{\varepsilon} - \frac{\pi}{2}|r| + {\cal O}(\varepsilon). \label{Eq:A-2}
\end{align}
Here, $\varepsilon$ is an infinitesimally small positive quantity associated with the principal-value integral in the first line. Since $\int \nabla\cdot v(x) dx=0$, the first term in the last line of Eq.~(\ref{Eq:A-2}) vanishes when substituted into Eq.~(\ref{Eq:A-1}), and we obtain the linear confining potential between magnetic monopole and anti-monopole,
\begin{align}
S_0 = \cdots -\frac{\pi}{2T} |x_{\rm am}-r_{\rm mm}|. 
\end{align}
A second-order spatial derivative of the linear confining potential yields a dipole-dipole interaction between  magnetic dipoles at $x_{\rm am}$ and $x_{\rm mm}$, 
\begin{align}
- \frac{\partial S_0}{\partial x_{{\rm am},\mu} \partial x_{{\rm mm},\nu}} 
= \frac{\pi}{2T} \Big(\frac{1-\hat{r}{\hat{r}^T}}{|r|}\Big), 
\end{align}
with $r \equiv x_{\rm am} - r_{\rm mm}$.

\section{renormalization to vortex fugacity parameter}
       In Sec.~IVC, we integrated out the short-loop DOF in Eq.~(\ref{Eq:IV-A-9}), and obtained the partition function for the long-loop DOF as in Eq.~(\ref{Eq:IV-C-23}) with the renormalized fugacity parameter $\overline{t}$, Berry phase term $\overline{\chi}$  and Coulomb interactions $\overline{s}_0(\Gamma_i,\Gamma_j)$. In this appendix, and the following two appendices, we calculate $\overline{t}$, $\overline{\chi}$ and $\overline{s}_0(\Gamma_i,\Gamma_j)$.  

        $\delta s_0(\Gamma_j)$ in Eq.~(\ref{Eq:IV-A-9}) gives rise to the renormalization of the vortex fugacity parameter $\ln t$. To see this, note that the distance between the two vortex-loop segments in $\delta s_0$ is a small quantity on the order of the lattice constant $a_0$. Thus, we expand $\delta s_0$ in the power of the distance $a_0$, keeping the leading order in the expansion,  
\begin{align}
&\delta s_0(\Gamma_j) =  \int^{l_j}_0 d\lambda \int_{a_0<|\lambda-\lambda^{\prime}|<a_0 b} 
d\lambda^{\prime} \nonumber \\
&\hspace{1.2cm} \Big\{\Omega_j(\lambda) \cdot \Omega_j (\lambda^{\prime})  \!\ 
F_1\big(x^j(\lambda) - x^j(\lambda^{\prime})\big) + \cdots \Big\} \nonumber \\
& \hspace{1.3cm} = a_0 \ln b \int^{l_j}_0 d\lambda \!\ \Big\{ \Omega_j(\lambda) \cdot 
\big( \Omega_j\big(\lambda + a_0\big) \nonumber \\
&\hspace{0.8cm} + \Omega_j\big(\lambda - a_0 \big) \big) \!\ 
F_1\big(\Omega_j(\lambda) \!\ a_0 +  {\cal O}(a^2_0)\big) + \cdots \Big\} \nonumber \\
&\hspace{1.3cm}= 2 a_0 \ln b \!\  
\int^{l_j}_{0} d\lambda  \!\ \Big\{ F_1\big(\Omega_j(\lambda) \!\ a_0\big)   \nonumber \\
&\hspace{1.6cm} + \Omega_j(\lambda)\cdot \eta_2 \cdot \Omega_j(\lambda) \!\ F_2\big(\Omega_j(\lambda) \!\ a_0\big) + \nonumber \\
& \hspace{1.4cm} + \Omega_j(\lambda)\cdot G\big(\Omega_j(\lambda)\!\ a_0\big)\cdot \Omega_j(\lambda) \Big\} + \cdots \label{Eq:IV-X-1}
\end{align}
Here ``$\cdots$" in the last line denotes higher-order gradient expansion terms in $a_0$, which renormalize other vortex loop parameters, e.g. the elastic energy parameter [See, for example, Section III].  Upon substitution into the third term in Eq.~(\ref{Eq:IV-A-9}), the leading-order expansion term can be included as the renormalization to the vortex fugacity parameter, 
\begin{align}
&\textsf{S}_n \!\  e^{-\sum^n_{j=1} s_1(\Gamma_j) - \sum^n_{i,j=1}s_0(\Gamma_i,\Gamma_j)} 
\bigg(1 - \sum^n_{k=1} \delta s(\Gamma_k) \bigg)  \nonumber \\
& \ = \textsf{S}_n \!\ e^{-\sum^n_{j=1} s_1(\Gamma_j) - \sum^n_{i,j=1}s_0(\Gamma_i,\Gamma_j)} 
\bigg( 1 - 2a_0 \ln b \sum^n_{k=1} \!\ l_k \!\ \nonumber \\
& \ \ \Big\{ \big\langle F_1(\Omega a_0)\big\rangle_n 
+ \big\langle  \Omega\cdot \eta \cdot \Omega \!\ F_2(\Omega a_0) \big\rangle_n
+ \big\langle \Omega \cdot G(\Omega a_0) \cdot \Omega\big\rangle_n \Big\} \bigg),  \label{Eq:IV-X-2}
\end{align}
with 
\begin{align}
\textsf{S}_n 
\equiv 
\prod^n_{j=1} \bigg(\int^{\infty}_{ab} dl_j \!\ t^{l_j} \int d^3R_j \int D\Omega_j(\lambda)\bigg). \label{Eq:IV-X-4}
\end{align}
Here $\langle \cdots \rangle_n$ denotes a statistical average over all possible $n$ vortex-loop configurations,  
\begin{align}
&\Big\langle f\big(\Omega\big) \Big\rangle_n \equiv \frac{\textsf{S}_n \!\ e^{-\sum^n_{j=1} s_1 - \sum^n_{i,j=1} s_0} f\big(\Omega_k (\lambda)\big)}{\textsf{S}_n \!\ e^{-\sum^n_{j=1} s_1 - \sum^n_{i,j=1} s_0}} 
\!\ . \label{Eq:IV-X-3} 
\end{align}
After the statistical average over the vortex-loop configurations, any function  $F(\Omega_k(\lambda) \!\ a_0)$ of a tangential vector $\Omega_k(\lambda)$ at $\lambda$  in the $k$th vortex loop becomes independent of $ \lambda$. Thus, the integral over $\lambda$ in the right-hand side of Eq.~(\ref{Eq:IV-X-1}) can be safely taken, yielding a factor of loop length [$l_k$] for each vortex loop. By the re-exponentiation, such a $\langle \delta s(\Gamma_k) \rangle_n$ renormalizes the vortex fugacity parameter, 
\begin{align}
&\textsf{S}_n \!\  e^{-\sum^n_{j=1} s_1(\Gamma_j) - \sum^n_{i,j=1}s_0(\Gamma_i,\Gamma_j)} 
\bigg(1 - \sum^n_{k=1} \delta s(\Gamma_k) \bigg) \nonumber \\
&=  \textsf{S}_n \!\ \!\ e^{\ln (\overline{t}/t) \sum^n_{j=1} l_j -\sum^n_{j=1} s_1(\Gamma_j) - \sum^n_{i,j=1}s_0(\Gamma_i,\Gamma_j) }. 
\label{Eq:IV-X-5}
\end{align}
Here the renormalized fugacity parameter $\overline{t}$ is given by
\begin{align}
&\ln \overline{t} = \ln t - 2a_0 \ln b \!\ \Big\{ \big\langle F_1(\Omega \!\ a_0)\big\rangle_n 
\nonumber \\
& \ \ \ + \big\langle  \Omega\cdot \eta \cdot \Omega \!\ F_2(\Omega \!\ a_0) \big\rangle_n 
+ \big\langle \Omega \cdot G(\Omega \!\ a_0) \cdot \Omega\big\rangle_n \Big\}. \label{Eq:IV-X-6}
\end{align}
The argument so far indicates that the renormalization to the fugacity parameter is dependent on a total number $n$ of vortex loops. As for its leading order expansion in the power of $a_0/T$, however, the renormalization becomes independent of $n$. Namely, as $F_1$, $F_2$, and $G$ are quantities on the order of $1/T$, they can be treated as small quantities, allowing a perturbative evaluation of the statistical average in the power of $a_0/T$, 
 \begin{align}
\big\langle f(\Omega) \big\rangle_n  &= \frac{\textsf{S}_n \!\ e^{-\sum^n_{j=1} s_1(\Gamma_j) } f\big(\Omega_k (\lambda) \big)}{\textsf{S}_n \!\ e^{-\sum^n_{j=1} s_1(\Gamma_j) }} 
 + {\cal O}(a_0 F_j f) \nonumber \\
 &\equiv \big\langle f(\Omega) \big\rangle  + \cdots, 
 \end{align}
 with 
 \begin{align}
 \big\langle f(\Omega) \big\rangle &\equiv  \frac{\int dl \!\ t^{l} \int d^3R  \int D\Omega(\lambda) \!\ e^{-s_1(\Gamma) } \!\ f\big(\Omega (\lambda)\big)}{\int dl \!\ t^{l} \int d^3R  \int D\Omega(\lambda) \!\ e^{-s_1(\Gamma) } }  \nonumber \\
 &= \frac{\int dl \!\ t^{l}  \int D\Omega(\lambda) \!\ e^{-s_1(\Gamma) } \!\ f\big(\Omega (\lambda) \big)}{\int dl \!\ t^{l} \int D\Omega(\lambda) \!\ e^{-s_1(\Gamma) } }.  \label{Eq:IV-X-7-ap}
 \end{align}
Importantly, the leading order term, Eq.~(\ref{Eq:IV-X-7-ap}), is given by an average over the single loop configuration: it is independent of the number $n$. In terms of the single-loop average, the renormalization to the fugacity parameter is given by, 
\begin{align}
&\ln \overline{t} = \ln t - 2a_0 \ln b \!\ \Big\{ \big\langle F_1(\Omega \!\ a_0)\big\rangle
\nonumber \\
&\ \ + \big\langle  \Omega\cdot \eta \cdot \Omega \!\ F_2(\Omega \!\ a_0) \big\rangle 
+ \big\langle \Omega \cdot G(\Omega \!\ a_0) \cdot \Omega\big\rangle \Big\}. \label{Eq:IV-X-6a-ap}
\end{align}

\section{renormalization to 1D Berry phase term}
    After the integration over the short-loop DOF in Eq.~(\ref{Eq:IV-A-9}) [$\int_V d^3 R \int_{S_2} d^2\Omega$], the first-order expansion term in $s_0(\Gamma,\Gamma_j)$ in the last line of Eq.~(\ref{Eq:IV-A-9}) generates the 1-loop renormalization to the Berry phase term. To see this, we consider the short loop $\Gamma$ as the circular loop with its diameter $a_0$,  and expand $s_0(\Gamma,\Gamma_j)$ in the power of the diameter $a_0$,
\begin{align}
&s_0(\Gamma,\Gamma_j) = \frac{a^2_0 \pi}{4} \Omega_{\mu} \epsilon_{\mu\lambda\phi} \oint_{\Gamma_j} 
\Big\{ \!\ dy^j_{\phi} \!\ \nabla_{R_{\lambda}} F_1(R-y^j) \nonumber \\
& + dy^j_{\phi}  \!\ (\eta_2)_{\phi\phi} \!\ \nabla_{R_{\lambda}} F_2(R-y^j) 
+ dy^j_{\psi} \!\  \nabla_{R_{\lambda}}G_{\phi\psi}(R-y^j) \!\ \Big\}. 
\label{Eq:IV-B-1}
\end{align}
The coplanar and circular loop $\Gamma$ is parameterized by its center-of-mass coordinate $R$, and a unit vector $\Omega\equiv  (\cos\psi\sin\theta,\sin\psi\sin\theta,\cos\theta)$ normal to the coplanar plane. An integration of  Eq.~(\ref{Eq:IV-B-1}) over $\Omega$ gives out 
\begin{align}
&\int_{S_2} d^2\Omega \!\ e^{-s_1(\Gamma)}  \sum_{j} 2s_0(\Gamma,\Gamma_j)
\nonumber \\
& =  2i (a_0\pi)^2 \frac{\partial}{\partial \nu} 
\bigg(\frac{\sin \nu}{\nu}\bigg) \epsilon_{0\lambda\phi} 
\sum_j \oint_{\Gamma_j} \nabla_{R_{\lambda}} \Big( d y^j_{\phi}\!\ F_1(R-y^j)  \nonumber \\
&\hspace{1cm}+ d y^j_{\phi}\!\ (\eta_2)_{\phi\phi} F_2(R-y^j) + dy^j_{\psi} G_{\phi\psi}(R-y^j)  \Big)
\nonumber \\
&  =   2i(\cdots) \sum_{j} \epsilon_{0\lambda\phi}  \epsilon_{\xi\eta\psi} \int_{S_j} dn_{\xi} \!\ \nabla_{R_{\lambda}} \nabla_{y^j_{\eta}} \Big( \delta_{\phi\psi} 
 F_1(R-y^j) \nonumber \\
 &\hspace{1cm} +  (\eta_2)_{\phi\psi}  F_2(R-y^j) 
+ G_{\phi\psi}(R-y^j) \Big), \label{Eq:IV-B-2}
\end{align}
where $s_1(\Gamma) \equiv i\nu \cos\theta$  and $\nu \equiv \frac{\pi^2 a^2_0}{2} \chi$.  From the 2nd line to the third line, the line integral with respect to $y^j$ along $\Gamma_j $ is transformed into a surface integral with respect to $y^j$ over an open surface $S_j$ with $\Gamma_j \equiv \partial S_j$.  $dn$ is a vector normal to $S_j$ at $y^j$, where $dn$ and the vorticity of $\Gamma_j$ obey the right-handed rule.

 An integration over the center-of-mass coordinate $R$ makes  Eq.~(\ref{Eq:IV-B-2}) into a quantity proportional to the projected area  $S^j_{12}$ for each closed loop $\Gamma_j$,  
\begin{align}
&\int_V d^3 R \int_{S_2} d^2\Omega \!\ e^{-s_1(\Gamma)}  
\sum_j 2s_{0}(\Gamma,\Gamma_j) \nonumber \\
& =  - 2i(\cdots) \sum_{j} \epsilon_{0\lambda\phi}  \epsilon_{\xi\eta\psi} \int_{S_j} dn_{\xi} \!\ \int_V d^3R  \!\ \!\ \nabla_{R_{\lambda}} \nabla_{R_{\eta}}\nonumber \\
& \hspace{1cm} \Big( \delta_{\phi\psi} 
\big( F_1(R-y^j) + F_2(R-y^j) \big) + G_{\phi\psi}(R-y^j) \Big) \nonumber \\
& =  - 2i(\cdots) \sum_j \big(\delta_{0\xi} \delta_{\lambda\eta} 
- \delta_{0\eta}\delta_{\lambda\xi} \big) \int_{S_j} dn_{\xi} \!\ \int_V d^3 R \nonumber \\
& \hspace{2cm}\nabla_{R_{\lambda}} \nabla_{R_{\eta}} \big(F_1(R-y^j) + F_2(R-y^j)\big) \nonumber 
\end{align}
\begin{align}
& = - 2i(\cdots) \sum^n_{j=1}  \int_V d^3 R \nonumber \\   
& \ \Big( \int_{S_j} dn_0  \nabla^2_{R_{\lambda}} \big(F_1(R-y^j)+F_2(R-y^j)\big) \nonumber \\
&\hspace{0.2cm} -  \int_{S_j} dn_{\lambda} \!\ \nabla_{R_0} \nabla_{R_{\lambda}} 
\big(F_1(R-y^j)+F_2(R-y^j)\big) \Big) \nonumber \\
&=  -  2i (a_0 \pi)^2 
\frac{\partial}{\partial \nu} \bigg(\frac{\sin \nu}{\nu}\bigg)  \nonumber \\
& \hspace{0.5cm}\times \int_V d^3 R  \big(\nabla^2_{R_1} + \nabla^2_{R_2}\big)\big(F_1(R)+F_2(R)\big ) 
\!\  \sum^n_{j=1}  S^j_{12}. \label{Eq:IV-B-3}
\end{align}
From Eq.~(\ref{Eq:IV-B-2}) to the second line, we use $\nabla_{y}=-\nabla_{R}$ and $\epsilon_{0\lambda\phi}(\eta_2)_{\phi\psi}=\epsilon_{0\lambda\phi}\delta_{\phi\psi}$. From the second line to the third line,  $G$ term is dropped from the integrand. This is because for $\xi=0$, the Fourier-transform of the $G$ term is zero, 
\begin{align}
&\epsilon_{0\lambda\phi}\epsilon_{0\eta\psi} \nabla_{R_\lambda}\nabla_{R_{\eta}} 
G_{\phi\psi}(R) \nonumber \\
& = - \int \frac{d^3q}{(2\pi)^3} 
\epsilon_{0\lambda\phi}\epsilon_{0\eta\psi} q^2 \hat{q}_{\lambda}\hat{q}_{\eta} 
\hat{q}_{\phi} \hat{q}_{\psi} G(q) e^{-iqR} = 0.
\end{align}
For $\xi=1,2$, the $G$ term is odd under the $\pi$-rotation around the $x_0$ axis [Eq.~(\ref{Eq:IV-A-1})], and the $R$-integral of such term is zero. By the same reason,  $\nabla_{R_0}\nabla_{R_{\lambda}} (F_1+F_2)$ for $\lambda=1,2$ is also removed from the integrand in the fourth line. Finally, the right-handed rule between $dn$ and the vorticity of $\Gamma_j$ equates $\int_{S_j} dn_0$  with $ S^j_{12}$  in the last line.

Such an imaginary number in Eq.~(\ref{Eq:IV-B-3}) can be included as a renormalization to the 1D Berry term in the third term of Eq.~(\ref{Eq:IV-A-9}), 
\begin{widetext}
\begin{align}
{\sf S} \!\ e^{-\sum_j s_1(\Gamma_j)} & \!\ \Bigg\{1 - \bigg(\int^{ab}_{a} dl \!\ t^l \int_{V} d^3 R \int_{S_2} d^2\Omega \bigg) e^{-s_1(\Gamma)} \Big(\sum_{j} 2s_0(\Gamma,\Gamma_j)\Big) \Bigg\} 
= {\sf S} \!\ e^{-\sum_j \overline{s}_1(\Gamma_j)}, \label{Eq:IV-B-5}  
\end{align}
\end{widetext}
with 
\begin{align}
{\sf S} &\equiv  \sum^{\infty}_{n=1} 
\frac{1}{n!} \prod^n_{j=1} \bigg(\int^{\infty}_{ab} dl_j \!\ t^{l_j} \int d^3R_j \int D\Omega_j(\lambda)\bigg). \label{Eq:IV-B-5a}
\end{align}
Here $\overline{s}_1(\Gamma_j)=i2\pi \overline{\chi} S^j_{12}$, and 
\begin{align}
\overline{\chi} &\equiv \chi - \ln b \!\ \!\  a \!\ t^a \!\ a^2_0\pi \frac{\partial}{\partial \nu} \Big(\frac{\sin \nu}{\nu}\Big) \nonumber \\
&\int_{V} d^3 R \!\ \big(\nabla^2_{R_1} + \nabla^2_{R_2}\big) \big(F_1(R) + F_2(R)\big) , \label{Eq:IV-B-4-ap}  
\end{align}
with $a=a_0\pi$.

\section{renormalization to Coulomb interaction potentials }
  After the integration over the short-loop DOF in Eq.~(\ref{Eq:IV-A-9}) [$\int_V d^3 R \int_{S_2} d^2\Omega$],  the second-order expansion term in $s_0(\Gamma,\Gamma_j)$ in the last line of Eq.~(\ref{Eq:IV-A-9}) generates 1-loop renormalization to the three types of the interaction potentials,
\begin{widetext}
\begin{align}
{\sf S} \!\  e^{-\sum^n_{i,j=1} s_0(\Gamma_i,\Gamma_j)} \Bigg\{ 1 + 
\bigg( \int^{ab}_{a} dl \!\ t^l \int_V d^3 R \int_{S_2} d^2 \Omega\bigg) e^{-s_1(\Gamma)}
\!\ \!\  \frac{1}{2} \Big( \sum^n_{j=1} 
2 s_0(\Gamma,\Gamma_j) \Big)^2 \Bigg\} =  {\sf S} \!\ e^{-\sum_{i,j} \overline{s}_0(\Gamma_i,\Gamma_j)}. \label{Eq:IV-C-0}
\end{align}
and 
\begin{align}
s_0(\Gamma_i,\Gamma_j) &\equiv \oint_{\Gamma_i}  \oint^{a_0b<|x^i-y^j|}_{\Gamma_j}  
\Big \{ d x^i \cdot dy^j \!\ F_1(x^i-y^j) + 
d x^i \cdot \eta_2 \cdot dy^j \!\ F_2(x^i-y^j) + dx^i \cdot G(x^i-y^j) \cdot dy^j\Big\}.  \label{Eq:IV-A-10-ap}
\end{align}
\end{widetext}

To determined $\overline{s}_0(\Gamma_i,\Gamma_j)$ thus introduced, let us first take the  $\Omega$-integral of a square of Eq.~(\ref{Eq:IV-B-1}) with $s_1(\Gamma)= i\nu \cos\theta$.  The squaring yields 6 different couplings among $F_1$, $F_2$, and $G$ , where all the couplings share the same $\Omega$-integration,  
\begin{align}
&\int_{S_2} d^2\Omega \!\ e^{-i\nu \cos\theta} \!\ 
\Omega_{\mu} \Omega_{\psi} \!\ \epsilon_{\mu\lambda\phi} 
\epsilon_{\psi\rho\sigma} \nonumber \\
& =\frac{4\pi}{3} \epsilon_{\mu\lambda\phi} \epsilon_{\mu\rho\sigma} 
\big( d_1 \delta_{\mu\mu} + d_2 (\eta_2)_{\mu\mu} \big) \nonumber \\
&=\frac{4\pi}{3} d_1 \big(\delta_{\lambda\rho} \delta_{\phi\sigma} 
- \delta_{\lambda\sigma} \delta_{\phi\rho}\big) + 
\frac{4\pi}{3} d_2 \big(-(\eta_2)_{\lambda\rho} \delta_{\phi\sigma} 
\nonumber \\
& \ \ \ - \delta_{\lambda\rho} (\eta_2)_{\phi\sigma} 
+ (\eta_2)_{\lambda\sigma} \delta_{\phi\rho} + \delta_{\lambda\sigma} 
(\eta_2)_{\phi\rho}\big). \label{Eq:IV-C-1}
\end{align}
Here  $d_1$ and $d_2$ are functions of the normalized 1D Berry phase parameter $\nu\equiv \frac{a^2_0\pi^2}{2}\chi$,  
\begin{align}
\begin{aligned}
&d_1  \equiv \frac{\sin \nu}{\nu}, \\
&d_2 \equiv - \frac{\sin \nu}{\nu} + \frac{3 (\sin \nu - \nu \cos\nu)}{\nu^3}.  
\end{aligned}\label{Eq:IV-C-2-ap}
\end{align}
In the non-topological limit ($\nu=0$) , $d_1=1$ and $d_2=0$. For small $\nu$, $0<d_1<1$ and $d_2>0$.

In the next subsection, we will first calculate the square of $F_1(r)$, and show that due to the 1D Berry phase term, screening effects of unpolarized vortex loops are suppressed, and the $F_1\times F_1$  screening effect of the polarized vortex loop induces spatial anisotropy in the $F_1(r)$ interaction, such that a characteristic length scale in the topological $(x_0)$ direction becomes stretched more than the length scale along the other two directions. We also show that the $F_1\times F_1$ screening effect of the polarized loops generates the $F_2(r)$ interaction.    

\subsection{screening effect mediated by the two $F_1$ interactions} 

A coupling between $\nabla_{R_{\lambda}}F_1(R-x^i)$ and $\nabla_{R_{\rho}} F_1(R-y^j)$ in $\frac{1}{2} \big( \sum_{j} 2s_0(\Gamma,\Gamma_j)\big)^2$ is given by 
\begin{align}
 \int_V d^3 R &\int_{S_2} d^2\Omega \!\ e^{-s_1(\Gamma)}
\!\ \sum^n_{i=1} \sum^n_{j=1} \oint_{\Gamma_i} dx^i_{\phi}\oint_{\Gamma_j} dy^j_{\sigma} 
\frac{a^4_0\pi^2}{8} \nonumber \\
&  \Omega_{\mu} \Omega_{\psi} \!\ \epsilon_{\mu\lambda\phi} 
\epsilon_{\psi\rho\sigma} 
\Big(\nabla_{R_{\lambda}} F_1(R-x^i)\Big)
\Big(\nabla_{R_{\rho}} F_1(R-y^j) \Big)  \nonumber \\
&\hspace{-1cm} = \frac{a^4_0 \pi^3}{6} 
\sum_{i,j} \epsilon_{\mu\lambda\phi} \epsilon_{\mu\rho\sigma} 
\big( d_1 \delta_{\mu\mu} + d_2 (\eta_2)_{\mu\mu} \big) 
\int_V d^3 R \nonumber \\
& \hspace{-0.5cm}\oint_{\Gamma_i} dx^i_{\phi} \oint_{\Gamma_j}  dy^j_{\sigma} 
\Big(\nabla_{R_{\lambda}} F_1(R-x^i)\Big)
\Big(\nabla_{R_{\rho}} F_1(R-y^j) \Big). \label{Eq:IV-C-3}
\end{align}
When using the second line of Eq.~(\ref{Eq:IV-C-1}), note that terms with a factor of $\delta_{\lambda\sigma}$ or $\delta_{\phi\rho}$ in its right-hand side reduce to zero in the partition function with closed vortex loops. This is because they all appear in the action with the divergence of the vortex vector. For example, a term with the factor of $\delta_{\lambda\sigma}$ in Eq.~(\ref{Eq:IV-C-1}) reduces to zero after integration by parts,    
\begin{align}
&\int_V d^3 R \oint_{\Gamma_i} dx^i_{\phi} \Big(\nabla_{R_{\sigma}} F_1(R-x^i)\Big) \nonumber \\
& \hspace{1cm}\times \oint_{\Gamma_j} dy^j_{\sigma} \!\ 
\Big(\nabla_{R_{\rho}} F_1(R-y^j) \Big) \nonumber \\
&= \int_V d^3 R \oint_{\Gamma_i} dx^i_{\phi} F_1(R-x^i) \nonumber \\
& \times \oint_{\Gamma_j} dy^j_{\sigma} \!\ \nabla_{y^{j}_{\sigma}} \Big( 
\Big(\nabla_{R_{\rho}} (F_1(R-y^j)) \Big) = 0.
\end{align}
The rest of terms in Eq.~(\ref{Eq:IV-C-3}) are summarized into two types of the interaction potentials, 
\begin{align}
& \int_V d^3 R \int_{S_2} d^2\Omega \!\  e^{-s_1(\Gamma)} \!\  
\sum_{i,j} \oint_{\Gamma_i} dx^i_{\phi}\oint_{\Gamma_j} dy^j_{\sigma} 
\frac{a^4_0\pi^2}{8} \nonumber \\
&\hspace{0.4cm} \Omega_{\mu} \Omega_{\psi} \!\ \epsilon_{\mu\lambda\phi} 
\epsilon_{\psi\rho\sigma} 
\Big(\nabla_{R_{\lambda}} F_1(R-x^i)\Big)
\Big(\nabla_{R_{\rho}} F_1(R-y^j) \Big)  \nonumber \\
 & \ \ =    \sum^n_{i,j=1} 
\oint_{\Gamma_i} \oint_{\Gamma_j} \Big( 
dx^i \cdot dy^j \!\ \!\ c_{111}(x^i-y^j) \nonumber \\
&\hspace{1.5cm} + dx^i \cdot \eta_2 \cdot dy^j \!\ \!\  c_{112}(x^i-y^j) \Big),  
\label{Eq:IV-C-4}
\end{align}
with 
\begin{align}
c_{111}(x^i-y^j) &\equiv \frac{a^4_0 \pi^3}{6} \Big(d_1 \delta_{\lambda\lambda}
- d_2 (\eta_2)_{\lambda\lambda} \Big)\int_V d^3 R \nonumber \\
& \Big(\nabla_{R_{\lambda}} F_1(R-x^i)\Big)
\Big(\nabla_{R_{\lambda}} F_1(R-y^j) \Big), \label{Eq:IV-C-5} \\ 
c_{112}(x^i-y^j) &\equiv \frac{a^4_0 \pi^3}{6} (-d_2) \delta_{\lambda\lambda} 
\int_V d^3 R \nonumber \\
& \Big(\nabla_{R_{\lambda}} F_1(R-x^i)\Big)
\Big(\nabla_{R_{\lambda}} F_1(R-y^j) \Big) \label{Eq:IV-C-6}. 
\end{align}
Upon a substitution of Eq.~(\ref{Eq:IV-C-4}) into Eq.(\ref{Eq:IV-C-0}),  $c_{111}(r)$ and $c_{112}(r)$ generate renormalization to $F_1(r)$ and $F_2(r)$ in $\overline{s}_0(\Gamma_i,\Gamma_j)$, respectively,  
\begin{align}
\overline{F}_1(r) & \equiv F_1(r) - a \!\ t^a \!\ \big(c_{111}(r) + \cdots \big) \!\ \ln b, \\ 
\overline{F}_2(r) & \equiv F_2(r) - a\!\ t^a \!\ \big(c_{112}(r) + \cdots \big) \!\ \ln b.  
\end{align}
Note that the convolution  in Eqs.~(\ref{Eq:IV-C-5},\ref{Eq:IV-C-6}) are expressed by products in the $q$ space, 
\begin{align}
c_{111}(q) &= \frac{a^4_0 \pi^3}{6} \Big(d_1 q^2
- d_2 (\eta_2)_{\lambda\lambda}  \!\ q^2_{\lambda}\Big) \!\ F^2_1(q), \label{Eq:IV-C-5a} \\ 
c_{112}(q) &= \frac{a^4_0 \pi^3}{6} (-d_2)  
\!\  q^2  \!\ F^2_1(q)  \label{Eq:IV-C-6a}, 
\end{align}
with $F_1(-q)=F_1(q)$. So is the 1-loop renormalization to $F_1$ and $F_2$,    
\begin{align}
\overline{F}_1(q) & \equiv F_1(q) - a \!\ t^a \!\ \big(c_{111}(q) + \cdots \big) \!\ \ln b, \\ 
\overline{F}_2(q) & \equiv F_2(q) - a\!\ t^a \!\ \big(c_{112}(q) + \cdots \big) \!\ \ln b.  
\end{align}
In the limit of $\nu=0$, the first equation reduces to Eq.~(\ref{Eq:III-11}) with $F_1(q)=2\pi^2/(Tq^2)$. In the presence of small $\nu$,  $c_{111}(q)$ induces spatial anisotropy in the $F_1$ interaction, e.g. $F^{-1}_1(q) = Tq^2/(2\pi^2)\rightarrow Tq^2/(2\pi^2) + C\ln b ((d_1-d_2)q^2_{\perp} + (d_1+2d_2) q^2_0)$ with a positive constant 
$C=a_0 \pi^4 t^a/6$ [Here we also recovered the normalization factor of the $l_j$-integral and $R$-integral]. Since $d_2>0$ for smaller $\nu$, the positive $C$ suggests an enhancement of a ratio of a length scale along the $x_0$ direction to a length scale along the others. This indicates that the small 1D Berry phase term elongates the characteristic length scale along the $x_0$ axis relative to the length scale along the other two.  For the small $\nu$, $c_{112}(q)$ in the second equation induces the $F_2(r)$ interaction with a positive coefficient and $1/r$ decay.  A short-distance part of such $F_2(r)$ interaction favors vortex-loop segments polarized along the $x_0$ axis.

In the next subsection, we will calculate the coupling between $F_1(r)$  and $F_2(r)$, and show that the $F_1\times F_2$ screening effect of polarized vortex loops induces the dipolar-type $G(r)$ interaction, helping to confine vortex loops within planes parallel to the topological ($x_0$) direction. We also show that the $F_1\times F_2$ screening effect of the polarized vortex loops induces the spatial anisotropy in the $F_2(r)$ interaction, such that a characteristic length scale along the topological $(x_0)$ direction becomes longer than the other length scale.

 \subsection{screening effect mediated by the $F_1$ and $F_2$ interactions} 
 A coupling between $\nabla_{R_{\lambda}}F_1(R-x^i)$ and $\nabla_{R_{\rho}} F_2(R-y^j)$ in $\frac{1}{2} \big( \sum_{j} 2s_0(\Gamma,\Gamma_j)\big)^2$ is summarized by 
\begin{align}
\int_V d^3 R& \int_{S_2} d^2\Omega \!\ e^{-s_1(\Gamma)} \!\ \sum^n_{i,j=1} \oint_{\Gamma_i} dx^i_{\phi}\oint_{\Gamma_j} dy^j_{\sigma} (\eta_2)_{\sigma\sigma}
  \frac{a^4_0\pi^2}{4}  \nonumber \\
  & \hspace{-0.8cm}\Omega_{\mu} \Omega_{\psi} \!\ \epsilon_{\mu\lambda\phi} 
\epsilon_{\psi\rho\sigma} 
\Big(\nabla_{R_{\lambda}} F_1(R-x^i)\Big)
\Big(\nabla_{R_{\rho}} F_2(R-y^j) \Big)  \nonumber \\
& = \frac{a^4_0 \pi^3}{3} 
\sum_{i,j} \epsilon_{\mu\lambda\phi} \epsilon_{\mu\rho\sigma} 
\big( d_1 \delta_{\mu\mu} + d_2 (\eta_2)_{\mu\mu} \big) \nonumber \\
&  \hspace{1.5cm} \int_V d^3 R \oint_{\Gamma_i} dx^i_{\phi} \oint_{\Gamma_j}  dy^j_{\sigma} (\eta_2)_{\sigma\sigma}  \nonumber \\
& \hspace{0.5cm}\Big(\nabla_{R_{\lambda}} F_1(R-x^i)\Big)
\Big(\nabla_{R_{\rho}} F_2(R-y^j) \Big) \label{Eq:IV-C-7}
\end{align}
Multiplication of Eq.~(\ref{Eq:IV-C-1}) by $\eta_2$ with $\eta^2_2=-\eta_2 + 2$ facilitates a subsequent calculation of Eq.~(\ref{Eq:IV-C-7}),  
\begin{align}
& \epsilon_{\mu\lambda\phi} \epsilon_{\mu\rho\sigma} 
\big( d_1 \delta_{\mu\mu} + d_2 (\eta_2)_{\mu\mu} \big) (\eta_2)_{\sigma\sigma}  \nonumber \\
&=  d_1 \big(\delta_{\lambda\rho} (\eta_2)_{\phi\sigma} 
- (\eta_2)_{\lambda\sigma} \delta_{\phi\rho}\big) + 
d_2 \big(-(\eta_2)_{\lambda\rho} (\eta_2)_{\phi\sigma}  \nonumber \\
&\hspace{1cm} - \delta_{\lambda\rho} (\eta^2_2)_{\phi\sigma} 
+ (\eta^2_2)_{\lambda\sigma} \delta_{\phi\rho} + (\eta_2)_{\lambda\sigma} 
(\eta_2)_{\phi\rho}\big) \nonumber \\
& = -2d_2 \big( \delta_{\lambda\rho} \delta_{\phi\sigma} -  \delta_{\lambda\sigma}\delta_{\phi\rho}\big)   \nonumber \\
& \hspace{0.4cm} + (d_1 + d_2) \big(\delta_{\lambda\rho} (\eta_2)_{\phi\sigma} - (\eta_2)_{\lambda\sigma} \delta_{\phi\rho}\big) \nonumber \\
& \hspace{0.8cm}  + d_2 \big(-(\eta_2)_{\lambda\rho} (\eta_2)_{\phi\sigma}  + (\eta_2)_{\lambda\sigma} 
(\eta_2)_{\phi\rho}\big). \label{Eq:IV-C-7a}
\end{align}
Note that terms with a factor of $\delta_{\lambda\sigma}$ or $\delta_{\phi\rho}$  in the right-hand side of Eq.~(\ref{Eq:IV-C-7a})  reduce to zero after the integration by parts. The rest of the terms in Eq.~(\ref{Eq:IV-C-7a}) are summarized into the three types of interaction potentials, 
\begin{align}
\int_V d^3 R& \int_{S_2} d^2\Omega \!\ e^{-s_1(\Gamma)} \!\ \sum^n_{i,j=1} \oint_{\Gamma_i} dx^i_{\phi}\oint_{\Gamma_j} dy^j_{\sigma} (\eta_2)_{\sigma\sigma}
\frac{a^4_0\pi^2}{4}  \nonumber \\
& \hspace{-0.9cm}\Omega_{\mu} \Omega_{\psi} \!\ \epsilon_{\mu\lambda\phi} 
\epsilon_{\psi\rho\sigma} 
\Big(\nabla_{R_{\lambda}} F_1(R-x^i)\Big)
\Big(\nabla_{R_{\rho}} F_2(R-y^j) \Big) \nonumber  \\
&= \sum_{i,j} 
\oint_{\Gamma_i} \oint_{\Gamma_j} \Big( 
dx^i \cdot dy^j \!\ \!\ c_{121}(x^i-y^j) +    \nonumber \\ 
&\hspace{1.4cm}  dx^i \cdot \eta_2 \cdot dy^j \!\ \!\  c_{122}(x^i-y^j) + \nonumber \\
& \hspace{1.0cm}  dx^i\cdot c_{120}(x^i-y^j) \cdot dy^j \Big),  
\label{Eq:IV-C-8}
\end{align}
with 
\begin{align}
c_{121}(x^i-y^j) &\equiv \frac{a^4_0 \pi^3}{3} \big(-2d_2 \big) \!\ \delta_{\lambda\lambda}
 \nonumber \\
&\hspace{-1.9cm} \int_V d^3 R \Big(\nabla_{R_{\lambda}} F_1(R-x^i)\Big)
\Big(\nabla_{R_{\lambda}} F_2(R-y^j) \Big), \label{Eq:IV-C-9} \\ 
c_{122}(x^i-y^j) &\equiv \frac{a^4_0 \pi^3}{3} \Big((d_1+d_2) \delta_{\lambda\lambda} 
- d_2 (\eta_2)_{\lambda\lambda} \Big)
 \nonumber \\
&\hspace{-1.9cm} \int_V d^3 R\Big(\nabla_{R_{\lambda}} F_1(R-x^i)\Big)
\Big(\nabla_{R_{\lambda}} F_2(R-y^j) \Big),\label{Eq:IV-C-10} \\
(c_{120})_{\phi\sigma}(x^i-y^j) & \equiv \frac{a^4_0 \pi^3}{3} d_2 (\eta_2)_{\phi\rho} (\eta_2)_{\lambda\sigma}  \nonumber \\
&\hspace{-1.9cm}  \int_V d^3 R \Big(\nabla_{R_{\lambda}} F_1(R-x^i)\Big)
\Big(\nabla_{R_{\rho}} F_2(R-y^j) \Big). \label{Eq:IV-C-11}
\end{align}
When Eq.~(\ref{Eq:IV-C-8}) is substituted into Eq.~(\ref{Eq:IV-C-0}),  $c_{121}(r)$ and $c_{122}(r)$ yield renormalizations to $F_1(r)$ and $F_2(r)$, while $c_{120}(r)$ induces a renormalization to $G(r)$.  In fact, the induced interactions take the following forms in the momentum space, 
\begin{align}
c_{121}(q )& = \frac{a^4_0 \pi^3}{3} \Big(-2d_2 \Big) q^2 F_1(q) F_2(q), \label{Eq:IV-C-9a} \\ 
c_{122}(q) &= \frac{a^4_0 \pi^3}{3} \Big(d_1 q^2 +  
3 d_2  \!\ q^2_{0} \Big) F_1(q) F_2(q),\label{Eq:IV-C-10a} \\
(c_{120})_{\phi\sigma}(q) & =  \frac{a^4_0 \pi^3}{3} d_2 \!\ q^2 F_1(q) F_2(q)\!\  
(\eta_2 \!\  \hat{q} \!\ \hat{q}^T \!\ \eta_2)_{\phi\sigma}\nonumber \\
&\equiv c_{120}(q) 
(\eta_2 \!\  \hat{q} \!\ \hat{q}^T \!\ \eta_2)_{\phi\sigma}. \label{Eq:IV-C-11a}
\end{align} 
Thus,
\begin{align}
\overline{F}_1(q) & \equiv F_1(q) - a \!\ t^a \!\ \big(c_{111}(q) + c_{121}(q) + \cdots \big) \!\ \ln b, \\ 
\overline{F}_2(q) & \equiv F_2(q) - a\!\ t^a \!\ \big(c_{112}(q) + c_{122}(q) + \cdots \big) \!\ \ln b, \\ 
\overline{G}_0(q) &= G_0(q)   - a \!\ t^a \!\ \big(c_{120}(q) + \cdots \big) \!\ \ln b. 
\end{align}
For $F_1(r)=F_2(r)=1/|r|$, the induced $G(r)$ potential takes a form of a dipole-dipole interaction modulated by $\eta_2$,  
\begin{align}
g(r)  &\propto - d_2 \int_{q \ne 0} \frac{d^3q}{(2\pi)^3} e^{-iqr}  \!\ 
\eta_2 \frac{\hat{q} \hat{q}^T}{q^2} 
\eta_2 \nonumber \\
&=-\frac{\pi d_2}{2}  \eta_2 \Big( \frac{1 - \hat{r}\hat{r}^T}{|r|} \Big)\eta_2.  
\end{align}
Note that two endpoints of an open vortex line are attracted by a linear confining potential [see Appendix A]~\cite{herbutModernApproachCritical2007}. Since the endpoint and vortex-loop segment can be regarded as magnetic monopoles and associated magnetic dipoles respectively, the form of $(1-\hat{r} \hat{r}^T)/|r|$ in the right-hand side can be interpreted as the dipole-dipole interaction derived from the derivatives of the linear confining potential. The induced $g(r)$ interaction characterizes the energetics of curved vortex loops. Due to the modulation factor $\eta_2$,  it favors those vortex loops curving within  the $x_1$-$x_0$ or $x_2$-$x_0$ planes, while disfavoring those loops  curving within the $x_1$-$x_2$ plane.  In other words, the induced $g(r)$ helps to confine the vortex loops in a plane parallel to the $x_0$ axis.   

In the following subsections, other couplings in $2(\sum_j s_0(\Gamma_i,\Gamma_j))^2$ are calculated in the same way. 
\subsection{screening effect mediated by the two $F_2$ interactions}
 A coupling between $\nabla_{R_{\lambda}}F_2(R-x^i)$ and $\nabla_{R_{\rho}} F_2(R-y^j)$  in $2\big(\sum_{j}s_0(\Gamma,\Gamma_j)\big)^2$ is given by
\begin{align}
&\sum^n_{i=1} \sum^n_{j=1} \oint_{\Gamma_i} dx^i_{\phi} (\eta_2)_{\phi\phi} \oint_{\Gamma_j} dy^j_{\sigma} (\eta_2)_{\sigma\sigma}
\int_V d^3 R \int_{S_2} d^2\Omega e^{-i\nu \cos\theta} \nonumber \\
& \frac{a^4_0\pi^2}{8} 
\Omega_{\mu} \Omega_{\psi} \!\ \epsilon_{\mu\lambda\phi} 
\epsilon_{\psi\rho\sigma} 
\Big(\nabla_{R_{\lambda}} F_2(R-x^i)\Big)
\Big(\nabla_{R_{\rho}} F_2(R-y^j) \Big)  \nonumber \\
& = \frac{a^4_0 \pi^3}{6} 
\sum_{i,j} \epsilon_{\mu\lambda\phi} \epsilon_{\mu\rho\sigma} 
\big( d_1 \delta_{\mu\mu} + d_2 (\eta_2)_{\mu\mu} \big)  \nonumber \\
&\hspace{1.2cm} \int_V d^3 R \oint_{\Gamma_i} dx^i_{\phi} (\eta_2)_{\phi\phi} \oint_{\Gamma_j}  dy^j_{\sigma} (\eta_2)_{\sigma\sigma} \nonumber \\ 
&\hspace{1.6cm} \Big(\nabla_{R_{\lambda}} F_2(R-x^i)\Big)
\Big(\nabla_{R_{\rho}} F_2(R-y^j) \Big). \label{Eq:IV-C-12}
\end{align}
A multiplication of the second line of Eq.~(\ref{Eq:IV-C-1}) by $(\eta_2)_{\phi\phi}$ and $(\eta_{2})_{\sigma\sigma}$  facilitates a subsequent calculation of Eq.~(\ref{Eq:IV-C-12}) , 
\begin{align}
& \epsilon_{\mu\lambda\phi} \epsilon_{\mu\rho\sigma} 
\big( d_1 \delta_{\mu\mu} + d_2 (\eta_2)_{\mu\mu} \big) (\eta_2)_{\phi\phi} (\eta_2)_{\sigma\sigma}  \nonumber \\
&=  d_1 \big(\delta_{\lambda\rho} (\eta^2_2)_{\phi\sigma} 
- (\eta_2)_{\lambda\sigma} (\eta_2)_{\phi\rho}\big)  \nonumber \\
&\hspace{0.4cm} + d_2 \big(-(\eta_2)_{\lambda\rho} (\eta^2_2)_{\phi\sigma} 
- \delta_{\lambda\rho} (\eta^3_2)_{\phi\sigma}  \nonumber \\
& \hspace{0.8cm} + (\eta^2_2)_{\lambda\sigma} (\eta_2)_{\phi\rho} + (\eta_2)_{\lambda\sigma} 
(\eta^2_2)_{\phi\rho}\big) \nonumber \\
& = 2(d_1+d_2)\delta_{\lambda\rho} \delta_{\phi\sigma}  - (d_1 +3 d_2) \delta_{\lambda\rho} (\eta_2)_{\phi\sigma} \nonumber \\
& \hspace{0.7cm} - 2d_2 \big( (\eta_2)_{\lambda\rho} \delta_{\phi\sigma} -  (\eta_2)_{\lambda\sigma} \delta_{\phi\rho}  - \delta_{\lambda\sigma} (\eta_2)_{\phi\rho} \big) \nonumber \\
& \hspace{0.2cm} + d_2 
(\eta_2)_{\lambda\rho} (\eta_2)_{\phi\sigma} - (d_1+2d_2) 
(\eta_2)_{\lambda\sigma} 
(\eta_2)_{\phi\rho},   \label{Eq:IV-C-12a}
\end{align}
with $\eta^2_2=-\eta_2 + 2$ and $\eta^3_2=3\eta_2-2$. Dropping terms with a factor of either  $\delta_{\lambda\sigma}$ or $\delta_{\phi\rho}$, we obtain
\begin{align}
&\sum^n_{i=1} \sum^n_{j=1} \oint_{\Gamma_i} dx^i_{\phi} (\eta_2)_{\phi\phi} \oint_{\Gamma_j} dy^j_{\sigma} (\eta_2)_{\sigma\sigma}
\int_V d^3 R \int_{S_2} d^2\Omega e^{-i\nu \cos\theta}  \nonumber \\
& \hspace{0.6cm} \frac{a^4_0\pi^2}{8} 
\Omega_{\mu} \Omega_{\psi} \!\ \epsilon_{\mu\lambda\phi} 
\epsilon_{\psi\rho\sigma} 
\Big(\nabla_{R_{\lambda}} F_2(R-x^i)\Big)
\Big(\nabla_{R_{\rho}} F_2(R-y^j) \Big)  \nonumber \\
 &= \sum_{i,j} 
\oint_{\Gamma_i} \oint_{\Gamma_j} \Big( 
dx^i \cdot dy^j \!\ \!\ c_{221}(x^i-y^j) \nonumber \\
& \hspace{1.2cm} + dx^i \cdot \eta_2 \cdot dy^j \!\ \!\  c_{222}(x^i-y^j) \nonumber \\
&\hspace{2.4cm} + dx^i\cdot c_{220}(x^i-y^j) \cdot dy^j \Big),  
\label{Eq:IV-C-13}
\end{align}
with 
\begin{align}
c_{221}(x^i-y^j) &\equiv \frac{a^4_0 \pi^3}{6} \Big( 2 (d_1 + d_2) \delta_{\lambda\lambda}
 - 2d_2 (\eta_2)_{\lambda\lambda}\Big) \nonumber \\
 & \hspace{-2.1cm} \int_V d^3 R \Big(\nabla_{R_{\lambda}} F_2(R-x^i)
\Big(\nabla_{R_{\lambda}} F_2(R-y^j) \Big), \label{Eq:IV-C-14} \\ 
c_{222}(x^i-y^j) &\equiv \frac{a^4_0 \pi^3}{6} \Big(-(d_1+3d_2) \delta_{\lambda\lambda} 
+ d_2 (\eta_2)_{\lambda\lambda} \Big) \nonumber \\
& \hspace{-2.1cm} \int_V d^3 R \Big(\nabla_{R_{\lambda}} F_2(R-x^i)\Big)
\Big(\nabla_{R_{\lambda}} F_2(R-y^j) \Big),\label{Eq:IV-C-15} \\
(c_{220})_{\phi\sigma}(x^i-y^j) & \equiv \frac{a^4_0 \pi^3}{6} 
(-)(d_1+2d_2) (\eta_2)_{\phi\rho} (\eta_2)_{\lambda\sigma} \nonumber \\
&\hspace{-2.1cm}\int_V d^3 R \Big(\nabla_{R_{\lambda}} F_2(R-x^i)\Big)
\Big(\nabla_{R_{\rho}} F_2(R-y^j) \Big). \label{Eq:IV-C-16}
\end{align}
Their Fourier transforms can be included into renormalization of $F_1(q)$, $F_2(q)$ and $G_0(q)$,
\begin{align}
c_{221}(q) &= \frac{a^4_0 \pi^3}{6} \Big( 2 d_1  q^2
 + 6d_2  q^2_{0}\Big)\!\ F^2_2(q), \label{Eq:IV-C-14a} \\ 
c_{222}(q) &= \frac{a^4_0 \pi^3}{6} \Big(-(d_1+3d_2) q^2
+ d_2 (\eta_2)_{\lambda\lambda} q^2_{\lambda} \Big) F^2_2(q),\label{Eq:IV-C-15a} \\
c_{220}(q) & \equiv \frac{a^4_0 \pi^3}{6}  
(-)(d_1+2d_2) \!\ q^2 F^2_2(q) \!\ . \label{Eq:IV-C-16a}
\end{align}
with $(c_{220})_{\phi\sigma}(q)\equiv c_{220}(q)(\eta_2  \!\ \hat{q}  \!\ \hat{q}^T \!\ \eta_2)_{\phi\sigma}$.

\subsection{screening effect mediated by the $F_1$ and $G$ interactions}
A coupling between $\nabla_{R_{\lambda}}F_1(R-x^i)$ and $\nabla_{R_{\rho}} G(R-y^j)$ in $2\big(\sum_{j}s_0(\Gamma,\Gamma_j)\big)^2$ is summarized as follows 
\begin{align}
&\sum^n_{i=1} \sum^n_{j=1} \oint_{\Gamma_i} dx^i_{\phi}\oint_{\Gamma_j} dy^j_{\gamma} 
\int_V d^3 R \int_{S_2} d^2\Omega e^{-i\nu \cos\theta} \!\ \frac{a^4_0\pi^2}{4}  \nonumber \\
& \hspace{0.2cm}\Omega_{\mu} \Omega_{\psi} \!\ \epsilon_{\mu\lambda\phi} 
\epsilon_{\psi\rho\sigma} 
\Big(\nabla_{R_{\lambda}} F_1(R-x^i)\Big)
\Big(\nabla_{R_{\rho}} G_{\sigma\gamma}(R-y^j) \Big)  \nonumber \\
& = \frac{a^4_0 \pi^3}{3} 
\sum_{i,j} \epsilon_{\mu\lambda\phi} \epsilon_{\mu\rho\sigma} 
\big( d_1 \delta_{\mu\mu} + d_2 (\eta_2)_{\mu\mu} \big) \int_V d^3 R  \nonumber \\
& \oint_{\Gamma_i} dx^i_{\phi} \oint_{\Gamma_j}  dy^j_{\gamma} 
\Big(\nabla_{R_{\lambda}} F_1(R-x^i)\Big)
\Big(\nabla_{R_{\rho}} G_{\sigma\gamma}(R-y^j) \Big) \nonumber \\
&  \equiv \sum_{i,j} \oint_{\Gamma_i} \oint_{\Gamma_j} 
dx^i\cdot c_{100}(x^i-y^j) \cdot dy^j. \label{Eq:IV-C-17}
\end{align}
3 by 3 $c_{100}(x^i-y^j)$ can be included as renormalization to $G(x^i-y^j)$ in $\overline{s}_0(\Gamma_i,\Gamma_j)$ of Eq.~(\ref{Eq:IV-C-0}). One can see this, from the Fourier transform of $c_{110}(x^i-y^j)$,
\begin{align}
&(c_{100})_{\phi\gamma}(q) \nonumber \\
& = \frac{a^4_0\pi^3}{3}\epsilon_{\mu\lambda\phi} \epsilon_{\mu\rho\sigma}  \big( d_1 \delta_{\mu\mu} + d_2 (\eta_2)_{\mu\mu} \big)  \nonumber \\
& \hspace{1.cm} q_{\lambda} q_{\rho}  \!\ F_{1}(q) G_0(q) \!\  \!\ (\eta_2)_{\sigma\sigma} \hat{q}_{\sigma}
\hat{q}_{\gamma} (\eta_2)_{\gamma\gamma}, \nonumber \\
 &= \frac{a^4_0\pi^3}{3} \Big( -2d_2 \big( \delta_{\lambda\rho}\delta_{\phi\sigma} 
 - \delta_{\lambda\sigma}\delta_{\phi\rho}\big) \nonumber \\
 & \hspace{1.cm} + (d_1+d_2) 
 \big(\delta_{\lambda\rho}(\eta_2)_{\phi\sigma} - (\eta_2)_{\lambda\sigma} 
 \delta_{\phi\rho}\big) \nonumber \\
 & \hspace{1cm}  + d_2 \big( - (\eta_2)_{\lambda\rho} (\eta_2)_{\phi\sigma} 
 + (\eta_2)_{\lambda\sigma} (\eta_2)_{\phi\rho}\big) \Big) 
\nonumber \\
& \hspace{1.cm} q_{\lambda} q_{\rho}  \hat{q}_{\sigma}  \!\ F_{1}(q) G_0(q) \!\  \!\
\hat{q}_{\gamma} (\eta_2)_{\gamma\gamma}, \nonumber \\
&\equiv c_{100}(q) \!\ (\eta_2)_{\phi\sigma} \hat{q}_{\sigma} \hat{q}_{\gamma} (\eta_2)_{\gamma\gamma}, \nonumber
\end{align}
with 
\begin{align}
c_{100}(q) =  \frac{a^4_0\pi^3}{3} (d_1+d_2) q^2 \!\ F_{1}(q) G_0(q). \label{Eq:IV-C-18}
\end{align}
In the second line, we use Eq.~(\ref{Eq:IV-C-7a}). From the second line to the third line, we drop terms with $\delta_{\phi\lambda}$, $\delta_{\phi\rho}$, or $\delta_{\phi\sigma}$ by integration by parts. By the re-exponentiation, Eq.~(\ref{Eq:IV-C-18}) produces the renormalization of $G_0(q)$ in $\overline{s}_0(\Gamma_i,\Gamma_j)s$  in Eq.~(\ref{Eq:IV-C-0}).

\subsection{screening effect mediated by the $F_2$ and $G$ interactions}
 A coupling between $\nabla_{R_{\lambda}}F_2(R-x^i)$ and $\nabla_{R_{\rho}} G(R-y^j)$ in $2\big(\sum_{j}s_0(\Gamma,\Gamma_j)\big)^2$ is summarized by 
\begin{align}
& \sum^n_{i=1} \sum^n_{j=1} \oint_{\Gamma_i} dx^i_{\phi} (\eta_2)_{\phi\phi} \oint_{\Gamma_j} dy^j_{\gamma}  
\int_V d^3 R \int_{S_2} d^2\Omega e^{-i\nu \cos\theta}  \nonumber \\
&\hspace{0.2cm} \frac{a^4_0\pi^2}{4} \Omega_{\mu} \Omega_{\psi} \!\ \epsilon_{\mu\lambda\phi} 
\epsilon_{\psi\rho\sigma} 
\Big(\nabla_{R_{\lambda}} F_2(R-x^i)\Big)
\Big(\nabla_{R_{\rho}} G_{\sigma\gamma}(R-y^j) \Big)  \nonumber \\
& = \frac{a^4_0 \pi^3}{3} 
\sum_{i,j} \epsilon_{\mu\lambda\phi} \epsilon_{\mu\rho\sigma} 
\big( d_1 \delta_{\mu\mu} + d_2 (\eta_2)_{\mu\mu} \big) \nonumber \\
& \hspace{1.2cm}  \int_V d^3 R \oint_{\Gamma_i} dx^i_{\phi} (\eta_2)_{\phi\phi} \oint_{\Gamma_j}  dy^j_{\gamma} \nonumber \\ 
&\hspace{1.6cm} \Big(\nabla_{R_{\lambda}} F_2(R-x^i)\Big)
\Big(\nabla_{R_{\rho}} G_{\sigma\gamma}(R-y^j) \Big) \nonumber \\
& \equiv \sum_{i,j} \oint_{\Gamma_i} \oint_{\Gamma_j} 
dx^i \cdot c_{200}(x^i-y^j) \cdot dy^j.  \label{Eq:IV-C-19}
\end{align}
The Fourier transform $c_{200}(q)$ of the 3 by 3 $c_{200}(x^i-y^j)$ yields the renormalization to $G_0(q)$ in $\overline{s}_0(\Gamma_i,\Gamma_j)s$  in Eq.~(\ref{Eq:IV-C-0}), 
\begin{widetext}
\begin{align}
&(c_{200})_{\phi\gamma}(q)  = \frac{a^4_0\pi^3}{3} \epsilon_{\mu\lambda\phi} \epsilon_{\mu\rho\sigma}  \big( d_1 \delta_{\mu\mu} + d_2 (\eta_2)_{\mu\mu} \big) 
(\eta_2)_{\phi\phi} q_{\lambda} q_{\rho}  \!\ F_{2}(q) G_0(q) \!\  \!\ (\eta_2)_{\sigma\sigma} \hat{q}_{\sigma}
\hat{q}_{\gamma} (\eta_2)_{\gamma\gamma} \nonumber \\
 &= \frac{a^4_0\pi^3}{3} \Big(2(d_1+d_2)\delta_{\lambda\rho} \delta_{\phi\sigma}  - (d_1 +3 d_2) \delta_{\lambda\rho} (\eta_2)_{\phi\sigma}  - 2d_2 \big( (\eta_2)_{\lambda\rho} \delta_{\phi\sigma} \nonumber \\
& \hspace{0.5cm} 
-  (\eta_2)_{\lambda\sigma} \delta_{\phi\rho}  - \delta_{\lambda\sigma} (\eta_2)_{\phi\rho} \big) + d_2 (\eta_2)_{\lambda\rho} (\eta_2)_{\phi\sigma} - (d_1+2d_2) 
(\eta_2)_{\lambda\sigma} 
(\eta_2)_{\phi\rho}  \Big)  q_{\lambda} q_{\rho}  \hat{q}_{\sigma}  \!\ F_{2}(q) G_0(q) \!\  \!\
\hat{q}_{\gamma} (\eta_2)_{\gamma\gamma} \nonumber \\
&=   c_{200}(q)\!\ (\eta_2)_{\phi\sigma} \hat{q}_{\sigma} \hat{q}_{\gamma} (\eta_2)_{\gamma\gamma}, \nonumber 
\end{align}
with 
\begin{align}
c_{200}(q) = \frac{a^4_0\pi^3}{3} (-)(d_1+d_2) \!\ \big( q^2 + (\eta_2)_{\lambda\lambda} q^2_{\lambda}\big)  \!\ F_{2}(q) G_0(q). \label{Eq:IV-C-20}
\end{align}
Here we use Eq.~(\ref{Eq:IV-C-12a}), and drop terms with factors of $\delta_{\phi\lambda}$, $\delta_{\phi\rho}$  or $\delta_{\phi\sigma}$ through integration by parts.

\subsection{screening effect mediated by the two $G$ interactions}
A coupling between $\nabla_{R_{\lambda}}G(R-x^i)$ and $\nabla_{R_{\rho}} G(R-y^j)$  in $2\big(\sum_j s_0(\Gamma,\Gamma_j)\big)^2$ is given by
\begin{align}
&\sum^n_{i=1} \sum^n_{j=1} \oint_{\Gamma_i} dx^i_{\kappa}  \oint_{\Gamma_j} dy^j_{\gamma}
\int_V d^3 R \int_{S_2} d^2\Omega e^{-i\nu \cos\theta}  \frac{a^4_0\pi^2}{8} 
\Omega_{\mu} \Omega_{\psi} \!\ \epsilon_{\mu\lambda\phi} 
\epsilon_{\psi\rho\sigma} 
\Big(\nabla_{R_{\lambda}} G_{\phi\kappa}(R-x^i)\Big)
\Big(\nabla_{R_{\rho}} G_{\sigma\gamma}(R-y^j) \Big)  \nonumber \\
& = \frac{a^4_0 \pi^3}{6} 
\sum_{i,j} \epsilon_{\mu\lambda\phi} \epsilon_{\mu\rho\sigma} 
\big( d_1 \delta_{\mu\mu} + d_2 (\eta_2)_{\mu\mu} \big)  \int_V d^3 R \oint_{\Gamma_i} dx^i_{\kappa} \oint_{\Gamma_j}  dy^j_{\gamma} 
\Big(\nabla_{R_{\lambda}} G_{\phi\kappa}(R-x^i)\Big)
\Big(\nabla_{R_{\rho}} G_{\sigma\gamma}(R-y^j) \Big) \nonumber \\
& \equiv \sum_{i,j} \oint_{\Gamma_i} \oint_{\Gamma_j} 
dx^i \cdot c_{000}(x^i-y^j) \cdot dy^j. \label{Eq:IV-C-21} 
\end{align}
The Fourier transform of $c_{000}(x^i-y^j)$ is calculated as follows,  
\begin{align}
(c_{000})_{\kappa\gamma}(q) &= \frac{a^4_0\pi^3}{6} \epsilon_{\mu\lambda\phi} \epsilon_{\mu\rho\sigma}  \big( d_1 \delta_{\mu\mu} + d_2 (\eta_2)_{\mu\mu} \big) 
 q_{\lambda} q_{\rho}  \!\ G_0(q) G_0(q) \!\ \!\  (\eta_2)_{\phi\phi} \hat{q}_{\phi}
\hat{q}_{\kappa} (\eta_2)_{\kappa\kappa} \!\  \!\ (\eta_2)_{\sigma\sigma} \hat{q}_{\sigma}
\hat{q}_{\gamma} (\eta_2)_{\gamma\gamma} \nonumber \\
 &= \frac{a^4_0\pi^3}{6} \Big(2(d_1+d_2)\delta_{\lambda\rho} \delta_{\phi\sigma}  - (d_1 +3 d_2) \delta_{\lambda\rho} (\eta_2)_{\phi\sigma}  - 2d_2 \big( (\eta_2)_{\lambda\rho} \delta_{\phi\sigma} -  (\eta_2)_{\lambda\sigma} \delta_{\phi\rho}  - \delta_{\lambda\sigma} (\eta_2)_{\phi\rho} \big) \nonumber \\
& \hspace{1cm} + d_2 
(\eta_2)_{\lambda\rho} (\eta_2)_{\phi\sigma} - (d_1+2d_2) 
(\eta_2)_{\lambda\sigma} 
(\eta_2)_{\phi\rho}  \Big)  q_{\lambda} q_{\rho}  \hat{q}_{\phi} 
\hat{q}_{\sigma}  \!\ G_0(q) G_0(q) \!\  \!\ \hat{q}_{\kappa} (\eta_2)_{\kappa\kappa}
\hat{q}_{\gamma} (\eta_2)_{\gamma\gamma} \nonumber \\
&=  c_{000}(q)\!\ (\eta_2)_{\kappa\kappa} \hat{q}_{\kappa} \hat{q}_{\gamma} (\eta_2)_{\gamma\gamma}, \nonumber 
\end{align}
with 
\begin{align}
&c_{000}(q) = \frac{a^4_0\pi^3}{6}  (d_1+d_2) \!\  q^2  \Big( 2 -  (\eta_2)_{\lambda\lambda}\hat{q}^2_{\lambda} - \big((\eta_2)_{\lambda\lambda}\hat{q}^2_{\lambda}\big)^2 \Big)  \!\ G_0(q) G_0(q).  \label{Eq:IV-C-22}
\end{align}

\section{Integrals of Eq.~(\ref{Eq:IV-E-13})}
Here we outline an integral calculation of Eq.~(\ref{Eq:IV-E-13}) as a memorandum,  
\begin{align}
&\big\langle \Omega \cdot Y_0 (\Omega)\cdot \Omega \big\rangle = \int_{S_2} d^2 \Omega \int \frac{d^3q}{(2\pi)^3} e^{iq \Omega-\alpha |q|} \frac{1}{q^2}\frac{A_0}{(\hat{q}^2_{\perp}+ b_0 \hat{q}^2_0)^2} \big(\Omega \cdot \eta_2 \cdot \hat{q}\big)^2. \label{Eq:C-1}
\end{align}
An infinitesimally small and positive $\alpha$ in the right-hand side controls the 
$q$-integral in the UV regime. The convergence factor is taken to zero in the end. With  $\Omega \equiv (\sin\varphi\cos\psi,\sin\varphi \sin\psi,\cos\varphi)$, $q  \equiv |q|(\sin\theta\cos\phi,\sin\theta\sin\phi,\cos\theta) \equiv |q| \hat{q}$,  integrals over $|q|$ and $\psi-\phi$ yield,
\begin{align}
 \big\langle \Omega \cdot Y_0 (\Omega)\cdot \Omega \big\rangle = & -\frac{A_0}{(2\pi)^2} 
\int^1_{-1} d(\cos\varphi) \int^1_{-1} d(\cos\theta) \frac{1}{(\sin^2 \theta + b_0 \cos^2\theta)^2} \nonumber \\
& \times \int^{2\pi}_{0} d\phi \frac{4\cos^2\theta \cos^2\varphi - 4\cos\theta\sin\theta \cos\varphi\sin\varphi \cos\phi + \sin^2\theta \sin^2\varphi \cos^2\phi}{i(\cos\theta\cos\varphi + \sin\theta\sin\varphi\cos\phi)-\alpha}. \label{Eq:C-2}
\end{align}
The right-hand side after the $\theta$-integral is an even function in $t \equiv \cos\varphi$. Thus, we consider only $0<t<1$ ($0<\varphi<\frac{\pi}{2}$). For the last term in the integrand, we integrate over $\phi$ first and then over $\theta$,  
\begin{align}
&\lim_{\alpha \rightarrow +0} 
\int^{1}_{-1} d(\cos\theta) \frac{1}{(\sin^2\theta + b_0 \cos^2\theta)^2} \int^{2\pi}_{0} d\phi \!\ \frac{ \sin^2\theta \sin^2\varphi \cos^2\phi}{i(\cos\theta\cos\varphi + \sin\theta\sin\varphi\cos\phi)-\alpha} \nonumber \\
 & =
 \lim_{\alpha\rightarrow +0} \frac{\sin\varphi}{2i} \int^{+\infty}_{-\infty} ds 
 \frac{1}{(1+b_0 s^2)^2}  \oint \frac{dz}{i} \frac{z^2+z^{-2}+2}{z^2+ 2(X+iY) z + 1} \nonumber \\
 & = - \frac{4\pi \sin^2\varphi}{\cos\varphi} \int^1_{0} dX  \frac{1}{(1+b_0 \tan^2 \varphi \!\   \!\ X^2)^2}  \frac{X^2}{\sqrt{1-X^2}} = - \pi^2 \frac{\sin^2\varphi}{\cos\varphi} \frac{1}{\sqrt{(1+b_0 \tan^2\varphi)^3}}, \label{Eq:C-3}
\end{align}
with $s\equiv \cot\theta$, $z\equiv e^{i\phi}$, $X\equiv s \cot \varphi$, $Y \equiv \alpha\!\ (\sqrt{1+s^2} /\sin\varphi)$. In the $z$-integral along the unit circle in the second line,  the pole at $z=0$ does not contribute to the integral, while poles at 
$z=z_{\pm} \equiv -(X+iY) \pm \sqrt{|W|} e^{\frac{i}{2}\arg W}$ contribute to the integral for $0<s<\tan\varphi$ and for $-\tan\varphi<s<0$, respectively. Here $W \equiv (X+iY)^2-1 \simeq X^2-1 + 2iXY$.  We take the same integrals for the other two in a similar way,
\begin{align}
&\lim_{\alpha \rightarrow +0} 
\int^{1}_{-1} d(\cos\theta) \frac{1}{\sin^2\theta + b_0 \cos^2\theta}  \int^{2\pi}_{0} d\phi \!\ \frac{ 4\cos^2\theta \cos^2\varphi -4\cos\theta\sin\theta\cos\varphi\sin\varphi\cos\phi}{i(\cos\theta\cos\varphi + \sin\theta\sin\varphi\cos\phi)-\alpha} \nonumber \\
 &= - 8\pi^2 \frac{\sin^2\varphi}{\cos\varphi} \frac{1}{\sqrt{(1+b_0 \tan^2\varphi)^3}} \label{Eq:C-4}
\end{align}
Finally, we take an integral over $t=\cos\varphi$, and obtain, 
\begin{align}
&\big\langle \Omega \cdot Y_0 (\Omega)\cdot \Omega \big\rangle = \frac{9A_0}{4} 
\int^1_{0} dt \frac{t^2(1-t^2)}{\sqrt{(b_0 -  (b_0-1) t^2)^3}}  \nonumber \\ 
&= \begin{cases}
 \frac{9A_0}{4(b_0-1)^2}\bigg(-5 + 8\sqrt{b_0} + \frac{2-5b_0}{\sqrt{1-b_0}}
 {\rm ArcSinh}\Big[\sqrt{\frac{1-b_0}{b_0}}\Big]\bigg) &  0< b_0 < 1,  \\  
 \frac{9A_0}{4(b_0-1)^2}\bigg(-5 + 8\sqrt{b_0} + \frac{2-5b_0}{\sqrt{-1+b_0}}{\rm ArcSin}\Big[\sqrt{\frac{b_0-1}{b_0}}\Big]\bigg)  &   1< b_0.
 \end{cases}
\end{align}
\end{widetext}
\bibliography{ref.bib}

\begin{thebibliography}{59}%
\makeatletter
\providecommand \@ifxundefined [1]{%
 \@ifx{#1\undefined}
}%
\providecommand \@ifnum [1]{%
 \ifnum #1\expandafter \@firstoftwo
 \else \expandafter \@secondoftwo
 \fi
}%
\providecommand \@ifx [1]{%
 \ifx #1\expandafter \@firstoftwo
 \else \expandafter \@secondoftwo
 \fi
}%
\providecommand \natexlab [1]{#1}%
\providecommand \enquote  [1]{``#1''}%
\providecommand \bibnamefont  [1]{#1}%
\providecommand \bibfnamefont [1]{#1}%
\providecommand \citenamefont [1]{#1}%
\providecommand \href@noop [0]{\@secondoftwo}%
\providecommand \href [0]{\begingroup \@sanitize@url \@href}%
\providecommand \@href[1]{\@@startlink{#1}\@@href}%
\providecommand \@@href[1]{\endgroup#1\@@endlink}%
\providecommand \@sanitize@url [0]{\catcode `\\12\catcode `\$12\catcode `\&12\catcode `\#12\catcode `\^12\catcode `\_12\catcode `\%12\relax}%
\providecommand \@@startlink[1]{}%
\providecommand \@@endlink[0]{}%
\providecommand \url  [0]{\begingroup\@sanitize@url \@url }%
\providecommand \@url [1]{\endgroup\@href {#1}{\urlprefix }}%
\providecommand \urlprefix  [0]{URL }%
\providecommand \Eprint [0]{\href }%
\providecommand \doibase [0]{https://doi.org/}%
\providecommand \selectlanguage [0]{\@gobble}%
\providecommand \bibinfo  [0]{\@secondoftwo}%
\providecommand \bibfield  [0]{\@secondoftwo}%
\providecommand \translation [1]{[#1]}%
\providecommand \BibitemOpen [0]{}%
\providecommand \bibitemStop [0]{}%
\providecommand \bibitemNoStop [0]{.\EOS\space}%
\providecommand \EOS [0]{\spacefactor3000\relax}%
\providecommand \BibitemShut  [1]{\csname bibitem#1\endcsname}%
\let\auto@bib@innerbib\@empty
\bibitem [{\citenamefont {Popov}(1973)}]{popov1973}%
  \BibitemOpen
  \bibfield  {author} {\bibinfo {author} {\bibfnamefont {V.~N.}\ \bibnamefont {Popov}},\ }\bibfield  {title} {\bibinfo {title} {Quantum vortices and phase transitions in bose systems},\ }\href@noop {} {\bibfield  {journal} {\bibinfo  {journal} {Soviet Physics -- JETP}\ }\textbf {\bibinfo {volume} {37}},\ \bibinfo {pages} {341} (\bibinfo {year} {1973})}\BibitemShut {NoStop}%
\bibitem [{\citenamefont {Wiegel}(1973)}]{wiegel1973}%
  \BibitemOpen
  \bibfield  {author} {\bibinfo {author} {\bibfnamefont {F.}~\bibnamefont {Wiegel}},\ }\bibfield  {title} {\bibinfo {title} {Vortex-ring model of bose condensation},\ }\href {https://doi.org/https://doi.org/10.1016/0031-8914(73)90348-0} {\bibfield  {journal} {\bibinfo  {journal} {Physica}\ }\textbf {\bibinfo {volume} {65}},\ \bibinfo {pages} {321} (\bibinfo {year} {1973})}\BibitemShut {NoStop}%
\bibitem [{\citenamefont {Banks}\ \emph {et~al.}(1977)\citenamefont {Banks}, \citenamefont {Myerson},\ and\ \citenamefont {Kogut}}]{banks1977}%
  \BibitemOpen
  \bibfield  {author} {\bibinfo {author} {\bibfnamefont {T.}~\bibnamefont {Banks}}, \bibinfo {author} {\bibfnamefont {R.}~\bibnamefont {Myerson}},\ and\ \bibinfo {author} {\bibfnamefont {J.}~\bibnamefont {Kogut}},\ }\bibfield  {title} {\bibinfo {title} {Phase transitions in abelian lattice gauge theories},\ }\href {https://doi.org/https://doi.org/10.1016/0550-3213(77)90129-8} {\bibfield  {journal} {\bibinfo  {journal} {Nuclear Physics B}\ }\textbf {\bibinfo {volume} {129}},\ \bibinfo {pages} {493} (\bibinfo {year} {1977})}\BibitemShut {NoStop}%
\bibitem [{\citenamefont {Nelson}\ and\ \citenamefont {Toner}(1981)}]{nelson1981}%
  \BibitemOpen
  \bibfield  {author} {\bibinfo {author} {\bibfnamefont {D.~R.}\ \bibnamefont {Nelson}}\ and\ \bibinfo {author} {\bibfnamefont {J.}~\bibnamefont {Toner}},\ }\bibfield  {title} {\bibinfo {title} {Bond-orientational order, dislocation loops, and melting of solids and smectic-$a$ liquid crystals},\ }\href {https://doi.org/10.1103/PhysRevB.24.363} {\bibfield  {journal} {\bibinfo  {journal} {Phys. Rev. B}\ }\textbf {\bibinfo {volume} {24}},\ \bibinfo {pages} {363} (\bibinfo {year} {1981})}\BibitemShut {NoStop}%
\bibitem [{\citenamefont {Williams}(1987)}]{williams1987}%
  \BibitemOpen
  \bibfield  {author} {\bibinfo {author} {\bibfnamefont {G.~A.}\ \bibnamefont {Williams}},\ }\bibfield  {title} {\bibinfo {title} {Vortex-ring model of the superfluid \ensuremath{\lambda} transition},\ }\href {https://doi.org/10.1103/PhysRevLett.59.1926} {\bibfield  {journal} {\bibinfo  {journal} {Phys. Rev. Lett.}\ }\textbf {\bibinfo {volume} {59}},\ \bibinfo {pages} {1926} (\bibinfo {year} {1987})}\BibitemShut {NoStop}%
\bibitem [{\citenamefont {Einhorn}\ and\ \citenamefont {Savit}(1978)}]{einhorn1978}%
  \BibitemOpen
  \bibfield  {author} {\bibinfo {author} {\bibfnamefont {M.~B.}\ \bibnamefont {Einhorn}}\ and\ \bibinfo {author} {\bibfnamefont {R.}~\bibnamefont {Savit}},\ }\bibfield  {title} {\bibinfo {title} {Topological excitations in the abelian higgs model},\ }\href {https://doi.org/10.1103/PhysRevD.17.2583} {\bibfield  {journal} {\bibinfo  {journal} {Phys. Rev. D}\ }\textbf {\bibinfo {volume} {17}},\ \bibinfo {pages} {2583} (\bibinfo {year} {1978})}\BibitemShut {NoStop}%
\bibitem [{\citenamefont {Peskin}(1978)}]{peskin1978}%
  \BibitemOpen
  \bibfield  {author} {\bibinfo {author} {\bibfnamefont {M.~E.}\ \bibnamefont {Peskin}},\ }\bibfield  {title} {\bibinfo {title} {Mandelstam-'t hooft duality in abelian lattice models},\ }\href {https://doi.org/https://doi.org/10.1016/0003-4916(78)90252-X} {\bibfield  {journal} {\bibinfo  {journal} {Annals of Physics}\ }\textbf {\bibinfo {volume} {113}},\ \bibinfo {pages} {122} (\bibinfo {year} {1978})}\BibitemShut {NoStop}%
\bibitem [{\citenamefont {Savit}(1980)}]{savit1980}%
  \BibitemOpen
  \bibfield  {author} {\bibinfo {author} {\bibfnamefont {R.}~\bibnamefont {Savit}},\ }\bibfield  {title} {\bibinfo {title} {Duality in field theory and statistical systems},\ }\href {https://doi.org/10.1103/RevModPhys.52.453} {\bibfield  {journal} {\bibinfo  {journal} {Rev. Mod. Phys.}\ }\textbf {\bibinfo {volume} {52}},\ \bibinfo {pages} {453} (\bibinfo {year} {1980})}\BibitemShut {NoStop}%
\bibitem [{\citenamefont {Dasgupta}\ and\ \citenamefont {Halperin}(1981)}]{dasgupta1981}%
  \BibitemOpen
  \bibfield  {author} {\bibinfo {author} {\bibfnamefont {C.}~\bibnamefont {Dasgupta}}\ and\ \bibinfo {author} {\bibfnamefont {B.~I.}\ \bibnamefont {Halperin}},\ }\bibfield  {title} {\bibinfo {title} {Phase transition in a lattice model of superconductivity},\ }\href {https://doi.org/10.1103/PhysRevLett.47.1556} {\bibfield  {journal} {\bibinfo  {journal} {Phys. Rev. Lett.}\ }\textbf {\bibinfo {volume} {47}},\ \bibinfo {pages} {1556} (\bibinfo {year} {1981})}\BibitemShut {NoStop}%
\bibitem [{\citenamefont {Blatter}\ \emph {et~al.}(1994)\citenamefont {Blatter}, \citenamefont {Feigel'man}, \citenamefont {Geshkenbein}, \citenamefont {Larkin},\ and\ \citenamefont {Vinokur}}]{blatter1994}%
  \BibitemOpen
  \bibfield  {author} {\bibinfo {author} {\bibfnamefont {G.}~\bibnamefont {Blatter}}, \bibinfo {author} {\bibfnamefont {M.~V.}\ \bibnamefont {Feigel'man}}, \bibinfo {author} {\bibfnamefont {V.~B.}\ \bibnamefont {Geshkenbein}}, \bibinfo {author} {\bibfnamefont {A.~I.}\ \bibnamefont {Larkin}},\ and\ \bibinfo {author} {\bibfnamefont {V.~M.}\ \bibnamefont {Vinokur}},\ }\bibfield  {title} {\bibinfo {title} {Vortices in high-temperature superconductors},\ }\href {https://doi.org/10.1103/RevModPhys.66.1125} {\bibfield  {journal} {\bibinfo  {journal} {Rev. Mod. Phys.}\ }\textbf {\bibinfo {volume} {66}},\ \bibinfo {pages} {1125} (\bibinfo {year} {1994})}\BibitemShut {NoStop}%
\bibitem [{\citenamefont {Nguyen}\ and\ \citenamefont {Sudb\o{}}(1999)}]{nguyen1999}%
  \BibitemOpen
  \bibfield  {author} {\bibinfo {author} {\bibfnamefont {A.~K.}\ \bibnamefont {Nguyen}}\ and\ \bibinfo {author} {\bibfnamefont {A.}~\bibnamefont {Sudb\o{}}},\ }\bibfield  {title} {\bibinfo {title} {Topological phase fluctuations, amplitude fluctuations, and criticality in extreme type-ii superconductors},\ }\href {https://doi.org/10.1103/PhysRevB.60.15307} {\bibfield  {journal} {\bibinfo  {journal} {Phys. Rev. B}\ }\textbf {\bibinfo {volume} {60}},\ \bibinfo {pages} {15307} (\bibinfo {year} {1999})}\BibitemShut {NoStop}%
\bibitem [{\citenamefont {Nguyen}\ and\ \citenamefont {Sudb\o{}}(1998{\natexlab{a}})}]{nguyen1998a}%
  \BibitemOpen
  \bibfield  {author} {\bibinfo {author} {\bibfnamefont {A.~K.}\ \bibnamefont {Nguyen}}\ and\ \bibinfo {author} {\bibfnamefont {A.}~\bibnamefont {Sudb\o{}}},\ }\bibfield  {title} {\bibinfo {title} {Onsager loop transition and first-order flux-line lattice melting in high-${T}_{c}$ superconductors},\ }\href {https://doi.org/10.1103/PhysRevB.57.3123} {\bibfield  {journal} {\bibinfo  {journal} {Phys. Rev. B}\ }\textbf {\bibinfo {volume} {57}},\ \bibinfo {pages} {3123} (\bibinfo {year} {1998}{\natexlab{a}})}\BibitemShut {NoStop}%
\bibitem [{\citenamefont {Nguyen}\ and\ \citenamefont {Sudb\o{}}(1998{\natexlab{b}})}]{nguyen1998b}%
  \BibitemOpen
  \bibfield  {author} {\bibinfo {author} {\bibfnamefont {A.~K.}\ \bibnamefont {Nguyen}}\ and\ \bibinfo {author} {\bibfnamefont {A.}~\bibnamefont {Sudb\o{}}},\ }\bibfield  {title} {\bibinfo {title} {Phase coherence and the boson analogy of vortex liquids},\ }\href {https://doi.org/10.1103/PhysRevB.58.2802} {\bibfield  {journal} {\bibinfo  {journal} {Phys. Rev. B}\ }\textbf {\bibinfo {volume} {58}},\ \bibinfo {pages} {2802} (\bibinfo {year} {1998}{\natexlab{b}})}\BibitemShut {NoStop}%
\bibitem [{\citenamefont {Ryu}\ and\ \citenamefont {Stroud}(1998)}]{ryu1998}%
  \BibitemOpen
  \bibfield  {author} {\bibinfo {author} {\bibfnamefont {S.}~\bibnamefont {Ryu}}\ and\ \bibinfo {author} {\bibfnamefont {D.}~\bibnamefont {Stroud}},\ }\bibfield  {title} {\bibinfo {title} {Nature of the low-field transition in the mixed state of high-temperature superconductors},\ }\href {https://doi.org/10.1103/PhysRevB.57.14476} {\bibfield  {journal} {\bibinfo  {journal} {Phys. Rev. B}\ }\textbf {\bibinfo {volume} {57}},\ \bibinfo {pages} {14476} (\bibinfo {year} {1998})}\BibitemShut {NoStop}%
\bibitem [{\citenamefont {Gade}(1993)}]{gadeAndersonLocalizationSublattice1993}%
  \BibitemOpen
  \bibfield  {author} {\bibinfo {author} {\bibfnamefont {R.}~\bibnamefont {Gade}},\ }\bibfield  {title} {\bibinfo {title} {Anderson localization for sublattice models},\ }\href {https://doi.org/10.1016/0550-3213(93)90601-K} {\bibfield  {journal} {\bibinfo  {journal} {Nuclear Physics B}\ }\textbf {\bibinfo {volume} {398}},\ \bibinfo {pages} {499} (\bibinfo {year} {1993})}\BibitemShut {NoStop}%
\bibitem [{\citenamefont {Gade}\ and\ \citenamefont {Wegner}(1991)}]{gadeReplicaLimitModels1991}%
  \BibitemOpen
  \bibfield  {author} {\bibinfo {author} {\bibfnamefont {R.}~\bibnamefont {Gade}}\ and\ \bibinfo {author} {\bibfnamefont {F.}~\bibnamefont {Wegner}},\ }\bibfield  {title} {\bibinfo {title} {The $n = 0$ replica limit of {U}($n$) and {U}($n$){SO}($n$) models},\ }\href {https://doi.org/10.1016/0550-3213(91)90401-I} {\bibfield  {journal} {\bibinfo  {journal} {Nuclear Physics B}\ }\textbf {\bibinfo {volume} {360}},\ \bibinfo {pages} {213} (\bibinfo {year} {1991})}\BibitemShut {NoStop}%
\bibitem [{\citenamefont {K{\"o}nig}\ \emph {et~al.}(2012)\citenamefont {K{\"o}nig}, \citenamefont {Ostrovsky}, \citenamefont {Protopopov},\ and\ \citenamefont {Mirlin}}]{konigMetalinsulatorTransitionTwodimensional2012}%
  \BibitemOpen
  \bibfield  {author} {\bibinfo {author} {\bibfnamefont {E.~J.}\ \bibnamefont {K{\"o}nig}}, \bibinfo {author} {\bibfnamefont {P.~M.}\ \bibnamefont {Ostrovsky}}, \bibinfo {author} {\bibfnamefont {I.~V.}\ \bibnamefont {Protopopov}},\ and\ \bibinfo {author} {\bibfnamefont {A.~D.}\ \bibnamefont {Mirlin}},\ }\bibfield  {title} {\bibinfo {title} {Metal-insulator transition in two-dimensional random fermion systems of chiral symmetry classes},\ }\href {https://doi.org/10.1103/PhysRevB.85.195130} {\bibfield  {journal} {\bibinfo  {journal} {Physical Review B}\ }\textbf {\bibinfo {volume} {85}},\ \bibinfo {pages} {195130} (\bibinfo {year} {2012})}\BibitemShut {NoStop}%
\bibitem [{\citenamefont {Altland}\ and\ \citenamefont {Merkt}(2001)}]{altlandSpectralTransportProperties2001}%
  \BibitemOpen
  \bibfield  {author} {\bibinfo {author} {\bibfnamefont {A.}~\bibnamefont {Altland}}\ and\ \bibinfo {author} {\bibfnamefont {R.}~\bibnamefont {Merkt}},\ }\bibfield  {title} {\bibinfo {title} {Spectral and transport properties of quantum wires with bond disorder},\ }\href {https://doi.org/10.1016/S0550-3213(01)00209-7} {\bibfield  {journal} {\bibinfo  {journal} {Nuclear Physics B}\ }\textbf {\bibinfo {volume} {607}},\ \bibinfo {pages} {511} (\bibinfo {year} {2001})}\BibitemShut {NoStop}%
\bibitem [{\citenamefont {Altland}\ \emph {et~al.}(2014)\citenamefont {Altland}, \citenamefont {Bagrets}, \citenamefont {Fritz}, \citenamefont {Kamenev},\ and\ \citenamefont {Schmiedt}}]{altlandQuantumCriticalityQuasiOneDimensional2014}%
  \BibitemOpen
  \bibfield  {author} {\bibinfo {author} {\bibfnamefont {A.}~\bibnamefont {Altland}}, \bibinfo {author} {\bibfnamefont {D.}~\bibnamefont {Bagrets}}, \bibinfo {author} {\bibfnamefont {L.}~\bibnamefont {Fritz}}, \bibinfo {author} {\bibfnamefont {A.}~\bibnamefont {Kamenev}},\ and\ \bibinfo {author} {\bibfnamefont {H.}~\bibnamefont {Schmiedt}},\ }\bibfield  {title} {\bibinfo {title} {Quantum {{Criticality}} of {{Quasi-One-Dimensional Topological Anderson Insulators}}},\ }\href {https://doi.org/10.1103/PhysRevLett.112.206602} {\bibfield  {journal} {\bibinfo  {journal} {Physical Review Letters}\ }\textbf {\bibinfo {volume} {112}},\ \bibinfo {pages} {206602} (\bibinfo {year} {2014})}\BibitemShut {NoStop}%
\bibitem [{\citenamefont {Altland}\ \emph {et~al.}(2015)\citenamefont {Altland}, \citenamefont {Bagrets},\ and\ \citenamefont {Kamenev}}]{altlandTopologyAndersonLocalization2015}%
  \BibitemOpen
  \bibfield  {author} {\bibinfo {author} {\bibfnamefont {A.}~\bibnamefont {Altland}}, \bibinfo {author} {\bibfnamefont {D.}~\bibnamefont {Bagrets}},\ and\ \bibinfo {author} {\bibfnamefont {A.}~\bibnamefont {Kamenev}},\ }\bibfield  {title} {\bibinfo {title} {Topology versus {{Anderson}} localization: {{Nonperturbative}} solutions in one dimension},\ }\href {https://doi.org/10.1103/PhysRevB.91.085429} {\bibfield  {journal} {\bibinfo  {journal} {Physical Review B}\ }\textbf {\bibinfo {volume} {91}},\ \bibinfo {pages} {085429} (\bibinfo {year} {2015})}\BibitemShut {NoStop}%
\bibitem [{\citenamefont {Luo}\ \emph {et~al.}(2020)\citenamefont {Luo}, \citenamefont {Xu}, \citenamefont {Ohtsuki},\ and\ \citenamefont {Shindou}}]{luoCriticalBehaviorAnderson2020}%
  \BibitemOpen
  \bibfield  {author} {\bibinfo {author} {\bibfnamefont {X.}~\bibnamefont {Luo}}, \bibinfo {author} {\bibfnamefont {B.}~\bibnamefont {Xu}}, \bibinfo {author} {\bibfnamefont {T.}~\bibnamefont {Ohtsuki}},\ and\ \bibinfo {author} {\bibfnamefont {R.}~\bibnamefont {Shindou}},\ }\bibfield  {title} {\bibinfo {title} {Critical behavior of {{Anderson}} transitions in three-dimensional orthogonal classes with particle-hole symmetries},\ }\href {https://doi.org/10.1103/PhysRevB.101.020202} {\bibfield  {journal} {\bibinfo  {journal} {Physical Review B}\ }\textbf {\bibinfo {volume} {101}},\ \bibinfo {pages} {020202} (\bibinfo {year} {2020})}\BibitemShut {NoStop}%
\bibitem [{\citenamefont {Wang}\ \emph {et~al.}(2021)\citenamefont {Wang}, \citenamefont {Ohtsuki},\ and\ \citenamefont {Shindou}}]{wang2021}%
  \BibitemOpen
  \bibfield  {author} {\bibinfo {author} {\bibfnamefont {T.}~\bibnamefont {Wang}}, \bibinfo {author} {\bibfnamefont {T.}~\bibnamefont {Ohtsuki}},\ and\ \bibinfo {author} {\bibfnamefont {R.}~\bibnamefont {Shindou}},\ }\bibfield  {title} {\bibinfo {title} {Universality classes of the anderson transition in the three-dimensional symmetry classes aiii, bdi, c, d, and ci},\ }\href {https://doi.org/10.1103/PhysRevB.104.014206} {\bibfield  {journal} {\bibinfo  {journal} {Phys. Rev. B}\ }\textbf {\bibinfo {volume} {104}},\ \bibinfo {pages} {014206} (\bibinfo {year} {2021})}\BibitemShut {NoStop}%
\bibitem [{\citenamefont {Luo}\ \emph {et~al.}(2022)\citenamefont {Luo}, \citenamefont {Xiao}, \citenamefont {Kawabata}, \citenamefont {Ohtsuki},\ and\ \citenamefont {Shindou}}]{luoUnifyingAndersonTransitions2022}%
  \BibitemOpen
  \bibfield  {author} {\bibinfo {author} {\bibfnamefont {X.}~\bibnamefont {Luo}}, \bibinfo {author} {\bibfnamefont {Z.}~\bibnamefont {Xiao}}, \bibinfo {author} {\bibfnamefont {K.}~\bibnamefont {Kawabata}}, \bibinfo {author} {\bibfnamefont {T.}~\bibnamefont {Ohtsuki}},\ and\ \bibinfo {author} {\bibfnamefont {R.}~\bibnamefont {Shindou}},\ }\bibfield  {title} {\bibinfo {title} {Unifying the {{Anderson}} transitions in {{Hermitian}} and non-{{Hermitian}} systems},\ }\href {https://doi.org/10.1103/PhysRevResearch.4.L022035} {\bibfield  {journal} {\bibinfo  {journal} {Physical Review Research}\ }\textbf {\bibinfo {volume} {4}},\ \bibinfo {pages} {L022035} (\bibinfo {year} {2022})}\BibitemShut {NoStop}%
\bibitem [{\citenamefont {Karcher}\ \emph {et~al.}(2023{\natexlab{a}})\citenamefont {Karcher}, \citenamefont {Gruzberg},\ and\ \citenamefont {Mirlin}}]{karcher2023a}%
  \BibitemOpen
  \bibfield  {author} {\bibinfo {author} {\bibfnamefont {J.~F.}\ \bibnamefont {Karcher}}, \bibinfo {author} {\bibfnamefont {I.~A.}\ \bibnamefont {Gruzberg}},\ and\ \bibinfo {author} {\bibfnamefont {A.~D.}\ \bibnamefont {Mirlin}},\ }\bibfield  {title} {\bibinfo {title} {Metal-insulator transition in a two-dimensional system of chiral unitary class},\ }\href {https://doi.org/10.1103/PhysRevB.107.L020201} {\bibfield  {journal} {\bibinfo  {journal} {Phys. Rev. B}\ }\textbf {\bibinfo {volume} {107}},\ \bibinfo {pages} {L020201} (\bibinfo {year} {2023}{\natexlab{a}})}\BibitemShut {NoStop}%
\bibitem [{\citenamefont {Karcher}\ \emph {et~al.}(2023{\natexlab{b}})\citenamefont {Karcher}, \citenamefont {Gruzberg},\ and\ \citenamefont {Mirlin}}]{karcher2023b}%
  \BibitemOpen
  \bibfield  {author} {\bibinfo {author} {\bibfnamefont {J.~F.}\ \bibnamefont {Karcher}}, \bibinfo {author} {\bibfnamefont {I.~A.}\ \bibnamefont {Gruzberg}},\ and\ \bibinfo {author} {\bibfnamefont {A.~D.}\ \bibnamefont {Mirlin}},\ }\bibfield  {title} {\bibinfo {title} {Generalized multifractality in two-dimensional disordered systems of chiral symmetry classes},\ }\href {https://doi.org/10.1103/PhysRevB.107.104202} {\bibfield  {journal} {\bibinfo  {journal} {Phys. Rev. B}\ }\textbf {\bibinfo {volume} {107}},\ \bibinfo {pages} {104202} (\bibinfo {year} {2023}{\natexlab{b}})}\BibitemShut {NoStop}%
\bibitem [{\citenamefont {{Mondragon-Shem}}\ \emph {et~al.}(2014)\citenamefont {{Mondragon-Shem}}, \citenamefont {Hughes}, \citenamefont {Song},\ and\ \citenamefont {Prodan}}]{mondragon-shemTopologicalCriticalityChiralSymmetric2014}%
  \BibitemOpen
  \bibfield  {author} {\bibinfo {author} {\bibfnamefont {I.}~\bibnamefont {{Mondragon-Shem}}}, \bibinfo {author} {\bibfnamefont {T.~L.}\ \bibnamefont {Hughes}}, \bibinfo {author} {\bibfnamefont {J.}~\bibnamefont {Song}},\ and\ \bibinfo {author} {\bibfnamefont {E.}~\bibnamefont {Prodan}},\ }\bibfield  {title} {\bibinfo {title} {Topological {{Criticality}} in the {{Chiral-Symmetric AIII Class}} at {{Strong Disorder}}},\ }\href {https://doi.org/10.1103/PhysRevLett.113.046802} {\bibfield  {journal} {\bibinfo  {journal} {Physical Review Letters}\ }\textbf {\bibinfo {volume} {113}},\ \bibinfo {pages} {046802} (\bibinfo {year} {2014})}\BibitemShut {NoStop}%
\bibitem [{\citenamefont {Claes}\ and\ \citenamefont {Hughes}(2020)}]{claes2020}%
  \BibitemOpen
  \bibfield  {author} {\bibinfo {author} {\bibfnamefont {J.}~\bibnamefont {Claes}}\ and\ \bibinfo {author} {\bibfnamefont {T.~L.}\ \bibnamefont {Hughes}},\ }\bibfield  {title} {\bibinfo {title} {Disorder driven phase transitions in weak aiii topological insulators},\ }\href {https://doi.org/10.1103/PhysRevB.101.224201} {\bibfield  {journal} {\bibinfo  {journal} {Phys. Rev. B}\ }\textbf {\bibinfo {volume} {101}},\ \bibinfo {pages} {224201} (\bibinfo {year} {2020})}\BibitemShut {NoStop}%
\bibitem [{\citenamefont {Xiao}\ \emph {et~al.}(2023)\citenamefont {Xiao}, \citenamefont {Kawabata}, \citenamefont {Luo}, \citenamefont {Ohtsuki},\ and\ \citenamefont {Shindou}}]{xiaoAnisotropicTopologicalAnderson2023}%
  \BibitemOpen
  \bibfield  {author} {\bibinfo {author} {\bibfnamefont {Z.}~\bibnamefont {Xiao}}, \bibinfo {author} {\bibfnamefont {K.}~\bibnamefont {Kawabata}}, \bibinfo {author} {\bibfnamefont {X.}~\bibnamefont {Luo}}, \bibinfo {author} {\bibfnamefont {T.}~\bibnamefont {Ohtsuki}},\ and\ \bibinfo {author} {\bibfnamefont {R.}~\bibnamefont {Shindou}},\ }\bibfield  {title} {\bibinfo {title} {Anisotropic {{Topological Anderson Transitions}} in {{Chiral Symmetry Classes}}},\ }\href {https://doi.org/10.1103/PhysRevLett.131.056301} {\bibfield  {journal} {\bibinfo  {journal} {Physical Review Letters}\ }\textbf {\bibinfo {volume} {131}},\ \bibinfo {pages} {056301} (\bibinfo {year} {2023})}\BibitemShut {NoStop}%
\bibitem [{\citenamefont {Zhao}\ \emph {et~al.}(2024)\citenamefont {Zhao}, \citenamefont {Xiao}, \citenamefont {Zhang},\ and\ \citenamefont {Shindou}}]{pwz2024}%
  \BibitemOpen
  \bibfield  {author} {\bibinfo {author} {\bibfnamefont {P.}~\bibnamefont {Zhao}}, \bibinfo {author} {\bibfnamefont {Z.}~\bibnamefont {Xiao}}, \bibinfo {author} {\bibfnamefont {Y.}~\bibnamefont {Zhang}},\ and\ \bibinfo {author} {\bibfnamefont {R.}~\bibnamefont {Shindou}},\ }\bibfield  {title} {\bibinfo {title} {Topological effect on the anderson transition in chiral symmetry classes},\ }\href {https://doi.org/10.1103/PhysRevLett.133.226601} {\bibfield  {journal} {\bibinfo  {journal} {Phys. Rev. Lett.}\ }\textbf {\bibinfo {volume} {133}},\ \bibinfo {pages} {226601} (\bibinfo {year} {2024})}\BibitemShut {NoStop}%
\bibitem [{\citenamefont {Herbut}(2007)}]{herbutModernApproachCritical2007}%
  \BibitemOpen
  \bibfield  {author} {\bibinfo {author} {\bibfnamefont {I.}~\bibnamefont {Herbut}},\ }\href@noop {} {\emph {\bibinfo {title} {A Modern Approach to Critical Phenomena}}}\ (\bibinfo  {publisher} {{Cambridge University Press}},\ \bibinfo {year} {2007})\BibitemShut {NoStop}%
\bibitem [{\citenamefont {Tanaka}\ and\ \citenamefont {Takayoshi}(2015)}]{tanakaShortGuideTopological2015}%
  \BibitemOpen
  \bibfield  {author} {\bibinfo {author} {\bibfnamefont {A.}~\bibnamefont {Tanaka}}\ and\ \bibinfo {author} {\bibfnamefont {S.}~\bibnamefont {Takayoshi}},\ }\bibfield  {title} {\bibinfo {title} {A short guide to topological terms in the effective theories of condensed matter},\ }\href {https://doi.org/10.1088/1468-6996/16/1/014404} {\bibfield  {journal} {\bibinfo  {journal} {Science and Technology of Advanced Materials}\ }\textbf {\bibinfo {volume} {16}},\ \bibinfo {pages} {014404} (\bibinfo {year} {2015})}\BibitemShut {NoStop}%
\bibitem [{\citenamefont {Fisher}\ \emph {et~al.}(1989)\citenamefont {Fisher}, \citenamefont {Weichman}, \citenamefont {Grinstein},\ and\ \citenamefont {Fisher}}]{fisher1989}%
  \BibitemOpen
  \bibfield  {author} {\bibinfo {author} {\bibfnamefont {M.~P.~A.}\ \bibnamefont {Fisher}}, \bibinfo {author} {\bibfnamefont {P.~B.}\ \bibnamefont {Weichman}}, \bibinfo {author} {\bibfnamefont {G.}~\bibnamefont {Grinstein}},\ and\ \bibinfo {author} {\bibfnamefont {D.~S.}\ \bibnamefont {Fisher}},\ }\bibfield  {title} {\bibinfo {title} {Boson localization and the superfluid-insulator transition},\ }\href {https://doi.org/10.1103/PhysRevB.40.546} {\bibfield  {journal} {\bibinfo  {journal} {Phys. Rev. B}\ }\textbf {\bibinfo {volume} {40}},\ \bibinfo {pages} {546} (\bibinfo {year} {1989})}\BibitemShut {NoStop}%
\bibitem [{\citenamefont {Fallani}\ \emph {et~al.}(2007)\citenamefont {Fallani}, \citenamefont {Lye}, \citenamefont {Guarrera}, \citenamefont {Fort},\ and\ \citenamefont {Inguscio}}]{fallani2007}%
  \BibitemOpen
  \bibfield  {author} {\bibinfo {author} {\bibfnamefont {L.}~\bibnamefont {Fallani}}, \bibinfo {author} {\bibfnamefont {J.~E.}\ \bibnamefont {Lye}}, \bibinfo {author} {\bibfnamefont {V.}~\bibnamefont {Guarrera}}, \bibinfo {author} {\bibfnamefont {C.}~\bibnamefont {Fort}},\ and\ \bibinfo {author} {\bibfnamefont {M.}~\bibnamefont {Inguscio}},\ }\bibfield  {title} {\bibinfo {title} {Ultracold atoms in a disordered crystal of light: Towards a bose glass},\ }\href {https://doi.org/10.1103/PhysRevLett.98.130404} {\bibfield  {journal} {\bibinfo  {journal} {Phys. Rev. Lett.}\ }\textbf {\bibinfo {volume} {98}},\ \bibinfo {pages} {130404} (\bibinfo {year} {2007})}\BibitemShut {NoStop}%
\bibitem [{\citenamefont {yoon Choi}\ \emph {et~al.}(2016)\citenamefont {yoon Choi}, \citenamefont {Hild}, \citenamefont {Zeiher}, \citenamefont {Schauß}, \citenamefont {Rubio-Abadal}, \citenamefont {Yefsah}, \citenamefont {Khemani}, \citenamefont {Huse}, \citenamefont {Bloch},\ and\ \citenamefont {Gross}}]{choi2016}%
  \BibitemOpen
  \bibfield  {author} {\bibinfo {author} {\bibfnamefont {J.}~\bibnamefont {yoon Choi}}, \bibinfo {author} {\bibfnamefont {S.}~\bibnamefont {Hild}}, \bibinfo {author} {\bibfnamefont {J.}~\bibnamefont {Zeiher}}, \bibinfo {author} {\bibfnamefont {P.}~\bibnamefont {Schauß}}, \bibinfo {author} {\bibfnamefont {A.}~\bibnamefont {Rubio-Abadal}}, \bibinfo {author} {\bibfnamefont {T.}~\bibnamefont {Yefsah}}, \bibinfo {author} {\bibfnamefont {V.}~\bibnamefont {Khemani}}, \bibinfo {author} {\bibfnamefont {D.~A.}\ \bibnamefont {Huse}}, \bibinfo {author} {\bibfnamefont {I.}~\bibnamefont {Bloch}},\ and\ \bibinfo {author} {\bibfnamefont {C.}~\bibnamefont {Gross}},\ }\bibfield  {title} {\bibinfo {title} {Exploring the many-body localization transition in two dimensions},\ }\href {https://doi.org/10.1126/science.aaf8834} {\bibfield  {journal} {\bibinfo  {journal} {Science}\ }\textbf {\bibinfo {volume} {352}},\ \bibinfo {pages} {1547} (\bibinfo {year} {2016})}\BibitemShut {NoStop}%
\bibitem [{\citenamefont {Griffiths}(1969)}]{griffiths1969}%
  \BibitemOpen
  \bibfield  {author} {\bibinfo {author} {\bibfnamefont {R.~B.}\ \bibnamefont {Griffiths}},\ }\bibfield  {title} {\bibinfo {title} {Nonanalytic behavior above the critical point in a random ising ferromagnet},\ }\href {https://doi.org/10.1103/PhysRevLett.23.17} {\bibfield  {journal} {\bibinfo  {journal} {Phys. Rev. Lett.}\ }\textbf {\bibinfo {volume} {23}},\ \bibinfo {pages} {17} (\bibinfo {year} {1969})}\BibitemShut {NoStop}%
\bibitem [{\citenamefont {Bray}(1987)}]{bray1987}%
  \BibitemOpen
  \bibfield  {author} {\bibinfo {author} {\bibfnamefont {A.~J.}\ \bibnamefont {Bray}},\ }\bibfield  {title} {\bibinfo {title} {Nature of the griffiths phase},\ }\href {https://doi.org/10.1103/PhysRevLett.59.586} {\bibfield  {journal} {\bibinfo  {journal} {Phys. Rev. Lett.}\ }\textbf {\bibinfo {volume} {59}},\ \bibinfo {pages} {586} (\bibinfo {year} {1987})}\BibitemShut {NoStop}%
\bibitem [{\citenamefont {Fisher}(1992)}]{dsfisher1992}%
  \BibitemOpen
  \bibfield  {author} {\bibinfo {author} {\bibfnamefont {D.~S.}\ \bibnamefont {Fisher}},\ }\bibfield  {title} {\bibinfo {title} {Random transverse field ising spin chains},\ }\href {https://doi.org/10.1103/PhysRevLett.69.534} {\bibfield  {journal} {\bibinfo  {journal} {Phys. Rev. Lett.}\ }\textbf {\bibinfo {volume} {69}},\ \bibinfo {pages} {534} (\bibinfo {year} {1992})}\BibitemShut {NoStop}%
\bibitem [{\citenamefont {Vojta}(2006)}]{Vojta2006}%
  \BibitemOpen
  \bibfield  {author} {\bibinfo {author} {\bibfnamefont {T.}~\bibnamefont {Vojta}},\ }\bibfield  {title} {\bibinfo {title} {Rare region effects at classical, quantum and nonequilibrium phase transitions},\ }\href {https://doi.org/10.1088/0305-4470/39/22/R01} {\bibfield  {journal} {\bibinfo  {journal} {Journal of Physics A: Mathematical and General}\ }\textbf {\bibinfo {volume} {39}},\ \bibinfo {pages} {R143} (\bibinfo {year} {2006})}\BibitemShut {NoStop}%
\bibitem [{\citenamefont {Vazquez}\ \emph {et~al.}(2011)\citenamefont {Vazquez}, \citenamefont {Bonachela}, \citenamefont {L\'opez},\ and\ \citenamefont {Mu\~noz}}]{vazquez2011}%
  \BibitemOpen
  \bibfield  {author} {\bibinfo {author} {\bibfnamefont {F.}~\bibnamefont {Vazquez}}, \bibinfo {author} {\bibfnamefont {J.~A.}\ \bibnamefont {Bonachela}}, \bibinfo {author} {\bibfnamefont {C.}~\bibnamefont {L\'opez}},\ and\ \bibinfo {author} {\bibfnamefont {M.~A.}\ \bibnamefont {Mu\~noz}},\ }\bibfield  {title} {\bibinfo {title} {Temporal griffiths phases},\ }\href {https://doi.org/10.1103/PhysRevLett.106.235702} {\bibfield  {journal} {\bibinfo  {journal} {Phys. Rev. Lett.}\ }\textbf {\bibinfo {volume} {106}},\ \bibinfo {pages} {235702} (\bibinfo {year} {2011})}\BibitemShut {NoStop}%
\bibitem [{\citenamefont {Kiometzis}\ \emph {et~al.}(1995)\citenamefont {Kiometzis}, \citenamefont {Kleinert},\ and\ \citenamefont {Schakel}}]{kiometzis95}%
  \BibitemOpen
  \bibfield  {author} {\bibinfo {author} {\bibfnamefont {M.}~\bibnamefont {Kiometzis}}, \bibinfo {author} {\bibfnamefont {H.}~\bibnamefont {Kleinert}},\ and\ \bibinfo {author} {\bibfnamefont {A.~M.~J.}\ \bibnamefont {Schakel}},\ }\bibfield  {title} {\bibinfo {title} {Dual description of the superconducting phase transition},\ }\href {https://doi.org/https://doi.org/10.1002/prop.2190430803} {\bibfield  {journal} {\bibinfo  {journal} {Fortschritte der Physik/Progress of Physics}\ }\textbf {\bibinfo {volume} {43}},\ \bibinfo {pages} {697} (\bibinfo {year} {1995})}\BibitemShut {NoStop}%
\bibitem [{\citenamefont {Kiometzis}\ \emph {et~al.}(1994)\citenamefont {Kiometzis}, \citenamefont {Kleinert},\ and\ \citenamefont {Schakel}}]{kiometzis1994}%
  \BibitemOpen
  \bibfield  {author} {\bibinfo {author} {\bibfnamefont {M.}~\bibnamefont {Kiometzis}}, \bibinfo {author} {\bibfnamefont {H.}~\bibnamefont {Kleinert}},\ and\ \bibinfo {author} {\bibfnamefont {A.~M.~J.}\ \bibnamefont {Schakel}},\ }\bibfield  {title} {\bibinfo {title} {Critical exponents of the superconducting phase transition},\ }\href {https://doi.org/10.1103/PhysRevLett.73.1975} {\bibfield  {journal} {\bibinfo  {journal} {Phys. Rev. Lett.}\ }\textbf {\bibinfo {volume} {73}},\ \bibinfo {pages} {1975} (\bibinfo {year} {1994})}\BibitemShut {NoStop}%
\bibitem [{\citenamefont {Kosterlitz}\ and\ \citenamefont {Thouless}(1973)}]{kosterlitz1973}%
  \BibitemOpen
  \bibfield  {author} {\bibinfo {author} {\bibfnamefont {J.~M.}\ \bibnamefont {Kosterlitz}}\ and\ \bibinfo {author} {\bibfnamefont {D.~J.}\ \bibnamefont {Thouless}},\ }\bibfield  {title} {\bibinfo {title} {Ordering, metastability and phase transitions in two-dimensional systems},\ }\href {https://doi.org/10.1088/0022-3719/6/7/010} {\bibfield  {journal} {\bibinfo  {journal} {Journal of Physics C: Solid State Physics}\ }\textbf {\bibinfo {volume} {6}},\ \bibinfo {pages} {1181} (\bibinfo {year} {1973})}\BibitemShut {NoStop}%
\bibitem [{\citenamefont {Kosterlitz}(1974)}]{kosterlitz1974}%
  \BibitemOpen
  \bibfield  {author} {\bibinfo {author} {\bibfnamefont {J.~M.}\ \bibnamefont {Kosterlitz}},\ }\bibfield  {title} {\bibinfo {title} {The critical properties of the two-dimensional xy model},\ }\href {https://doi.org/10.1088/0022-3719/7/6/005} {\bibfield  {journal} {\bibinfo  {journal} {Journal of Physics C: Solid State Physics}\ }\textbf {\bibinfo {volume} {7}},\ \bibinfo {pages} {1046} (\bibinfo {year} {1974})}\BibitemShut {NoStop}%
\bibitem [{\citenamefont {Berezinskii}(1971)}]{Berezinskii71}%
  \BibitemOpen
  \bibfield  {author} {\bibinfo {author} {\bibfnamefont {V.~L.}\ \bibnamefont {Berezinskii}},\ }\bibfield  {title} {\bibinfo {title} {Destruction of long-range order in one-dimensional and two-dimensional systems having a continuous symmetry group i. classical systems},\ }\href {https://jetp.ras.ru/cgi-bin/dn/e_032_03_0493.pdf} {\bibfield  {journal} {\bibinfo  {journal} {Soviet Journal of Experimental and Theoretical Physics}\ }\textbf {\bibinfo {volume} {32}},\ \bibinfo {pages} {493} (\bibinfo {year} {1971})}\BibitemShut {NoStop}%
\bibitem [{\citenamefont {Berezinskii}(1972)}]{Berezinskii72}%
  \BibitemOpen
  \bibfield  {author} {\bibinfo {author} {\bibfnamefont {V.~L.}\ \bibnamefont {Berezinskii}},\ }\bibfield  {title} {\bibinfo {title} {Destruction of long-range order in one-dimensional and two-dimensional systems having a continuous symmetry group ii. classical systems},\ }\href {https://jetp.ras.ru/cgi-bin/dn/e_032_03_0493.pdf} {\bibfield  {journal} {\bibinfo  {journal} {Soviet Journal of Experimental and Theoretical Physics}\ }\textbf {\bibinfo {volume} {34}},\ \bibinfo {pages} {610} (\bibinfo {year} {1972})}\BibitemShut {NoStop}%
\bibitem [{\citenamefont {Savit}(1978)}]{savit1978}%
  \BibitemOpen
  \bibfield  {author} {\bibinfo {author} {\bibfnamefont {R.}~\bibnamefont {Savit}},\ }\bibfield  {title} {\bibinfo {title} {Vortices and the low-temperature structure of the $x\ensuremath{-}y$ model},\ }\href {https://doi.org/10.1103/PhysRevB.17.1340} {\bibfield  {journal} {\bibinfo  {journal} {Phys. Rev. B}\ }\textbf {\bibinfo {volume} {17}},\ \bibinfo {pages} {1340} (\bibinfo {year} {1978})}\BibitemShut {NoStop}%
\bibitem [{\citenamefont {Shenoy}(1989)}]{shenoy1989}%
  \BibitemOpen
  \bibfield  {author} {\bibinfo {author} {\bibfnamefont {S.~R.}\ \bibnamefont {Shenoy}},\ }\bibfield  {title} {\bibinfo {title} {Vortex-loop scaling in the three-dimensional xy ferromagnet},\ }\href {https://doi.org/10.1103/PhysRevB.40.5056} {\bibfield  {journal} {\bibinfo  {journal} {Phys. Rev. B}\ }\textbf {\bibinfo {volume} {40}},\ \bibinfo {pages} {5056} (\bibinfo {year} {1989})}\BibitemShut {NoStop}%
\bibitem [{\citenamefont {Fradkin}\ \emph {et~al.}(1978)\citenamefont {Fradkin}, \citenamefont {Huberman},\ and\ \citenamefont {Shenker}}]{fradkin1978}%
  \BibitemOpen
  \bibfield  {author} {\bibinfo {author} {\bibfnamefont {E.}~\bibnamefont {Fradkin}}, \bibinfo {author} {\bibfnamefont {B.~A.}\ \bibnamefont {Huberman}},\ and\ \bibinfo {author} {\bibfnamefont {S.~H.}\ \bibnamefont {Shenker}},\ }\bibfield  {title} {\bibinfo {title} {Gauge symmetries in random magnetic systems},\ }\href {https://doi.org/10.1103/PhysRevB.18.4789} {\bibfield  {journal} {\bibinfo  {journal} {Phys. Rev. B}\ }\textbf {\bibinfo {volume} {18}},\ \bibinfo {pages} {4789} (\bibinfo {year} {1978})}\BibitemShut {NoStop}%
\bibitem [{\citenamefont {Jos\'e}\ \emph {et~al.}(1977)\citenamefont {Jos\'e}, \citenamefont {Kadanoff}, \citenamefont {Kirkpatrick},\ and\ \citenamefont {Nelson}}]{jose1977}%
  \BibitemOpen
  \bibfield  {author} {\bibinfo {author} {\bibfnamefont {J.~V.}\ \bibnamefont {Jos\'e}}, \bibinfo {author} {\bibfnamefont {L.~P.}\ \bibnamefont {Kadanoff}}, \bibinfo {author} {\bibfnamefont {S.}~\bibnamefont {Kirkpatrick}},\ and\ \bibinfo {author} {\bibfnamefont {D.~R.}\ \bibnamefont {Nelson}},\ }\bibfield  {title} {\bibinfo {title} {Renormalization, vortices, and symmetry-breaking perturbations in the two-dimensional planar model},\ }\href {https://doi.org/10.1103/PhysRevB.16.1217} {\bibfield  {journal} {\bibinfo  {journal} {Phys. Rev. B}\ }\textbf {\bibinfo {volume} {16}},\ \bibinfo {pages} {1217} (\bibinfo {year} {1977})}\BibitemShut {NoStop}%
\bibitem [{\citenamefont {Cardy}(1996)}]{cardy1996}%
  \BibitemOpen
  \bibfield  {author} {\bibinfo {author} {\bibfnamefont {J.}~\bibnamefont {Cardy}},\ }\href@noop {} {\emph {\bibinfo {title} {Scaing and Renormalization in Statistical Physics}}}\ (\bibinfo  {publisher} {Cambridge University Press},\ \bibinfo {address} {Cambridge},\ \bibinfo {year} {1996})\BibitemShut {NoStop}%
\bibitem [{\citenamefont {Nelson}\ and\ \citenamefont {Fisher}(1977)}]{nelson1977}%
  \BibitemOpen
  \bibfield  {author} {\bibinfo {author} {\bibfnamefont {D.~R.}\ \bibnamefont {Nelson}}\ and\ \bibinfo {author} {\bibfnamefont {D.~S.}\ \bibnamefont {Fisher}},\ }\bibfield  {title} {\bibinfo {title} {Dynamics of classical $\mathrm{XY}$ spins in one and two dimensions},\ }\href {https://doi.org/10.1103/PhysRevB.16.4945} {\bibfield  {journal} {\bibinfo  {journal} {Phys. Rev. B}\ }\textbf {\bibinfo {volume} {16}},\ \bibinfo {pages} {4945} (\bibinfo {year} {1977})}\BibitemShut {NoStop}%
\bibitem [{\citenamefont {Williams}(1993)}]{williams1993}%
  \BibitemOpen
  \bibfield  {author} {\bibinfo {author} {\bibfnamefont {G.~A.}\ \bibnamefont {Williams}},\ }\bibfield  {title} {\bibinfo {title} {Vortex dynamics and superfluid relaxation near the $^{4}\mathrm{He}$ \ensuremath{\lambda} transition},\ }\href {https://doi.org/10.1103/PhysRevLett.71.392} {\bibfield  {journal} {\bibinfo  {journal} {Phys. Rev. Lett.}\ }\textbf {\bibinfo {volume} {71}},\ \bibinfo {pages} {392} (\bibinfo {year} {1993})}\BibitemShut {NoStop}%
\bibitem [{\citenamefont {Williams}(1999)}]{williams1999}%
  \BibitemOpen
  \bibfield  {author} {\bibinfo {author} {\bibfnamefont {G.~A.}\ \bibnamefont {Williams}},\ }\bibfield  {title} {\bibinfo {title} {Vortex-loop phase transitions in liquid helium, cosmic strings, and high- ${T}_{c}$ superconductors},\ }\href {https://doi.org/10.1103/PhysRevLett.82.1201} {\bibfield  {journal} {\bibinfo  {journal} {Phys. Rev. Lett.}\ }\textbf {\bibinfo {volume} {82}},\ \bibinfo {pages} {1201} (\bibinfo {year} {1999})}\BibitemShut {NoStop}%
\bibitem [{\citenamefont {Williams}(2004)}]{williams2004}%
  \BibitemOpen
  \bibfield  {author} {\bibinfo {author} {\bibfnamefont {G.~A.}\ \bibnamefont {Williams}},\ }\bibfield  {title} {\bibinfo {title} {Vortex fluctuations in the critical casimir effect of superfluid and superconducting films},\ }\href {https://doi.org/10.1103/PhysRevLett.92.197003} {\bibfield  {journal} {\bibinfo  {journal} {Phys. Rev. Lett.}\ }\textbf {\bibinfo {volume} {92}},\ \bibinfo {pages} {197003} (\bibinfo {year} {2004})}\BibitemShut {NoStop}%
\bibitem [{\citenamefont {Antunes}\ and\ \citenamefont {Bettencourt}(1998)}]{antunes1998a}%
  \BibitemOpen
  \bibfield  {author} {\bibinfo {author} {\bibfnamefont {N.~D.}\ \bibnamefont {Antunes}}\ and\ \bibinfo {author} {\bibfnamefont {L.~M.~A.}\ \bibnamefont {Bettencourt}},\ }\bibfield  {title} {\bibinfo {title} {The length distribution of vortex strings in $\mathit{U}(1)$ equilibrium scalar field theory},\ }\href {https://doi.org/10.1103/PhysRevLett.81.3083} {\bibfield  {journal} {\bibinfo  {journal} {Phys. Rev. Lett.}\ }\textbf {\bibinfo {volume} {81}},\ \bibinfo {pages} {3083} (\bibinfo {year} {1998})}\BibitemShut {NoStop}%
\bibitem [{\citenamefont {Antunes}\ \emph {et~al.}(1998)\citenamefont {Antunes}, \citenamefont {Bettencourt},\ and\ \citenamefont {Hindmarsh}}]{antunes1998b}%
  \BibitemOpen
  \bibfield  {author} {\bibinfo {author} {\bibfnamefont {N.~D.}\ \bibnamefont {Antunes}}, \bibinfo {author} {\bibfnamefont {L.~M.~A.}\ \bibnamefont {Bettencourt}},\ and\ \bibinfo {author} {\bibfnamefont {M.}~\bibnamefont {Hindmarsh}},\ }\bibfield  {title} {\bibinfo {title} {Thermodynamics of cosmic string densities in u(1) scalar field theory},\ }\href {https://doi.org/10.1103/PhysRevLett.80.908} {\bibfield  {journal} {\bibinfo  {journal} {Phys. Rev. Lett.}\ }\textbf {\bibinfo {volume} {80}},\ \bibinfo {pages} {908} (\bibinfo {year} {1998})}\BibitemShut {NoStop}%
\bibitem [{\citenamefont {Wilson}\ and\ \citenamefont {Fisher}(1972)}]{wilson1972}%
  \BibitemOpen
  \bibfield  {author} {\bibinfo {author} {\bibfnamefont {K.~G.}\ \bibnamefont {Wilson}}\ and\ \bibinfo {author} {\bibfnamefont {M.~E.}\ \bibnamefont {Fisher}},\ }\bibfield  {title} {\bibinfo {title} {Critical exponents in 3.99 dimensions},\ }\href {https://doi.org/10.1103/PhysRevLett.28.240} {\bibfield  {journal} {\bibinfo  {journal} {Phys. Rev. Lett.}\ }\textbf {\bibinfo {volume} {28}},\ \bibinfo {pages} {240} (\bibinfo {year} {1972})}\BibitemShut {NoStop}%
\bibitem [{\citenamefont {Wilson}\ and\ \citenamefont {Kogut}(1974)}]{wilson1974}%
  \BibitemOpen
  \bibfield  {author} {\bibinfo {author} {\bibfnamefont {K.~G.}\ \bibnamefont {Wilson}}\ and\ \bibinfo {author} {\bibfnamefont {J.}~\bibnamefont {Kogut}},\ }\bibfield  {title} {\bibinfo {title} {The renormalization group and the $\epsilon$ expansion},\ }\href {https://doi.org/https://doi.org/10.1016/0370-1573(74)90023-4} {\bibfield  {journal} {\bibinfo  {journal} {Physics Reports}\ }\textbf {\bibinfo {volume} {12}},\ \bibinfo {pages} {75} (\bibinfo {year} {1974})}\BibitemShut {NoStop}%
\bibitem [{\citenamefont {Jackson}(1999)}]{jackson1999}%
  \BibitemOpen
  \bibfield  {author} {\bibinfo {author} {\bibfnamefont {J.~D.}\ \bibnamefont {Jackson}},\ }\href@noop {} {\emph {\bibinfo {title} {classical Electrodynamics (Third Edition)}}}\ (\bibinfo  {publisher} {{Wiley}},\ \bibinfo {year} {1999})\BibitemShut {NoStop}%
\end{thebibliography}%

\end{document}